\documentclass[twocolumn,tighten]{aastex61}

\input epsf.tex

\begin{document}
\title{Galaxy populations in massive $z=0.2-0.9$ clusters: I. Analysis of spectroscopy}
\author{Inger J{\o}rgensen$^1$, Kristin Chiboucas}
\affil{Gemini Observatory, 670 N.\ A`ohoku Pl., Hilo, HI 96720, USA}
\author{Emily Berkson}
\affil{Harris Corporation, Rochester, NY, USA}
\author{Omega Smith}
\affil{Planetarium \& Visualization Theater, University of Alaska Anchorage, AK, USA}
\author{Marianne Takamiya}
\affil{University of Hawaii, Hilo, HI, USA}
\author{Alexa Villaume}
\affil{Department of Astronomy and Astrophysics, University of California, Santa Cruz, CA, USA}

\email{ijorgensen@gemini.edu, kchiboucas@gemini.edu, eberkson@harris.com,
osmith11@alaska.edu, takamiya@hawaii.edu, avillaum@ucsc.edu}

\accepted{October 25, 2017}
\submitjournal{Astronomical Journal}

\begin{abstract}
We present an analysis of stellar populations in passive galaxies in seven massive X-ray clusters at $z=0.19-0.89$.
Based on absorption line strengths measured from our high signal-to-noise spectra, the
data support primarily passive evolution of the galaxies.
We use the scaling relations between velocity dispersions and the absorption line strengths
to determine representative mean line strengths for the clusters. From the age determinations based
on the line strengths (and stellar population models), we find a formation redshift
of $z_{\rm form}=1.96_{-0.19}^{+0.24}$. 
Based on line strength measurements from high signal-to-noise composite spectra of our data, 
we establish the relations between velocity dispersion, ages, metallicities [M/H] and abundance ratios $\rm [\alpha /Fe]$
as a function of redshift.
The [M/H]--velocity dispersion and $\rm [\alpha /Fe]$--velocity dispersion relations are steep and tight.
The age--velocity dispersion relation is flat, with zero point changes reflecting passive evolution.
The scatter in all three parameters are within 0.08--0.15 dex at fixed
velocity dispersions, indicating a large degree of synchronization in the 
evolution of the galaxies.
We find indication of cluster-to-cluster differences in metallicities and abundance ratios.
However, variations in stellar populations with the cluster environment can only 
account for a very small fraction of the intrinsic scatter in the scaling relations.
Thus, within these very massive clusters the main driver of the properties of the stellar populations 
in passive galaxies appears to be the galaxy velocity dispersion.
\end{abstract}


\section{Introduction}

A fundamental question in many investigations of intermediate and high-redshift galaxies
is how these galaxies may evolve into their observed counter parts at lower redshifts
(e.g., Treu et al.\ 2005; Renzini 2006 and references therein; Faber et al.\ 2007; 
Saglia et al.\ 2010; Barro et al.\ 2014; Choi et al.\ 2014; J\o rgensen et al.\ 2014; Kriek et al.\ 2016).
Studies of galaxies in low redshift clusters are also often framed around the same question of 
how these galaxies came to have their current properties. As examples, Roediger et al.\ (2011) studied
stellar populations in the Virgo cluster based on deep photometry, while
McDermid et al.\ (2015) studied stellar populations of field and Virgo galaxies based
in the ATLAS-3D spectroscopic sample. Both studies aim to understand the 
evolutionary processes that lead to the properties of these local galaxies.

One of the main issues is to what extent all galaxies share a common 
evolutionary path and if so can we identify the main driver(s) of the evolution
and establish whether these are related to internal properties of the galaxies
or properties of the cluster environment (e.g., Mateus et al.\ 2007; Muzzin et al.\ 2012).
It is clear that the galaxy populations of the field and the dense cluster cores
are different as first established by Dressler (1980) in the now classical paper
on the morphology-density relation. It is also recognized that there is morphological
evolution as a function of redshift as spirals and irregulars are transformed
into passive bulge-dominated galaxies (Dressler et al.\ 1997; Smith et al.\ 2005).
However, if we focus on the already passive galaxies at all redshifts, 
the question remains to what extent are their properties driven by internal 
properties of the galaxies (e.g., mass or velocity dispersion, Muzzin et al.\ 2012; McDermid et al.\ 2015)
or by the environment in which they reside.

To properly compare galaxies at different redshifts
and use them as snapshots of a possible common evolutionary path, the high redshift
galaxies must be valid progenitors for the low redshift galaxies
(see van Dokkum \& Franx (2001) for a detailed discussion of progenitor bias). 
In addition, the 
environment of the high redshift galaxies must be a valid progenitor environment
for the environment of the low redshift galaxies.

Simulations predict that the masses of the clusters grow over time
(van den Bosch 2002; Fakhouri et al.\ 2010; Correa et al.\ 2015), a fact
that also has to be taken into account when attempting to study galaxies in intermediate
redshift clusters that are meant to be valid progenitor environments for the environments
found in low redshift massive clusters like the Coma and Perseus clusters.
Galaxies in the cores of massive clusters can be expected to remain in the
clusters as both the clusters and the galaxies evolve (e.g., Biviano \& Poggianti 2009).
Thus, by studying galaxies in cores of massive clusters we partially solve one of
the progenitor issues of tying together progenitors at high redshift with resulting 
galaxies and environments at lower redshift.
However, some galaxies found in the cores of low redshift clusters (or in projection) may have entered
the cluster later.
Thus, not all galaxies in the cores of low redshift clusters may be the product
of galaxies that were also in the cores of clusters at $z \sim 1$.

Once a galaxy is bulge-dominated it is likely to remain bulge-dominated,
as it is very difficult to disrupt a high-density bulge (Brooks \& Christensen 2016
and references therein).
Thus, bulge-dominated
high-redshift galaxies must be the progenitors of some of the bulge-dominated
galaxies at lower redshift. However, the progenitors of some 
low redshift bulge-dominated galaxies were most likely disk galaxies at higher redshifts
and only recently at $z<0.5$ became passive bulge-dominated galaxies on
the red sequence, see e.g., Saglia et al.\ (2010, EDisCS).

Scaling relations between absorption line strengths and velocity dispersions of passive bulge-dominated
galaxies offer tools to study the evolution with redshift of these galaxies
as well as how the stellar populations are related to their velocity dispersions (or masses).
Studies of local samples of passive galaxies show
tight correlations with velocity dispersions both for the line indices and 
the derived ages, metallicities [M/H] and abundance ratios $\rm [\alpha /Fe]$, 
see, e.g., J\o rgensen (1999) and Trager et al.\ (2000) 
for the first comprehensive results in this area.

\begin{deluxetable*}{llrrrrrr}
\tablecaption{Cluster Properties \label{tab-clusters} }
\tablewidth{0pt}
\tabletypesize{\scriptsize}
\tablehead{
\colhead{Cluster} & \colhead{Redshift} & \colhead{$\sigma _{\rm cluster}$} & 
\colhead{$L_{500}$} & \colhead{$M_{500}$} & \colhead{$R_{500}$} & \colhead{N$_{\rm member}$}  & \colhead{Ref.}\\
 & & $\rm km~s^{-1}$ & $10^{44} \rm{erg\,s^{-1}}$ & $10^{14}M_{\sun}$ & Mpc  & \\
\colhead{(1)} & \colhead{(2)} & \colhead{(3)} & \colhead{(4)} & \colhead{(5)} & \colhead{(6)} & \colhead{(7)} & \colhead{(8)}
}
\startdata
Perseus = Abell 426         & 0.0179 & $1277_{-78}^{+95}$ &   6.217 & 6.151 & 1.286 &  63  & Z1990 \\
Abell 194\tablenotemark{a}  & 0.0180 &  $480_{-38}^{+48}$ &   0.070 & 0.398 & 0.516 &  17  & Z1990 \\
Coma = Abell 1656           & 0.0231 & $1010_{-44}^{+51}$ &   3.456 & 4.285 & 1.138 & 116  & Z1990 \\
RXJ1311.4--0120 = Abell 1689 & $0.1865\pm 0.0010$ & $2182_{-163}^{+150}$ & 12.524 & 8.392 & 1.350 &  72 & J2017 \\ 
RXJ0056.2+2622 = Abell 115  & $0.1922\pm 0.0008$ & $1444_{-140}^{+119}$ &  7.485 & 6.068 & 1.206 &  58 & J2017 \\ 
RXJ0056.2+2622N\tablenotemark{b} & $0.1932\pm 0.0010$ & $1328_{-334}^{+213}$ &  3.935 & 4.100 & 1.058 &  12 & J2017 \\ 
RXJ0056.2+2622S\tablenotemark{b} & $0.1929\pm 0.0010$ & $1218_{-206}^{+164}$ &  4.094 & 4.200 & 1.067 &  22 & J2017 \\ 
RXJ0027.6+2616              & $0.3650\pm 0.0009$ & $1232_{-165}^{+122}$ &  8.376 & 5.684 & 1.108 &  34 & J2017\\  
RXJ0027.6+2616 group        & $0.3404\pm 0.0003$ & $172_{-47}^{+29} $ & \nodata & \nodata & \nodata & 9 & J2017\\  
RXJ1347.5-1145\tablenotemark{c} & $0.4507\pm 0.0008$ & $1259_{-250}^{+210} $ & 8.278 & 5.264 & 1.046 &  43 & J2017 \\ 
MS0451.6--0305              & $0.5398\pm 0.0010$ & $1450_{-159}^{+105}$ & 15.352 & 7.134 & 1.118 &  47  & J2013 \\
RXJ0152.7--1357             & $0.8350\pm 0.0012$ & $1110_{-174}^{+147}$ &  6.291 & 3.222 & 0.763 &  29 & J2005 \\
RXJ0152.7--1357N\tablenotemark{c} & $0.8372\pm 0.0014$ & $681 \pm 232$ &  1.933 & 1.567 & 0.599 &  7 & J2005 \\
RXJ0152.7--1357S\tablenotemark{c} & $0.8349\pm 0.0020$ & $866 \pm 266$ &  2.961 & 2.043 & 0.657 &  6 & J2005 \\
RXJ1226.9+3332              & $0.8908\pm 0.0011$ & $1298_{-137}^{+122}$ & 11.253 & 4.386 & 0.827 &  55 & J2013 \\
\enddata
\tablecomments{Column 1: Galaxy cluster. Column 2: Cluster redshift. Column 3: Cluster velocity dispersion.
Column 4: X-ray luminosity in the 0.1--2.4 keV band within the radius $R_{500}$, from Piffaretti et al.\ (2011), except as noted.
Column 5: Cluster mass derived from X-ray data within the radius $R_{500}$, from Piffaretti et al., except as noted.
Column 6: Radius within which the mean over-density of the cluster is 500 times the critical density at the 
cluster redshift, from Piffaretti et al., except as noted.
Column 7: Number of member galaxies for which spectroscopy is used in this paper.
Column 8: References for redshifts and velocity dispersions:
Z1990: Zabludoff et al.\ (1990).
J2005: J\o rgensen et al.\ (2005). 
J2013: J\o rgensen \& Chiboucas (2013).
J2017: This paper.
}
\tablenotetext{a}{Abell 194 does not meet the X-ray luminosity selection criteria of the main cluster sample.}
\tablenotetext{b}{Re-calibrated X-ray data from Mahdavi et al.\ 2013, 2014; see Section \ref{SEC-CLUSTERZ}.}
\tablenotetext{c}{Re-calibrated X-ray data from Ettori et al.\ 2004; see Section \ref{SEC-CLUSTERZ}.}
\end{deluxetable*}

More recent investigations of passive $z\approx 0$ galaxies have also been focused on 
determining possible correlations between ages, [M/H] and  $\rm [ \alpha /Fe]$ 
and the velocity dispersion of the galaxies as well as the scatter of 
these correlations, 
e.g., Thomas et al.\ (2005, 2010), Harrison et al.\ (2011) and references therein.
In particular, these studies find correlations between ages and velocity dispersions
with fairly steep slopes.
Kelson et al.\ (2006) presented one of the earliest attempts to determine such detailed
parameters for stellar populations at intermediate redshifts using measurements
of line strengths of passive galaxies in a $z=0.33$ cluster. 
These authors found that only the metallicities
were strongly correlated with the velocity dispersions, while ages and abundance ratios did
not depend on the velocity dispersions.
More recently, Choi et al.\ (2014) investigated this issue using stacked spectra
of a large sample of galaxies covering from $z=0.7$ to the present. Their results
show a rather weak correlation between galaxy masses and metallicities, while
[Mg/Fe] (or equivalently $\rm [\alpha /Fe]$) depends strongly on the galaxy mass.
The mean ages are also correlated with galaxy mass. However, the samples
are too shallow to detect possible correlations beyond $z=0.4$.
Thus, we lack consensus on whether these relations are already in place early on in the 
evolution of the galaxies, and also whether these relations evolve with redshift.
In particular, if the steep age-velocity dispersion relation found at $z\approx 0$ is
evolved ''backwards'' to higher redshift, under the assumption of passive evolution,
then the prediction is an even steeper relation in the past. It has not yet
been tested if such a steep relation exists, and if not how the higher redshift
data and the presence of the low redshift age-velocity dispersion relation
may be reconciled.

The present paper is part of our series of papers based on
the data from our project the ''Gemini/HST Galaxy Cluster Project'' (GCP).
The GCP was designed to study the evolution of the bulge-dominated passive galaxies in
very massive clusters. The project covers fourteen clusters spanning the redshift
interval from $z=0.2$ to $z=1.0$. 
Using the X-ray data from Piffaretti et al.\ (2011),
the luminosity limit for the sample is $L_{500} = 10^{44}\,{\rm erg\,s^{-1}}$ in the 0.1-2.4 keV band 
and within the radius $R_{500}$ for a $\Lambda$CDM cosmology with 
$\rm H_0 = 70\,km\,s^{-1}\,Mpc^{-1}$, $\Omega_{\rm M}=0.3$, and $\Omega_{\rm \Lambda}=0.7$.
The radius $R_{500}$ is the radius within which the average cluster over-density is 500 times
the critical density of the Universe at the redshift of the cluster.

For each cluster, we obtain high S/N spectra for 30-60 candidate cluster members. 
This usually gives data for at least 20 passive members in each cluster.
The spectra typically have a S/N per {\AA}ngstrom in the rest frame of 20--40 for the highest redshift
galaxies, while we reach S/N=50--200 for $z=0.2-0.6$ galaxies, sufficient to 
reliably measure velocity dispersions and absorption line strength for individual galaxies.
Our samples reach from the brightest cluster galaxies with typical dynamical masses 
of ${\rm Mass} \approx 10^{12.6}\, M_{\sun}$ to galaxies with 
dynamically masses of ${\rm Mass} \approx 10^{10.3} M_{\sun}$, equivalent
to a velocity dispersion of about $\rm 100\, km\,s^{-1}$.
The project data also include high spatial resolution imaging of the clusters, primarily
obtained with the Advanced Camera for Surveys (ACS) or the Wide Field and Planetary Camera 2 
(WFPC2) on board {\it Hubble Space Telescope} ({\it HST}).
See J\o rgensen et al.\ (2005) for a more complete description of the observing
strategy for the project.

\begin{deluxetable*}{lrr}
\tablecaption{Gemini Instrumentation\label{tab-inst} }
\tabletypesize{\scriptsize}
\tablehead{Parameter & GMOS-N & GMOS-S }
\startdata
CCDs            & 3 $\times$ E2V 2048$\times$4608 & 3 $\times$ E2V 2048$\times$4608 \\
r.o.n.\tablenotemark{a}          & (3.5,3.3,3.0) e$^-$  & (4.0,3.7,3.3) e$^-$      \\
gain\tablenotemark{a}            & (2.04,2.3,2.19) e$^-$/ADU  &  (2.33,2.07,2.07) e$^-$/ADU \\
Pixel scale     & 0.0727arcsec/pixel  & 0.073arcsec/pixel \\
Field of view   & $5\farcm5\times5\farcm5$ & $5\farcm5\times5\farcm5$ \\
Imaging filters  & $g', r', i'$  & $g', r', i'$\\
Gratings        & B600\_G5303  & \\
                & R400\_G5304  & R400\_G5324 \\
Spectroscopic filter & none & GG455\_G0329 \\
Wavelength range\tablenotemark{b} & 4000-8100\AA &  4550-8200\AA  \\
\enddata
\tablenotetext{a}{Values for the three detectors in the array.}
\tablenotetext{b}{The exact wavelength range varies from slitlet to slitlet.}
\end{deluxetable*}

In our previous papers (J\o rgensen et al.\ 2005; J\o rgensen \& Chiboucas 2013) 
we studied the scaling relations for three clusters MS0451.6--0305, RXJ0152.7--1357, and RXJ1226.9+3332
at $z=0.54-0.89$ and compared these clusters to our local reference samples.
However, at the time we lacked coverage between $z=0.5$ and the present.
We also did not attempt to investigate how ages, metallicities [M/H], and
abundance ratios $\rm [ \alpha /Fe]$
might depend on the velocity dispersions of the galaxies.
In this paper we present the joint analysis of the spectroscopy of the three $z=0.54-0.89$ clusters
and four clusters at $z=0.19-0.45$, for which we present new data. 
These seven clusters are the most massive in the GCP at each redshift.
We will return to the less massive GCP clusters in subsequent papers.
We establish the scaling relations for the seven clusters and investigate how these
change with redshift in terms of zero points and possibly scatter. We also test
for dependence on the cluster environments.
Based on composite spectra of the galaxies (stacked in bins by velocity dispersion),
we derive ages, [M/H] and $\rm [ \alpha /Fe]$.
We investigate the relations between these parameters and the velocity dispersions, 
and to what extent the data are consistent with passive evolution with a common formation redshift. 
(The formation redshift should be understood as the approximate epoch of th last major star formation episode.)
It is beyond the scope of the current paper to attempt an investigation of the detailed
star formation histories of the galaxies. 
We will return to this issue in a future paper on the GCP data.

The observational data, with emphasis on the new data for the $z=0.19-0.45$ clusters, are 
described in Section \ref{SEC-DATA} and in the Appendix. 
In Section \ref{SEC-CLUSTERZ} we outline the cluster X-ray data used in the analysis. 
We establish the cluster redshifts and velocity dispersions, and discuss the presence
of sub-structure in some of the clusters.
In Section \ref{SEC-METHODSAMPLE} we define the final sample of bulge-dominated passive galaxies used
in the analysis and give an overview of the methods and models used throughout the paper.
The composite spectra are also described in this section.
Our main results are described in Sections \ref{SEC-SCALING} and \ref{SEC-INDICES}.
We discuss the results in Section \ref{SEC-DISCUSSION}.
The conclusions are summarized in Section \ref{SEC-CONCLUSION}.

Throughout this paper we adopt a $\Lambda$CDM cosmology with 
$\rm H_0 = 70\,km\,s^{-1}\,Mpc^{-1}$, $\Omega_{\rm M}=0.3$, and $\Omega_{\rm \Lambda}=0.7$.

\begin{deluxetable*}{llrrrrrrr}
\tablecaption{GMOS-N and GMOS-S Spectroscopic Data \label{tab-spdata} }
\tablewidth{0pc}
\tabletypesize{\scriptsize}
\tablehead{
\colhead{Cluster} & \colhead{Program ID and dates} & \colhead{Exposure time} & \colhead{$N_{\rm exp}$} & \colhead{FWHM} & \colhead{$\sigma _{\rm inst}$} & \colhead{Aperture} & \colhead{Slit lengths} & \colhead{S/N} \\ 
\colhead{(1)}        & \colhead{(2)} & \colhead{(3)} & \colhead{(4)} & \colhead{(5)} & \colhead{(6)} & \colhead{(7)} & \colhead{(8)} & \colhead{(9)} }
\startdata
Abell 1689      & GN-2001B-Q-10 & 14,400,13,800 & 20 & 0.76, 0.80 & 1.492\AA, 84 $\rm km\,s^{-1}$ & $0.75\times 2.0$, 0.70 & 6.0--15.0 & 217 \\ 
                & UT 2002 Jan 11--16 & + 9960,12000 \\
RXJ0056.2+2622  & GN-2003B-Q-21 & 28,800 + 28,800 & 24 & 0.75, 0.66 & 1.647\AA, 110 $\rm km\,s^{-1}$ & $0.75\times 1.4$, 0.59 & 4.0--10.0 & 146 \\ 
                & UT 2003 Jul 27--Oct 10 \\
RXJ0027.6+2616  & GN-2003B-Q-21 & 28,800 + 28,800 & 24 & 0.66, 0.56 & 2.518\AA, 123 $\rm km\,s^{-1}$ & $0.75\times 1.4$, 0.59 & 5.0--8.1 & 155 \\  
                & UT 2003 Aug 26--Sep 26 \\
RXJ1347.5--1145 & GS-2005A-Q-27 & 24,000 + 24,000 & 20 & 0.72, 0.90 & 2.476\AA, 119 $\rm km\,s^{-1}$ & $0.75\times 1.4$, 0.59 & 4.8--10.6 & 57 \\ 
                &  UT 2005 Mar 7--May 9 \\
\enddata
\tablecomments{Column 1: Galaxy cluster. 
Column 2: Gemini program ID and dates of observations. Program IDs starting with GN and GS were obtained with GMOS-N and GMOS-S, respectively.
Column 3: Exposure times in seconds. 
Column 4: Number of useful exposures. 
Column 5: Image quality for each mask measured as the average full-width-half-maximum (FWHM) in arcsec of the blue stars,
or for Abell 1689 and RXJ0056.2+2622 of the acquisition stars, included in the masks. 
The measurement was done at a wavelength corresponding to 5000{\AA} in the rest frame of the clusters. 
Column 6: Median instrumental resolution derived as sigma in Gaussian fits to the sky lines of the stacked 
spectra. The second entry is the equivalent resolution in $\rm km\,s^{-1}$ at 4500{\AA} in the rest frame of the cluster.
Column 7: Aperture size in arcsec. The first entry is the rectangular extraction aperture 
(slit width $\times$ extraction length). The second entry is the radius in an equivalent 
circular aperture, $r_{\rm ap}= 1.025 (\rm {length} \times \rm{width} / \pi)^{1/2}$, cf.\ J\o rgensen et al.\ (1995b).
Column 8: Slit lengths in arcsec.
Column 9: Median S/N per {\AA}ngstrom for the cluster members, in the rest frame of the clusters. 
}
\end{deluxetable*}

\section{Observational data \label{SEC-DATA}}

Our analysis is based on new data for four massive X-ray luminous clusters at $z=0.19-0.45$ as well as 
our previously published data for three clusters at $z=0.54-0.89$
(J\o rgensen et al.\ 2005; Chiboucas et al.\ 2009; J\o rgensen \& Chiboucas 2013). 
In those papers we presented ground-based imaging and spectroscopy as well as imaging with {\it HST}/ACS
of the clusters MS0451.6--0305 ($z=0.54$), RXJ0152.7--1357 ($z=0.83$), and RXJ1226.9+3332 ($z=0.89$).
Our new data cover the clusters Abell 1689 ($z=0.19$), RXJ0056.2+2622 ($z=0.19$), RXJ0027.6+2616
($z=0.36$) and RXJ1347.5--1145 ($z=0.45$).
Further, we use our data for Coma, Perseus and Abell 194, 
(J\o rgensen et al.\ 1995ab; J\o rgensen 1999; 
J\o rgensen \& Chiboucas 2013; J\o rgensen et al.\ in prep.), 
as our local $z\approx 0$ reference sample.
Table \ref{tab-clusters} lists the cluster properties for all the clusters.
 
Abell 1689 and Abell 115/RXJ0056.2+2622 are part of the Abell catalog of northern clusters  (Abell et al.\ 1989).
RXJ0027.6+2616 was first listed as a cluster in the Northern {\it ROSAT} all-sky (NORAS) X-ray survey by
B\"{o}hringer et al.\ (2000).
RXJ1347.5--1145  was discovered as the most X-ray luminous cluster of the {\it ROSAT} clusters 
(Schindler et al.\ 1997). The cluster has been extensively investigated since.

In Sections \ref{SEC-imaging} and \ref{SEC-spectroscopy} we summarize the new data
for the $z=0.19-0.45$ clusters. The Appendix contains additional details regarding the new data,
all derived photometric and spectroscopic parameters, grey scale images of the clusters with X-ray data 
overlaid, as well as spectral plots of the member galaxies.
Table \ref{tab-inst} summarizes the instrumentation used in the observations.

\subsection{Imaging of the $z=0.19-0.45$ Clusters \label{SEC-imaging}}

Imaging of Abell 1689, RXJ0056.2+2622 and \\ RXJ0027.6+2616 was obtained with the Gemini Multi-Object Spectrograph 
on Gemini North (GMOS-N), while RXJ1347.5--1145 was observed with the twin instrument GMOS-S on Gemini South. 
For a full description of GMOS-N and its twin, see Hook et al.\ (2004).
All four clusters were observed using the $g'$, $r'$, and $i'$ filters. 
The seeing for the observations varied between $\approx 0.5$ arcsec and $1.1$ arcsec.
The imaging was obtained primarily to aid the sample selection and mask design for the 
spectroscopic observations.
The observations were processed using the Gemini IRAF package. Processing includes
bias subtraction, flat fielding, correction of signal on all three detectors to electrons,
correction for fringing in $r'$- and $i'$-band observations,
mosaicing of the detectors and stacking of the dithered observations of each field.
The stacked images were then processed with SExtractor (Bertin \& Arnots 1996) as described
in J\o rgensen et al.\ (2005). 
The photometric calibration for the GMOS-N data relies on the zero points and color terms
established by J\o rgensen (2009), while the GMOS-S data were calibrated using observations of
a photometric standard star field obtained the same night as the RXJ1347.5--1145 imaging
as well as color terms from the Gemini web pages. 
The Appendix contains additional details on the imaging observations and the processing of these data,
as well as derived magnitudes and colors for the targets in the spectroscopic samples.

\begin{figure*}
\plotone{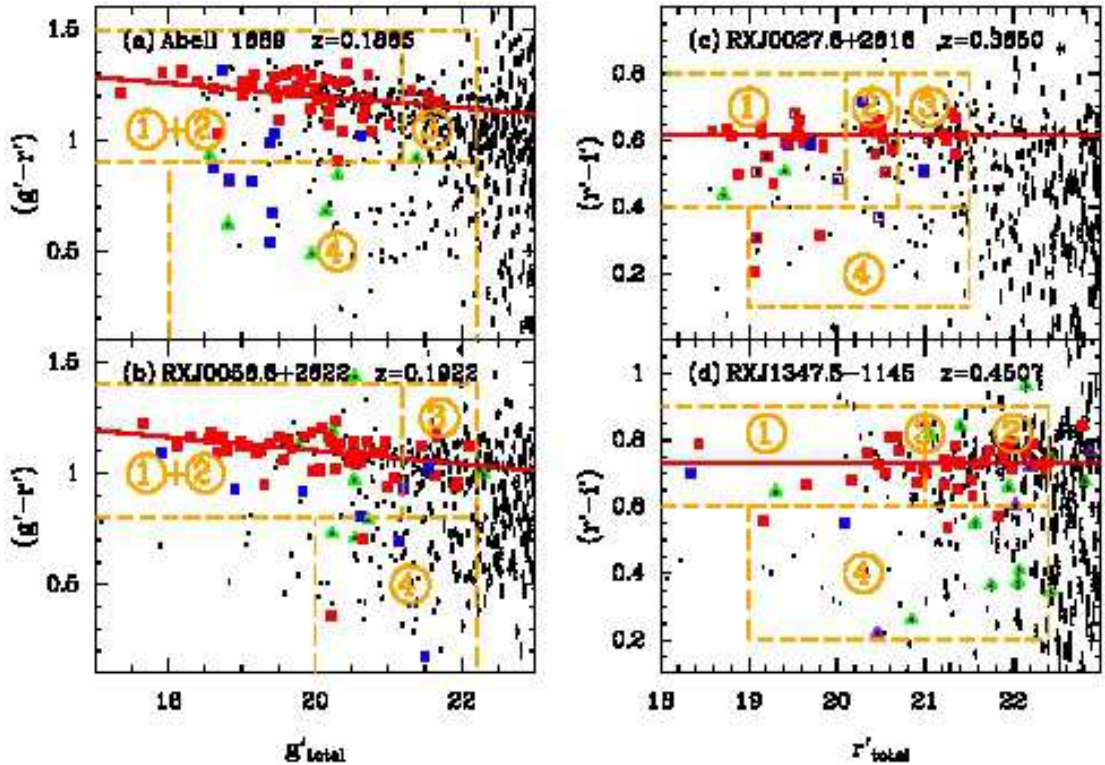}
\caption{
Color-magnitude relations for the four clusters at $z=0.19-0.45$.
The photometry shown on the figure has been corrected for Galactic extinction, 
cf.\ Table \ref{tab-imdata} in the Appendix.
Solid red squares -- bulge-dominated members of the cluster with EW[\ion{O}{2}]$\le 5${\AA}; 
solid blue squares -- members with EW[\ion{O}{2}] $> 5${\AA} and/or disk-dominated.
Open points on panel (c) for RXJ0027.6+2616 are members of the foreground group, see text.
Green open triangles -- confirmed non-members. 
Purple open triangles (panel d) -- galaxies for which the obtained spectra do not allow redshift determination. 
Small black points -- all galaxies in the field.
Red lines are the best fit relations for the bulge-dominated member galaxies, excluding 
galaxies 0.2 mag bluer than the red sequence and those with significant [\ion{O}{2}] emission. 
Dashed orange lines and numbers visualize the selection criteria for the spectroscopic samples,
see Table \ref{tab-spsel} in the Appendix.
The object class ``3'' for RXJ1347.5-1145 covers objects with the same magnitudes and colors as 
classes ``1'' and ``2'', but without prior redshift information at the time of the sample selection.
\label{fig-phot_CM} }
\end{figure*}

The cluster RXJ0056.2+2622 was also imaged with GMOS-N on Gemini North in the $r'$ filter
in seeing of $0.33-0.35$ arcsec. The observations were obtained with the purpose of 
determining effective radii and surface brightness of the galaxies. These data will 
be covered in our paper J\o rgensen et al.\ (in prep.) on the 2-dimensional photometry 
of the galaxies in the clusters. In the current paper we use preliminary
results to exclude disk-dominated galaxies from the analysis.
Imaging from {\it HST} exists for the three other clusters. We will also cover these
data in J\o rgensen et al.\ (in prep.), and use the data here only to exclude disk-dominated 
galaxies from the analysis.
The final sample selection is described in Section \ref{SEC-SAMPLES}.

\subsection{Spectroscopy of the $z=0.19-0.45$ Clusters \label{SEC-spectroscopy}}

The spectroscopic observations were obtained in multi-object spectroscopic mode with GMOS-N or GMOS-S,
see Table \ref{tab-spdata} for an overview of the observations.
Two fields were observed in Abell 1689, each with two masks. Two fields were observed in RXJ0056.2+2622 with one mask each. 
RXJ0027.6+2616 and RXJ1347.5--1145 each have one field observed with two masks. 
The sample selection for the spectroscopic observations was based on the photometry from the ground-based imaging.
Figure \ref{fig-phot_CM} shows the color-magnitude relations with the 
selection criteria overlaid.  The criteria are summarized in Table \ref{tab-spsel}, in the Appendix.
The selection method is similar to that used for MS0451.6--0305, RXJ0152.7--1357 and RXJ1226.9+3332 
(J\o rgensen \& Chiboucas 2013), optimizing the inclusion of member galaxies on the red
sequence as well as covering 3-4.5 magnitudes along the red sequence. 
For Abell 1689, RXJ0056.2+2622 and RXJ1347.5--1145 we also used published redshifts
to optimize inclusion of known members and exclude non-members. At the time of the sample
selection redshifts for Abell 1689 and RXJ0056.2+2622 were taken from 
NASA/IPAC Extragalactic Database (NED), while for RXJ1347.5--1145 we used redshifts 
from Cohen et al.\ (2002) and Ravindranath \& Ho (2002).
After inclusion of all possible targets of the highest priority (object classes 1--2 as
marked on Fig.\ \ref{fig-phot_CM}), remaining space in the mask was filled with fainter cluster members
and/or targets on the red sequence but without redshifts (object class 3) if possible, and otherwise with
bluer galaxies some of which can be expected not to be members (object class 4).
The spectroscopic samples are marked on grey scale figures of the clusters available
in the Appendix as Figures \ref{fig-A1689grey}--\ref{fig-RXJ1347grey}.

The spectroscopic observations were processed using the methods described in detail in 
J\o rgensen et al.\ (2005) and J\o rgensen \& Chiboucas (2013).  
The data processing results in 1-dimensional spectra calibrated to a relative flux scale. 
The spectra were used to derive redshifts, velocity dispersions and absorption line strengths.
The spectroscopic parameters were determined using the same methods as described in J\o rgensen et al.\ (2005).
In particular, the redshifts and velocity dispersions were determined by fitting the spectra with
a mix of three template stars (spectral types K0III, G1V, and B8V) using software made available by Karl Gebhardt
(Gebhardt et al.\ 2000, 2003). The galaxies that were found to be members of the clusters were fit in
the rest frame wavelength range 3750--5500{\AA}. 
The reader is referred to J\o rgensen \& Chiboucas (2013) for a discussion of the use of 
template stars rather than (a larger number of) model spectra for the fitting of the spectra.
For consistency with our previous work, we fit the spectra of the $z=0.19-0.45$ galaxies
using the same template stars as in J\o rgensen \& Chiboucas.
The results from the template fitting are available in the Appendix (Table \ref{tab-kin}).
The observations covered 34-72 member galaxies in the four clusters, cf.\ Table \ref{tab-clusters}.

We determined strengths of absorption lines using the Lick/IDS definitions (Worthey et al.\ 1994).
In addition we determined the indices for H$\delta _A$ and H$\gamma _A$ as defined by 
Worthey \& Ottaviani (1997), the H$\beta _G$ index as defined in Gonz\'{a}lez (1993) (see also,
J\o rgensen 1997 for the index definition),
CN3883 and CaHK as defined in Davidge \& Clark (1994), D4000 as defined by
Gorgas et al.\ (1999), and the high order Balmer line index H$\zeta _{\rm A}$ as defined by Nantais et al.\ (2013).
For galaxies with detectable  [\ion{O}{2}] emission the strength
of the [\ion{O}{2}]$\lambda \lambda$3726,3729 doublet was determined as an equivalent width, EW[\ion{O}{2}].
In the following we refer to the doublet as the ``[\ion{O}{2}] line''.
All measured absorption line indices and the [\ion{O}{2}] equivalent widths are available in the Appendix 
(Tables \ref{tab-line}--\ref{tab-EWOII}).
The typical measurement uncertainties were established based on internal comparisons,
and are listed in the Appendix, Table \ref{tab-intcomp}.
In our analysis we use the Balmer line indices H$\beta _G$ and 
$({\rm H\delta _A + H\gamma _A})' \equiv -2.5~\log \left ( 1.-({\rm H\delta _A + H\gamma _A})/(43.75+38.75) \right )$ \\
(Kuntschner 2000) as age sensitive indices, the 
iron indices Fe4383 and $\rm \langle Fe \rangle \equiv (Fe5270+Fe5335)/2$,
and the indices Mg$b$, CN3883 and C4668 all of which are sensitive to the abundance ratios
$\rm [\alpha /Fe]$, see Section \ref{SEC-MODELS}.
We also use the combination indices 
${\rm [MgFe] \equiv (Mg{\it b} \cdot \langle Fe \rangle)^{1/2}}$ (Gonz\'{a}lez 1993)
and
${\rm [C4668\,Fe4383] \equiv C4668 \cdot (Fe4383)^{1/3}}$ \\
(J\o rgensen \& Chiboucas 2013), both constructed to be independent of  $\rm [\alpha /Fe]$
for the stellar population models used in the analysis, see Section \ref{SEC-MODELS}.

\section{Cluster properties: Radii, masses, redshifts, velocity dispersions, and possible substructure \label{SEC-CLUSTERZ}}

As one of our goals is to investigate the possible effects of cluster environments on the stellar 
populations of the galaxies, we here establish consistent X-ray properties of the 
clusters, specifically the luminosities, radii and masses defined based on the radii $R_{\rm 500}$.
$R_{\rm 500}$ is the radius within which the average mass density of the cluster is 500 
times the critical density of the Universe at the cluster's redshift. 
We then address (1) whether the higher redshift clusters are viable progenitors for the lower redshift clusters
in terms of their masses, (2) whether the X-ray parameters and the cluster velocity dispersion follow expected
relations, and (3) whether the clusters show signs of sub-clustering or merging.

\begin{figure}
\epsfxsize 8.5cm
\epsfbox{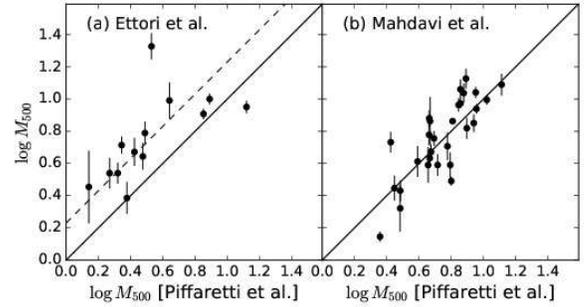}
\caption{Calibration of X-ray mass measurements, $M_{\rm 500}$, from Ettori et al.\ (2004)
and Mahdavi et al.\ (2013, 2014) to consistency with Piffaretti et al.\ (2011).
Solid lines -- one-to-one relations.
For data from  Ettori et al.\ we find a median offset in $\log M_{\rm 500}$ of 0.23,
with an rms of 0.21 for 14 clusters in common. The dashed line on panel (a) is the 
one-to-one line offset with 0.23.
The data from Mahdavi et al.\ are consistent with Piffaretti et al. The rms in 
$\log M_{\rm 500}$ is 0.14 for the 29 clusters in common. See text for details of the calibration. 
\label{fig-XrayM500_calibsmall} }
\end{figure}

\subsection{Adopted X-ray Data}

We take the cluster catalog from Piffaretti et al.\ (2011) as our main source of the
X-ray properties of the clusters.  This catalog covers most X-ray clusters at $z<1$.
However, it treats RXJ0056.2+2622 and RXJ0152.7--1357 as single clusters,
while the X-ray structure and the distribution of the galaxies show that they are
binary clusters (e.g., Barrena et al.\ 2007; Maughan et al.\ 2003). 
For these clusters we therefore use data from Mahdavi et al.\ (2013, 2014), 
who give X-ray parameters for the two sub-clusters
in RXJ0056.2+2622, and Ettori et al.\ (2004), who give X-ray parameters for the two
sub-clusters in RXJ0152.7--1357. We note that Ettori et al.\ (2009) also give values
for the RXJ0152.7--1357 sub-clusters, but these are significantly larger than
the values in Ettori et al.\ (2004) and the sum of the masses as well as the luminosities 
are inconsistent with the values from Piffaretti et al.\ (2011) for the full cluster. 
For other clusters listed in both papers by Ettori et al.\ the values are consistent.
It is beyond the scope of this paper to explore the reason for this apparent inconsistency.
Because of the increasing evidence that the mass of RXJ1347.5--1145 initially 
determined from X-ray data (and adopted by Piffaretti et al.) may be too large 
and that the cluster is experiencing interactions with an infalling sub-cluster (Kreisch et al.\ 2016)
we instead adopt the lower value from Ettori et al.\ (2004). These authors correct
the X-ray measurements for diffuse X-ray emission south-east of the main cluster
and associated with the infalling sub-cluster.

We have calibrated the X-ray data from Mahdavi et al.\ (2013, 2014) and Ettori et al.\ (2004) 
to reach consistency with Piffaretti et al.\ using the clusters in common. 
As necessary, we first convert the published X-ray 
parameters to $M_{500}$ using the equations from Piffaretti et al. We then 
determine the offset in $\log M_{ 500}$ to reach consistency with Piffaretti et al. 
When converting between radii $R_{500}$, masses $M_{500}$, and luminosities $L_{500}$ 
we use the relations given by Piffaretti et al.\ (their Equations 2 and 3). 
As needed we also adopt the following conversions from Piffaretti et al.\
$L_{500} = 0.91 L_{\rm total}$, $R_{200} = 1.52 R_{500}$, $L_{500}=0.96 L_{200}$.
The relation between $R_{200}$ and $R_{500}$ is equivalent to 
$M_{200} = 1.40 M_{500}$. 

We show the comparisons of $M_{\rm 500}$ from Piffaretti et al.\ with 
data from Ettori at al.\ (2004) and Mahdavi et al.\ (2013, 2014) in Figure \ref{fig-XrayM500_calibsmall}.
The $M_{\rm 500}$ values from Mahdavi et al.\ are consistent 
with those from Piffaretti et al., while for values from Ettori et al.\ we subtract 0.23
from $\log M_{\rm 500}$ to reach consistency with Piffaretti et al. Luminosities, $L_{\rm 500}$,
and radii, $R_{\rm 500}$, are then derived from $M_{\rm 500}$ using the equations 
from Piffaretti et al.

\subsection{Cluster Properties and Possible Sub-Structure}

\begin{figure}
\epsfxsize 8.5cm
\epsfbox{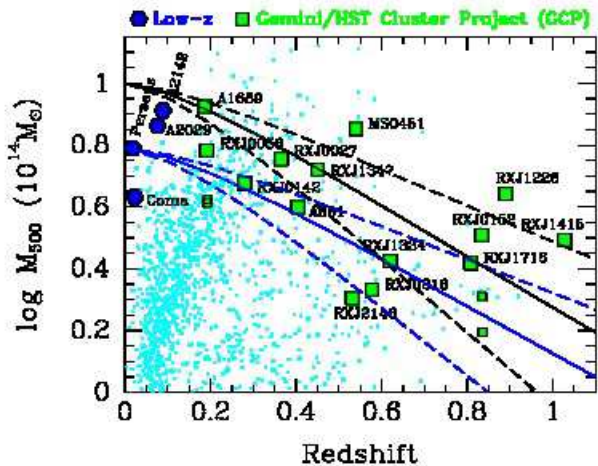}
\caption{The cluster masses, $M_{\rm 500}$, based on X-ray data versus the redshifts of the clusters.
Blue hexagons -- our local reference sample; green squares - the GCP $z=0.2-1$ sample.
The pairs of slightly smaller points at the same redshifts as RXJ0056.6+2622
and RXJ0152.7--1357 show the values for the sub-clusters of these binary clusters.
$M_{\rm 500}$ is from Piffaretti et al.\ (2011), except as noted in Table \ref{tab-clusters}.
Small cyan points -- all clusters from Piffaretti et al.\ shown for reference.
Blue and black lines -- mass development of clusters based on numerical simulations
by van den Bosch (2002).
The black lines terminate at Mass=$10^{15} M_{\sun}$ at $z=0$ roughly
matching the highest mass clusters at $z=0.1-0.2$.
The blue lines terminate at Mass=$10^{14.8} M_{\sun}$ at $z=0$
matching the mass of the Perseus cluster.
The dashed lines represent the typical uncertainty in the
mass development represented by the numerical simulations.
\label{fig-M500} }
\end{figure}

Figure \ref{fig-M500} shows the adopted masses, $M_{\rm 500}$, versus redshifts for the full GCP cluster
sample.
We show typical models for the growth of cluster masses with time, based on results
from van den Bosch (2002). 
These models are in general agreement with newer and more detailed analysis of the results from
the Millennium Simulations (Fakhouri et al.\ 2010).
The seven $z=0.19-0.89$ clusters analyzed in the present paper are all very massive
and represent the most massive clusters in the GCP at a given redshift.
With current models for mass evolution of clusters, their masses at $z\approx 0$ would be
higher than those of the Coma and Perseus clusters and close to the masses of the 
two very massive local clusters Abell 2029 and Abell 2142 (also marked on Fig.\ \ref{fig-M500}). 
We will in a future paper investigate those two local clusters to assess if the stellar 
populations of their passive galaxies are similar to those of the Coma and Perseus clusters. 
For now we will assume that the bulge-dominated passive galaxies in the seven $z=0.19-0.89$ clusters 
are viable progenitors for bulge-dominated passive galaxies in Coma and Perseus.

\begin{figure}
\epsfxsize 8.5cm
\epsfbox{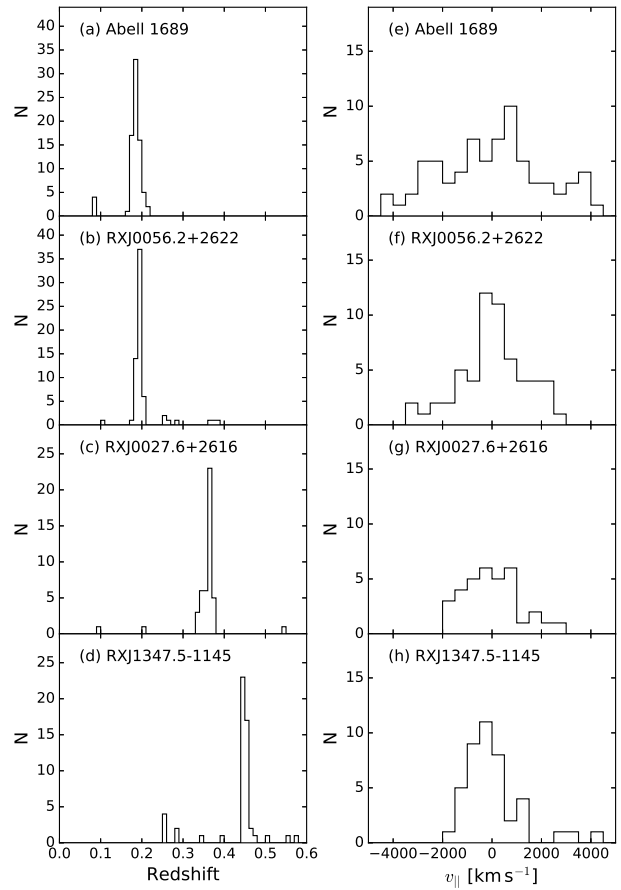}
\caption{(a)--(d) Redshift distributions of the spectroscopic samples.
(e)--(h) Distribution of the radial velocities (in the rest frames of the clusters) 
relative to the cluster redshifts for cluster members, $v_{\|} = c (z - z_{\rm cluster}) / (1+z_{\rm cluster})$.
\label{fig-zhist} }
\end{figure}

We determined the cluster redshifts and velocity dispersions for the $z=0.19-0.45$ clusters 
using our data and the bi-weight method by Beers et al.\ (1990).
Figure \ref{fig-zhist} shows the redshift distributions of the samples. 
In J\o rgensen \& Chiboucas (2013) we published similar results for the three higher redshift clusters.
Table \ref{tab-clusters} summarizes redshifts and velocity dispersions for the clusters.
Using Kolmogorov-Smirnov tests we found that for all the clusters the velocity distributions 
of the member galaxies are consistent with Gaussian distributions.

\begin{figure}
\epsfxsize 8.5cm
\epsfbox{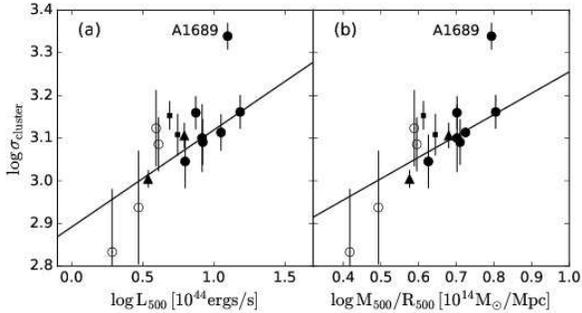}
\caption{ 
Cluster velocity dispersion versus (a) the X-ray luminosity $L_{\rm 500}$ and (b) $M _{\rm 500} \, R _{\rm 500}^{-1}$.
Data from Table \ref{tab-clusters}.
Large solid circles -- the seven $z=0.19-0.89$ clusters;
large open circles -- data for the binary clusters RXJ0056.2+2622 and RXJ0152.7--1357;
large triangles -- Coma and Perseus;
small squares -- other GCP clusters (Barr et al.\ 2005, Hibon et al.\ in prep.);
Solid lines show the relations $L \propto \sigma _{\rm cluster}^{4.4}$ (panel a) with the slope adopted from
Mahdavi \& Geller (2001), and 
$M_{\rm 500}\,R_{\rm 500}^{-1} \propto \sigma _{\rm cluster}^{2}$ (panel b).
In both cases, the median zero points are derived from the data, excluding Abell 1689 and the binary clusters. 
\label{fig-lsigma_cluster} }
\end{figure}

In Figure \ref{fig-lsigma_cluster} we show the cluster velocity dispersions versus the X-ray properties of the
clusters. 
At a velocity dispersion of $2182~{\rm km\,s^{-1}}$,
Abell 1689 deviates from the expected relations by $\sim 7$ times the 
uncertainty on the cluster velocity dispersion. 
The kinematic structure of the cluster has been studied using very large samples of redshift data.
Czoske (2004) concludes based on 525 member redshifts that the central velocity dispersion 
is $\approx 2100  {\rm km~s^{-1}}$ and that the central structure of the cluster is complex. 
Lemze et al.\ (2009) analyzed this data set further, confirming the high central velocity dispersion.
They also determined a virial cluster mass of $\approx 1.9 \cdot 10^{15} M_{\sun}$ (for our assumed cosmology),
equivalent to $M_{500} = 1.3 \cdot 10^{15} M_{\sun}$. A similar mass estimate is found by Umetsa et al.\ (2015)
based on lensing data. Umetsa et al.\ as well as Morandi et al.\ (2011) argue that the cluster is not 
spherical. Morandi et al.\ in particular state that earlier mass estimates from X-data were about a factor two
too small as spherical symmetry was assumed. 
Finally, Andersson \& Madejski (2004) used the temperature of the X-ray gas to argue that Abell 1689
is in fact a merger, something that cannot be detected from the X-ray morphology alone.
Thus, we conclude that Abell 1689 deviates from the relations shown on Figure \ref{fig-lsigma_cluster}
due to (1) the mass being underestimated and (2) the unrelaxed nature of the center of the cluster 
leading to a very high central cluster velocity dispersion.

Barrena et al.\ (2007) studied the dynamical status of RXJ0056.2+2622.
These authors find that the cluster consists of two structures in the plane of the sky, see also
the grey scale image with X-ray data overlaid available in the Appendix
Figure \ref{fig-RXJ0056grey}. 
Barrena et al.\ find a total cluster velocity dispersions of 
$\sigma _{\rm cluster} =1362_{-108}^{+126}\,{\rm km\,s^{-1}}$ and a redshift of $z=0.1929 \pm 0.0005$, 
both in agreement with our results within the uncertainties.
We attempt to determine the sub-cluster velocity dispersions by defining likely members of 
the two sub-clusters from the cluster center distances.
We include only galaxies that have cluster center distances to one or the other of 
the sub-cluster centers of less than half the distance between the centers.
The galaxies IDs 191 and 1654 are taken as defining the southern and northern sub-cluster center, respectively.
We then find cluster velocity dispersions of the sub-clusters of $\sigma _{\rm cluster} = 1200-1300\,{\rm km\,s^{-1}}$
cf.\ Table \ref{tab-clusters}, which
places the sub-clusters above the mean relations with the X-ray properties, but with rather
large uncertainties on $\sigma _{\rm cluster}$, see Figure \ref{fig-lsigma_cluster}. 

Our data show that the field of RXJ0027.6+2616 contains a foreground group at $z=0.3404$.
Our data include nine galaxies in this group mostly on the western edge of the observed field,
see Figure \ref{fig-RXJ0027grey} in the Appendix.
The velocity dispersion of this group as determined from the nine galaxies is very low, 
$\sigma = 172_{-47}^{+29}\,{\rm km\,s^{-1}}$.
These galaxies were not included in the analysis of the cluster members. 

The cluster velocity dispersion for RXJ1347.5--1145 was also measured by 
Cohen \& Kneib (2002) who found $\sigma_{\rm cluster}=910 \pm 130\,{\rm km\,s^{-1}}$ based on 47 members,
and by Lu et al.\ (2010) who based on additional spectroscopic observations found 
$\sigma_{\rm cluster}=1163 \pm 97\,{\rm km\,s^{-1}}$ for a sample of $\approx 95$ members 
(sample size evaluated from Fig.\ 2 in that paper).
The result from Lu et al.\ is in agreement with our result of $1259_{-250}^{+210}\,{\rm km\,s^{-1}}$.
Lu et al.\ state that the result from Cohen \& Kneib may be in error. 
We confirm that this indeed seems to be the case, 
as we find $\sigma_{\rm cluster}=1256_{-197}^{+115}\,{\rm km\,s^{-1}}$ when using the Cohen \& Kneib data. 
We caution that the results for this cluster are very sensitive to the exact choice of the upper limit
for the cluster redshift. Had we for example excluded the three highest redshift galaxies in our
sample, we would have found $\sigma_{\rm cluster}=791_{-88}^{+70}\,{\rm km\,s^{-1}}$.
The Lu et al.\ sample is significantly larger and less likely to be affected by small number statistics.
Thus, we evaluate that the joint data support a velocity dispersion of $\approx 1150-1250\,{\rm km\,s^{-1}}$ 
for this cluster. When combined with the X-ray data from Ettori et al.\ (2004), 
this places the cluster on the mean relations between the cluster velocity dispersions and the X-ray properties. 
Cohen \& Kneib propose that the cluster may be a merger.
Later investigations of the X-ray data of the cluster support this idea but for different 
reasons, see e.g., Ettori et al.\ (2004) for the discussion of the diffuse X-ray emission south-east 
of the main cluster, and the recent study by Kreisch et al.\ (2016) based on deep {\it Chandra} X-ray observations.

In summary, four of the seven clusters in our sample have evidence of non-relaxed structures and/or 
deviate from the expected relations between the cluster velocity dispersions and the X-ray parameters.
These clusters are Abell 1689, RXJ0056.2+2622, RXJ1347.5--1347, and RXJ0152.7--1347.
RXJ0027.6+2616 has a foreground group, but this may be at large enough distance in redshift space to not
affect the cluster itself.
MS0451.6--0305 and RXJ1226.9+3332 have no evidence sub-clustering and also follow the expected relations.
These complicated cluster properties should be kept in mind when we apply the fairly simple approach 
of using either the cluster center distances (in units of $R _{\rm 500}$)
or the product of the cluster center distances and the radial velocities relative to the cluster redshifts
as parameterization of the cluster environment for a given galaxy. 

\section{The final galaxy sample and the methods \label{SEC-METHODSAMPLE} }

We base our analysis on (1) empirical scaling relations between the velocity dispersions
and the strength of the absorption lines from the individual spectra,
and (2) ages, metallicities [M/H] and 
abundance ratios $\rm [\alpha /Fe]$ derived from composite spectra absorption line indices using
single stellar population models.
For the local reference sample we use luminosity weighted average line indices in place
of determinations from composite spectra.

\begin{deluxetable*}{lrrrrrr}
\tablecaption{Samples \label{tab-sample} }
\tablewidth{0pc}
\tabletypesize{\scriptsize}
\tablehead{ \colhead{} &  \colhead{} &  \colhead{} & \multicolumn{4}{c}{ Bulge-dominated } \\
\colhead{Cluster} & \colhead{$N_{\rm members}$} & \colhead{$N_{\rm disk}$} 
    & \colhead{$N_{\rm emis}$} & \colhead{$N_{\rm lowmass}$} & \colhead{$N_{\rm highmass}$} & \colhead{$N_{\rm sample}$} \\
\colhead{(1)} & \colhead{(2)} &\colhead{(3)} &\colhead{(4)} &\colhead{(5)} &\colhead{(6)} &\colhead{(7)} }
\startdata
Abell 1689       & 72 & 3\tablenotemark{a} & 6 & 15 & 48 & 48 \\
RXJ0056.2+2622   & 58 & 4 & 4\tablenotemark{b} & 11 & 39 & 38 \\
RXJ0027.6+2616   & 34 & 0 &  3\tablenotemark{c} & 10 & 21 & 21 \\
RXJ1347.5-1345   & 43 & 1\tablenotemark{d} & 1 & 6 & 35 & 31 \\
MS0451.6--0305   & 47 & 9 & 1 & 0 & 37 & 34 \\
RXJ0152.7--1357  & 29 & 3 & 3 & 2 & 21 & 21 \\
RXJ1226.9+3332  & 55 & 6 & 7 & 2 & 38 & 28 \\
\enddata
\tablecomments{Column 1: Galaxy cluster. Column 2: Number of members with available spectroscopy; Column 3: Number of disk galaxies.
Column 4: Number of bulge-dominated galaxies with $\rm Mass < 10^{10.3} M_{\sun}$ or, equivalently, $\log \sigma < 2$, see text,
the sample includes 3 E+A galaxies in Abell 1689, 1 E+A galaxy in  RXJ0056.2+2622.
Column 5: Number of bulge-dominated with emission with $\rm Mass_{\rm dyn} < 10^{10.3} M_{\sun}$ or, equivalently, $\log \sigma < 2$. 
Column 6: Number of bulge-dominated passive galaxies with $\rm Mass_{\rm dyn} \ge 10^{10.3} M_{\sun}$ or, equivalently, $\log \sigma \ge 2$.
Column 7: Number of bulge-dominated passive galaxies in the final samples, which exclude E+A galaxies, galaxies in RXJ1347.5-1345 with
S/N$<$25 per {\AA} in the rest frame, galaxies in MS0451.6--0305 for which the spectra a contaminated by signal from neighboring objects 
(cf.\ J\o rgensen \& Chiboucas 2013), and galaxies in RXJ1226.9+3332 with S/N$<$20.}
\tablenotetext{a}{Includes ID 972, ID 752, excludes the BCG ID 584.}
\tablenotetext{b}{ID 1296 has H$\beta$ and [\ion{O}{3}] emission. The spectrum does not cover [\ion{O}{2}].}
\tablenotetext{c}{IDs 51 and 545 have significant emission fill in of the H$\beta$ and are included in the emission line galaxies.}
\tablenotetext{d}{ID 195 based in visual inspection of imaging, see text.}
\end{deluxetable*}

\subsection{Samples \label{SEC-SAMPLES} }

Table \ref{tab-sample} summarizes the number of galaxies in each cluster for which we 
have data. The table details how many of these are bulge-dominated passive galaxies
with dynamical masses ${\rm Mass_{\rm dyn}} \ge 10^{10.3} {M_\sun}$ or, equivalently, $\log \sigma \ge 2.0$,
and therefore the primary focus of this paper.
We use $\log \sigma$ in place of the dynamical masses of the galaxies,
in our investigation of the possible primary driver of the galaxy properties.
We acknowledge that the two parameters are not fully interchangeable, and will return
to this issue in our upcoming paper in which we also make use of the 2-dimensional
photometry of all the galaxies.

The sample selection makes use of the S\'{e}rsic index, $n_{\rm ser}$, (S\'{e}rsic 1968) 
either from our photometry 
(Chiboucas et al.\ 2009; J\o rgensen \& Chiboucas 2013; J\o rgensen et al.\ in prep.),
or for Abell 1689 from Houghton et al.\ (2012).
We require $n_{\rm ser} \ge 1.5$ for a galaxy to be considered bulge-dominated.
Galaxies for which we have no $n_{\rm ser}$ measurement, but which appear to be spiral
galaxies based on imaging data, are considered disk-galaxies for the purpose of the
sample selection. 
Abell 1689 ID 584 is the brightest cluster galaxy and we include it with the bulge-dominated galaxies
even though Houghton et al.\ give a  S\'{e}rsic index of 1.2.
As in our previous papers, we consider galaxies with 
equivalent width of EW[\ion{O}{2}]$\le 5${\AA} passive (non-emission).
This criterion matches the spectral classifications defined by Dressler et al.\ (1999) and used 
by other researchers, e.g.\ Sato \& Martin (2006) and Saglia et al.\ (2010). 
If the [\ion{O}{2}] line is not included in the spectral coverage, we use the presence
or absence of emission in H$\beta$ and/or [\ion{O}{3}]$\lambda$5007 to evaluate if a galaxy is passive.

The galaxy RXJ0056.2+2622 ID 1170 was excluded from the analysis as it is an E+A galaxy 
dominated by extremely young stellar populations. 
Including it in the determination of especially the scaling relation 
scatter would bias the determination and not give a representative view of the typical
scatter of the line strengths for the cluster's galaxies.
Other E+A galaxies in the sample are excluded from the analysis as they have $\log \sigma < 2.0$.

Due to the strong fringing in the GMOS-S observations of RXJ1347.5--1145 and its effect 
on the sky subtraction we conservatively exclude galaxies with S/N$<25$ from the analysis. 
This removes IDs 444, 513, 657, and 919 from our analysis. The measurements are still
listed in the tables, but we caution that in particular the line index measurements are
expected to have larger uncertainties than otherwise adopted for the cluster.
In addition, we remove RXJ1347.5--1145 ID 743 and ID 1011 from the analysis. 
The GMOS-S imaging shows ID 743 as double; there is no {\it HST} imaging covering this galaxy. 
The galaxy also has $\log \sigma<2.0$. 
The spectrum of ID 1011 appears to be a BL Lac object with a featureless optical continuum superimposed
on a spectrum of an old stellar population.
Thus, the effect is that all the line indices are weaker than otherwise expected from an
old stellar population.
The object is within 8 arcsec of a point source in the Chandra X-ray catalog (Evans et al.\ 2010).

In the analysis we focus on the bulge-dominated passive galaxies with 
dynamical masses ${\rm Mass} \ge 10^{10.3} {M_\sun}$ or  $\log \sigma \ge 2.0$.
The lower limit is set by both the available local reference sample
and the available data for the intermediate redshift clusters.
The number of bulge-dominated galaxies with low masses (or velocity dispersions) 
are listed in Column (5) in  Table \ref{tab-sample}.
Column (7) in Table \ref{tab-sample} gives the number of galaxies in each cluster
that meets all criteria for being included in the analysis.
The sample selection for Coma, Perseus, Abell 194, MS0451.6--0305, RXJ0152.7--1357 
and RXJ1226.9+3332 is identical to the selection used in J\o rgensen \& Chiboucas (2013).
In samples for the two highest redshift clusters we reach cluster 
center distances of $R_{\rm cl}\approx 1.8 R_{500}$,
while all other cluster samples only reach $R_{\rm cl}/R_{500} = 0.8-1.0$.
For galaxies in the two double-clusters RXJ0056.2+2622 and RXJ0152.7--1157, we use $R_{\rm cl}/R _{\rm 500}$ 
derived relative to the sub-cluster center closest to the galaxy.

\subsection{Composite Spectra \label{SEC-COMPOSITES} }

In addition to the measurements for the individual galaxies, we construct composite
spectra with higher S/N by co-adding the spectra. 
The spectra are co-added by velocity dispersion. 
When possible given the number of observed galaxies, we use bins of 0.05 in $\log \sigma$.
When this results in less than three galaxies in a bin and/or low S/N, we use bins of 0.1 in $\log \sigma$.
The velocity dispersion bin with $\log \sigma < 2.0$ contains galaxies
that we otherwise omit from the analysis of the individual measurements, see
Section \ref{SEC-SAMPLES}.
Our analysis in Section \ref{SEC-INDICES} focuses on the blue indices measured from
the composites. These measurements and 
the average velocity dispersions of the galaxies included in each composite
are available in Table \ref{tab-composites} in the Appendix.
We do not attempt to measure the velocity dispersions directly from the composites.

For our local reference sample in Perseus, we use luminosity weighted average parameters for sub-samples
of galaxies using the same binning as described above for the higher redshift galaxies.
The average parameters for the Perseus sub-samples are
also listed in Table \ref{tab-composites} in the Appendix.

\subsection{Adopted Scaling Relations}

We adopt the empirical scaling relations and zero points for the local reference sample and the 
$z=0.5-0.9$ clusters from J\o rgensen \& Chiboucas (2013).
These relations were established using a fitting technique that is very robust to outliers.
It minimizes the sum of the absolute residuals perpendicular to the relation,
determines the zero points as the median, and uncertainties on the slopes using
a boot-strap method, see J\o rgensen et al.\ (2005) and J\o rgensen \& Chiboucas (2013)
for details. As in those papers, we then determine the 
random uncertainties on the zero point differences, $\Delta \gamma$, 
between the intermediate redshift and local reference samples as
\begin{equation}
\sigma _{\Delta \gamma} = \left ( {\rm rms}_{{\rm low-}z}^2/N_{{\rm low-}z}
  +  {\rm rms}_{{\rm int-}z}^2/N_{{\rm int-}z} \right )^{0.5}
\end{equation}
Subscripts ``low-$z$'' and ``int-$z$'' refer to the local reference sample
and one of the intermediate redshift clusters, respectively. 
Only the random uncertainties are shown on the figures in the following.
We expect the systematic uncertainties on the zero point differences to
be dominated by the possible inconsistency in the calibration of the 
velocity dispersions, 0.026 in $\log \sigma$, and may 
be estimated as 0.026 times the coefficient for $\log \sigma$, see
also J\o rgensen et al.\ (2005).

\begin{deluxetable*}{lrrr rrr rrr rrr}
\tablecaption{Scaling Relations \label{tab-relations} }
\tablewidth{0pc}
\tabletypesize{\scriptsize}
\tablehead{
\colhead{Relation} & \multicolumn{3}{c}{Abell 1689} & 
  \multicolumn{3}{c}{RXJ0056.2+2622} &\multicolumn{3}{c}{RXJ0027.6+2616} & \multicolumn{3}{c}{RXJ1347.5--1145} \\
 & \colhead{$\gamma$} & \colhead{$N_{\rm gal}$} & \colhead{rms} 
 & \colhead{$\gamma$} & \colhead{$N_{\rm gal}$} & \colhead{rms} &  \colhead{$\gamma$} & \colhead{$N_{\rm gal}$} & \colhead{rms} 
 & \colhead{$\gamma$} & \colhead{$N_{\rm gal}$} & \colhead{rms} \\
\colhead{(1)} & \colhead{(2)} & \colhead{(3)} & \colhead{(4)} 
& \colhead{(5)} & \colhead{(6)} & \colhead{(7)} & \colhead{(8)} & \colhead{(9)} & \colhead{(10)} 
& \colhead{(11)} & \colhead{(12)} & \colhead{(13)} 
}
\startdata
$(\rm{H\delta _A + H\gamma _A})' = (-0.085\pm 0.015) \log \sigma + \gamma$ 
                                                             & 0.096 & 46 & 0.02 & 0.105 & 36 & 0.02 & 0.122 & 21 & 0.01 & 0.127 & 31 & 0.02 \\ 
$\log \rm{H\beta _G} = (-0.24\pm 0.04) \log \sigma + \gamma$ & 0.826 & 46 & 0.04 & 0.837 & 38 & 0.04 & 0.817 & 21 & 0.05 & 0.856 & 31 & 0.08 \\ 
${\rm CN3883}  = (0.29\pm 0.06) \log \sigma + \gamma$       & -0.361 & 37 & 0.05 &-0.390 & 29 & 0.05 &-0.372 & 20 & 0.04 &-0.433 & 20 & 0.03 \\ 
$\log {\rm Fe4383} = (0.19\pm 0.07) \log \sigma + \gamma$   &  0.261 & 47 & 0.05 & 0.231 & 38 & 0.07 & 0.214 & 21 & 0.06 & 0.202 & 31 & 0.11 \\ 
$\log {\rm C4668}  = (0.33\pm 0.08) \log \sigma + \gamma$   &  0.107 & 40 & 0.06 & 0.094 & 26 & 0.09 & 0.073 & 21 & 0.10 & 0.054 & 31 & 0.10 \\ 
$\log \rm{Mg{\it b}}      = (0.294\pm 0.016) \log \sigma + \gamma$ & 0.010 & 48 & 0.04 &-0.015 & 37 & 0.05 & -0.014 & 21 & 0.04 & -0.034 & 31 & 0.05 \\ 
$\log \rm{\langle Fe \rangle}     = (0.118\pm 0.012) \log \sigma + \gamma$ & 0.193 & 48 & 0.07 & 0.197 & 37 & 0.07 & 0.142 & 21 & 0.06 & 0.183 & 29 & 0.07 \\ 
\enddata
\tablecomments{Column 1: Scaling relation adopted from J\o rgensen \& Chiboucas (2013). Column 2: Zero point for the Abell 1689 sample. Column 3: Number of galaxies
included from the Abell 1689 sample. Column 4: rms in the Y-direction of the scaling relation for the Abell 1689 sample.
Columns 5--7: Zero point, number of galaxies, rms in the Y-direction for the RXJ0056.2+2622 sample.
Columns 8--10: Zero point, number of galaxies, rms in the Y-direction for the RXJ0027.6+2616 sample.
Columns 11--13: Zero point, number of galaxies, rms in the Y-direction for the RXJ1347.5--1145 sample.}
\end{deluxetable*}

\subsection{Stellar Population Models \label{SEC-MODELS} }

We use single stellar population (SSP) models in the interpretation of the data. 
In J\o rgensen \& Chiboucas (2013) we discussed the differences and similarities of the models from 
Thomas et al.\ (2011) and Schiavon (2007). 
Since then, new models taking into
account variations in abundance ratios have been published by Vazdekis et al.\ (2015). 
Vazdekis assumes that at fixed metallicity, non-$\alpha$ elements, including C, N, and Na, 
decrease in abundance when $\rm [\alpha /Fe]$ increases. Magnesium is used as a proxy for all $\alpha$-elements. 
Thomas et al.\ on the other hand assume that the elements C, N and Na track the $\alpha$-elements in abundance ratios. 

We used the model spectral energy distributions (SEDs) available for the Vazdekis models to test
if these models for $\rm [\alpha /Fe] >0$ can reproduce the line strengths found in our 
cluster galaxy samples. Our test shows that the models fail to reproduce
the strong CN3883 and C4668 indices present in the galaxies that also have strong Mg$b$ indices.
Thus, the underlying assumption of a decrease in C and N with increasing Mg abundances
appears to not be supported by the data.
We therefore proceed as done in J\o rgensen \& Chiboucas (2013), using the models from Thomas et al.

In J\o rgensen \& Chiboucas (2013) we 
established an empirical relation between $\rm CN_2$ and CN3883,
Eq.\ 4 in that paper: $\rm CN3883 = (0.84 \pm 0.13)\, CN_2 + 0.146$. 
We have confirmed that this relation if also valid for the full sample used in the present paper.
As done in J\o rgensen \& Chiboucas (2013), we use the relation to transform the 
Thomas et al.\ model predictions for CN$_2$ to predictions for CN3883.
We use the models for a Salpeter (1955) initial mass function (IMF).
In J\o rgensen \& Chiboucas (2013) we derived model relations linear
in the logarithm of the age, the metallicity [M/H], and the abundance $\rm [\alpha /Fe]$. 
We use these relations (Table 9 in that paper) to aid our analysis in the following.
While traditionally, one may have considered the various indices as primarily tracking one
of the quantities age, [M/H] or $\rm [\alpha /Fe]$,
as shown by the relations, and first described by 
Worthey (1994) and Worthey \& Ottaviani (1997) (see also Tripicco \& Bell 1995), 
all the line indices depend on all three physical quantities.
The Balmer line strengths decrease with increasing ages, while all the metal line strengths
increase. The metal line strengths increase with increasing [M/H] while
the Balmer lines strengths decrease.
The dependence on $\rm [\alpha /Fe]$ is more complicated, with the iron line strengths decreasing with
increasing $\rm [\alpha /Fe]$, and all other metal line strengths increasing.
The Balmer line strengths also increase with $\rm [\alpha /Fe]$, but in general less so than the metal line
strengths.

Having established which set of SSP models we use, we next turn to models for the star formation
history. We evaluate models that assume passive evolution over the time period 
covered by the redshifts of the observed clusters. 
For our adopted cosmology, the look-back time to $z=0.89$, the highest redshift covered by our 
data, is 7.3 Gyr.
Thus, we assume that following the initial period of star formation, the 
bulge-dominated passive galaxies in our samples evolve passively, without any further star formation.
As is common practice, we parameterize these models using a formation
redshift $z_{\rm form}$, which should be understood to be the approximate epoch of the last 
major star formation episode.

\begin{figure*}
\plotone{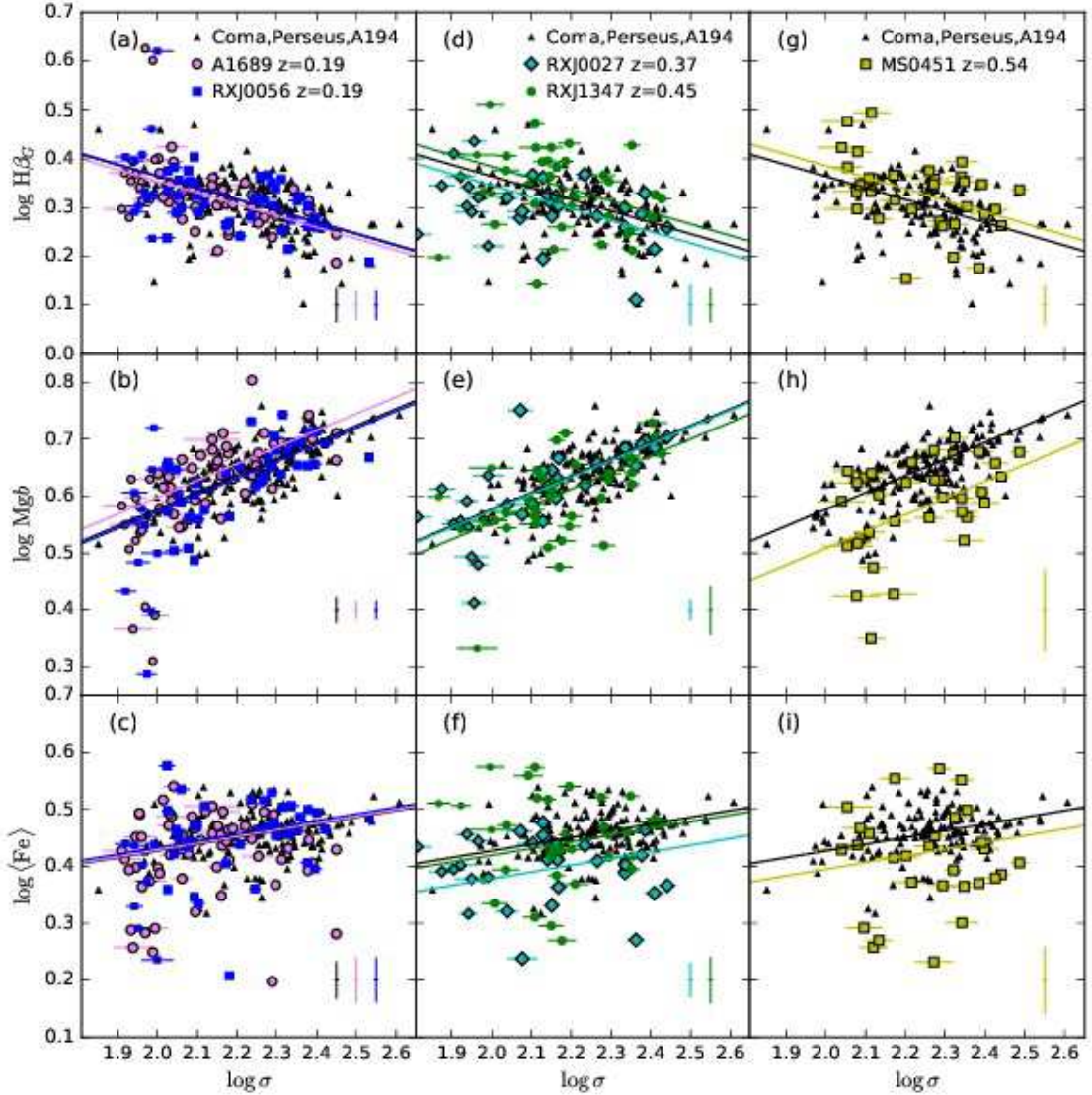}
\caption{ 
Absorption line strengths versus velocity dispersions for the line indices in the visible 
(H$\beta _G$, Mg$b$, and $\langle {\rm Fe} \rangle$).
The figure shows measurements for individual galaxies.
Panels (a)--(c) Pink circles --  Abell 1689; blue squares -- RXJ0056.2+2622.
Panels (d)--(f) Cyan diamonds -- RXJ0027.6+2616; green circles -- RXJ1347.5--1145.
Panels (g)--(i) Yellow squares -- MS0451.6--0305.
Smaller points show data for galaxies with $\log \sigma < 2.0$ (Abell 1689, RXJ0056.2+2622, 
RXJ0027.6+2616, RXJ1347.5--1145).
Black triangles -- The local reference sample shown on all panels for reference.
Typical errors bars are shown on the panels color coded to match the symbols.
Black lines -- best fits for the local reference sample (J\o rgensen \& Chiboucas 2013). 
Lines color-coded to match the symbols  -- the scaling relation offset to the median zero point of each cluster sample.
\label{fig-lsigma_visline} }
\end{figure*}

\begin{figure*}
\plotone{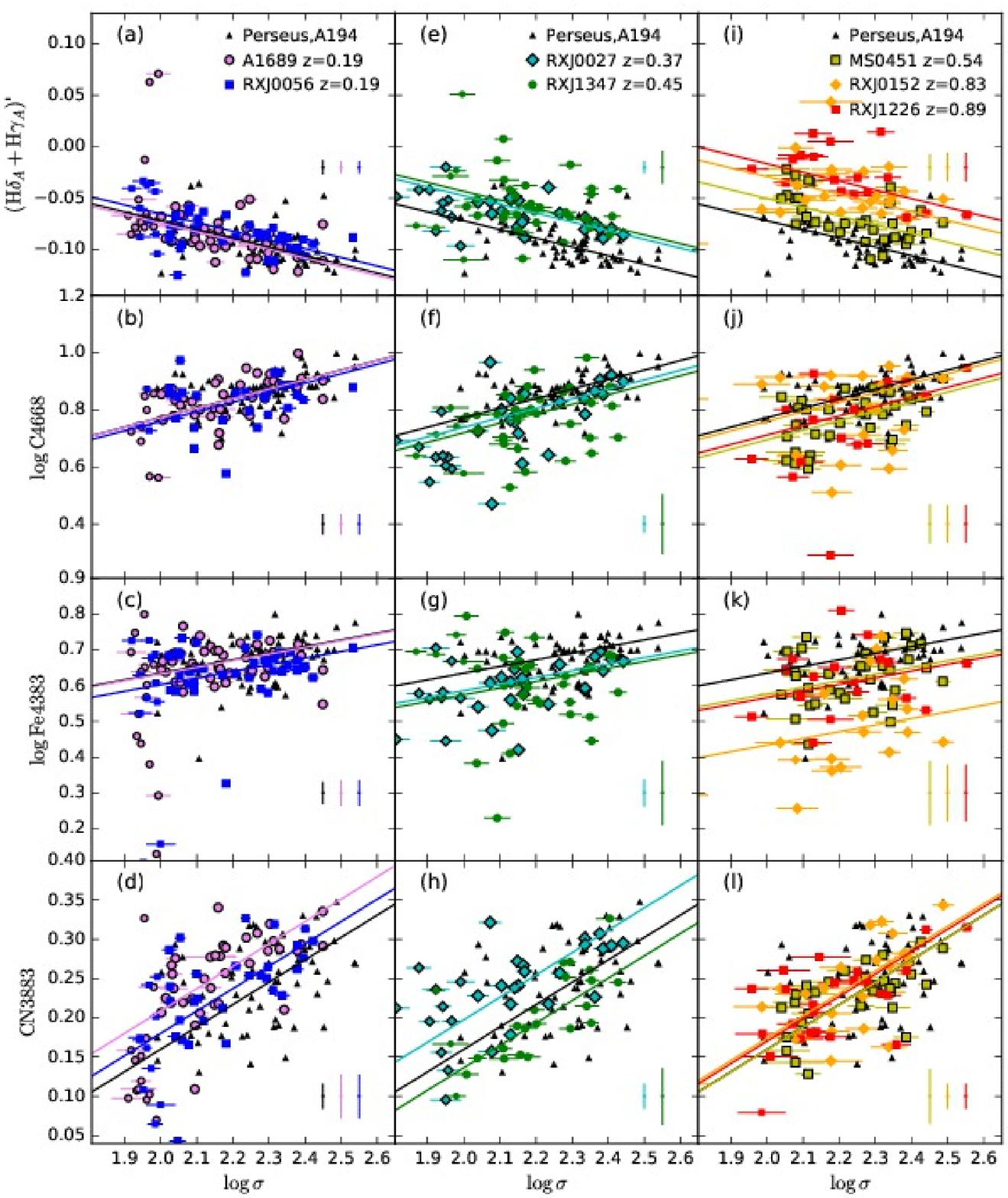}
\caption{ 
Absorption line strengths versus velocity dispersions for the line indices in the blue 
($(\rm{H\delta _A + H\gamma _A})'$, C4668, Fe4383, and CN3883).
The figure shows measurements for individual galaxies.
Panels (a)--(d) Pink circles --  Abell 1689; blue squares -- RXJ0056.2+2622.
Panels (e)--(h) Cyan diamonds -- RXJ0027.6+2616; green circles -- RXJ1347.5--1145.
Panels (i)--(l) Yellow squares -- MS0451.6--0305; orange diamonds -- RXJ0152.7--1357; red squares -- RXJ1226.9+3332.
Smaller points show data for galaxies with $\log \sigma < 2.0$ (Abell 1689, RXJ0056.2+2622, 
RXJ0027.6+2616, RXJ1347.5--1145, or Mass $< 10^{10.3} M_{\sun}$ (RXJ0152.7--1357, RXJ1226.9+3332).
Black triangles -- The local reference sample shown on all panels for reference.
Typical errors bars are shown on the panels color coded to match the symbols.
Black lines -- best fits for the local reference sample (J\o rgensen \& Chiboucas 2013). 
Lines color-coded to match the symbols  -- the scaling relation offset to the median zero point of each cluster sample.
\label{fig-lsigma_blueline} }
\end{figure*}


In our analysis we also use the results from Thomas et al.\ (2005, 2010) on
the relations between the velocity dispersions, ages, metallicities [M/H], and abundance ratios $\rm [\alpha / Fe]$.
In particular, as a consequence of the relation between velocity dispersions and ages at $z\approx 0$
the formation redshift $z_{\rm form}$ must depend on the galaxy velocity dispersion.
With that prediction we can then derive predictions for ages and line indices as a function of
redshift and galaxy velocity dispersion. We show those predictions in relevant figures in the following.

Our analysis implicitly assumes that the galaxies we observe
in the higher redshift clusters can be considered progenitors to the galaxies 
in the clusters at lower redshifts. 
As discussed in detail by van Dokkum \& Franx (2001) this may not be a valid assumption.
In Section \ref{SEC-DISCUSSION} we return to this issue of progenitor bias.

\section{The scaling relations: Line indices versus velocity dispersions \label{SEC-SCALING}}

In Figures \ref{fig-lsigma_visline} and \ref{fig-lsigma_blueline} we show the relations 
between the velocity dispersions and the indices 
in the visual (H$\beta _G$, Mg$b$, $\langle {\rm Fe} \rangle$) and in the blue 
($(\rm{H\delta _A + H\gamma _A})'$, C4668, Fe4383, CN3883), respectively. 
These figures are based on the measurements for the individual galaxies. 
For clarity, we have included on the figures all seven clusters and the local reference sample. 
The panels for MS0451.6--0305, RXJ0152.7--1357 and RXJ1226.9+3332 are similar to 
figures presented in J\o rgensen \& Chiboucas (2013) and show the same data as in that paper.
The zero points and scatter for the four clusters with new data are summarized in Table \ref{tab-relations}.
For the other clusters we refer to Table 10 in J\o rgensen \& Chiboucas (2013).
From Figures \ref{fig-lsigma_visline} and \ref{fig-lsigma_blueline} we find that
the $z=0.19-0.89$ clusters follow scaling relations between velocity dispersions and 
line indices with similar slopes as found for the local reference sample.
In the following, we summarize our results regarding the scatter relative to the
relations, we investigate possible dependence on the cluster environment, and
we establish the zero point differences as a function of redshift.

\subsection{Scatter in the Relations}

\begin{figure}
\begin{center}
\epsfxsize 8.5cm
\epsfbox{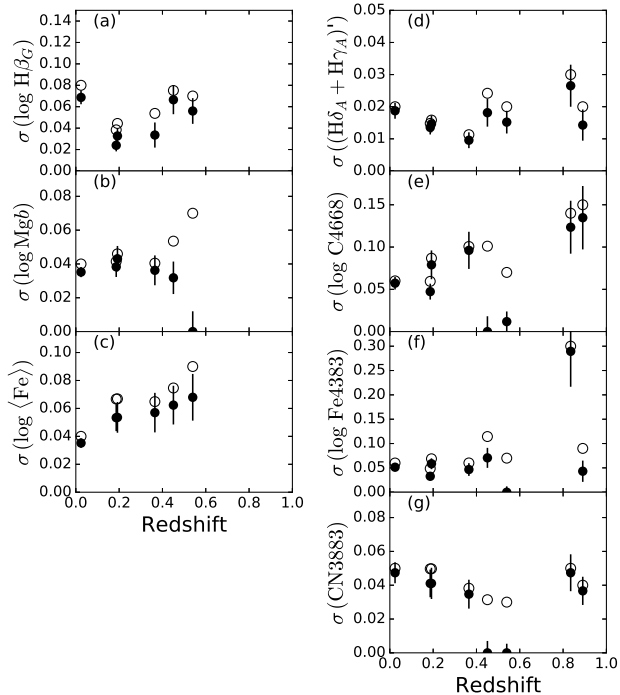}
\end{center}
\caption{
The scatter in the scaling relations for the clusters as a function of redshift. 
Open points -- measured scatter; solid points -- intrinsic scatter.
\label{fig-intrms} }
\end{figure}

Figure \ref{fig-intrms} shows the measured scatter as well as the intrinsic scatter
in the relations as a function of redshift. 
We derive the intrinsic scatter from the measured scatter by subtracting off in
quadrature the adopted typical measurement uncertainties for the line indices.
The uncertainties are given in the Appendix (Table \ref{tab-intcomp}) for the $z=0.19-0.45$ clusters and 
in J\o rgensen \& Chiboucas (2013) for the higher redshift clusters.
For the local reference sample, we adopt as typical measurement uncertainty for each
line index the median of the individual uncertainties.
As shown on Figure \ref{fig-intrms}, in a few cases the resulting intrinsic scatter in a relation is zero. 
This likely is due to overestimated measurement uncertainties.

From Figure \ref{fig-intrms} we conclude that all the relations have intrinsic scatter. 
The scatter in a given relation is similar for all clusters, except 
for the scatter in the Fe4383-velocity dispersion relation for RXJ0152.7--1357 at $z=0.83$. 
In J\o rgensen et al.\ (2005) we found this cluster to have very high
average abundance ratio $\rm [\alpha /Fe]$ due to galaxies with low Fe4383 indices.
These are the galaxies causing the high scatter in the Fe4383-velocity dispersion relation.
We used Kendall's $\tau$ rank order tests to evaluate if correlations may be present between
the intrinsic scatter in any of the relations and the redshifts of the clusters.
We omitted from the tests any measurements with an intrinsic scatter found to be zero,
as well as the very low scatter measurement for the MS0451.6--0305 C4668-velocity dispersion relation.
We find possible correlations with redshift for the scatter in $\rm \langle Fe \rangle$
and C4668, for which the probability of no correlation being present is 0.5\% and 1.5\%, 
respectively. It requires data for significantly more clusters to firmly establish these 
possible correlations. In all other cases no significant correlations were found.

\begin{figure*}
\epsfxsize 13cm
\begin{center}
\epsfbox{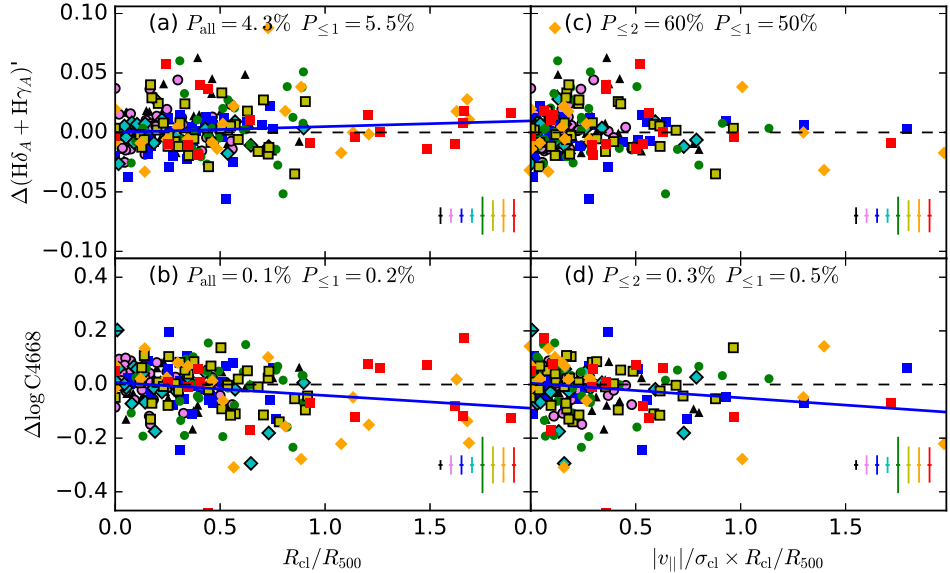}
\end{center}
\caption{ 
Residuals relative to the scaling relations shown for the line measurements based on the individual spectra,
as a function of the cluster center distances $R_{\rm cl}/R_{500}$
or the phase-space parameter $| v_{\rm ||}|/\sigma _{\rm cl} \cdot R_{\rm cl}/R_{500}$.
Symbols as on Figure \ref{fig-lsigma_blueline}.
Typical uncertainties on the residuals are shown on each panel, color coded to match the clusters. 
Blue lines -- best fits in cases where the correlations are significant at least at the 2-sigma level.
For reference, $R_{200}=1.52\, R_{500}$, cf.\ Section \ref{SEC-CLUSTERZ}.
\label{fig-zplsigmalines} }
\end{figure*}

The intrinsic scatter in the relations can in principle be used to set limits on
the scatter in ages and/or metallicities and abundance ratios of the stellar populations in the galaxies.
In doing so we use the model relationships between line index strengths, age, metallicity,
and abundance ratios established in J\o rgensen \& Chiboucas (2013).
If we assume that only the ages vary at a given velocity dispersion, then the scatter in 
the $(\rm{H\delta _A + H\gamma _A})'$-velocity dispersion relation of 0.015 translates to
an age scatter of only 0.12 dex.
Similarly, the scatter in the H$\beta _G$-velocity dispersion relation of 0.045 gives an age 
scatter of 0.20 dex.
These results point towards very large degree of synchronization of the star formation history
for passive galaxies at a given velocity dispersion.
The scatter in the Mg$b$ and Fe4383 relations imply similarly low age scatter.
However, the larger scatter in the other three metal-line relations ($\langle {\rm Fe} \rangle$, C4668, and CN3883)
combined with their dependence on age imply an age scatter of 0.45-0.55 dex.
The dependence on metallicity for these three indices is a factor 2--4 stronger than
their dependence on age (cf.\ J\o rgensen \& Chiboucas 2013).
Thus, the larger scatter in the relations for these indices may be due to scatter in metallicities
as well as ages at a given velocity dispersion.

\subsection{Cluster Environment Dependency of the Relations}

We have investigated the possible cluster environment dependency of the scaling relations.
We use both the cluster center distances, $R_{\rm cl}/R_{500}$, and the radial velocity of the galaxies 
relative to the clusters, $v_{\rm ||}/\sigma _{\rm cl}$ in this test. 
In particular, as shown by Haines et al.\ (2012, 2015) in their 
analysis of clusters extracted from the Millennium Simulations (Springel et al. 2005),
the phase-space parameter $|v_{\rm ||}|/\sigma _{\rm cl} \cdot R_{\rm cl}/R_{500}$ provides a
one-dimensional measure within the phase-space diagram expected to be directly related to
the accretion epoch of a galaxy onto the cluster. 

We use Spearman rank order tests to test for correlations between the residuals for all the scaling relations and 
both measures of the cluster environment.
The test is performed for both the full samples and samples limited to $R_{\rm cl}/R_{500}\le 1.0$,
and to $|v_{\rm ||}|/\sigma _{\rm cl} \cdot R_{\rm cl}/R_{500}\le 2$ or $\le 1$. 
This  is done to ensure that the results are not driven by the relatively few galaxies at values above one
in either environment parameter. 
Figure \ref{fig-zplsigmalines} shows the residuals of the scaling relations
for $(\rm{H\delta _A + H\gamma _A})'$ and C4668 versus the cluster center
distances, $R_{\rm cl}/R_{500}$, and versus $|v_{\rm ||}|/\sigma _{\rm cl} \cdot R_{\rm cl}/R_{500}$.
The panels are labeled with the probabilities that no correlations are present.
The best fit relations are shown.
In addition to the correlations shown on the figure, we also found similar shallow but significant
correlations for the residuals in Mg$b$.  No other correlations were significant.
In the cases of correlations between the residuals and environment parameters, the 
contribution to the intrinsic scatter in the relations originating from the environment
is very small as these relations are very shallow. For example, taking into account
the correlation with environment for the C4668-velocity dispersion relation reduces the scatter
in the relation by only 2 percent.
Thus, in our presentation of the results regarding the zero points (Section \ref{SEC-SCALINGZERO}) 
we do not take into account these very weak dependences on the environment.

\begin{figure}
\epsfxsize 8.5cm
\epsfbox{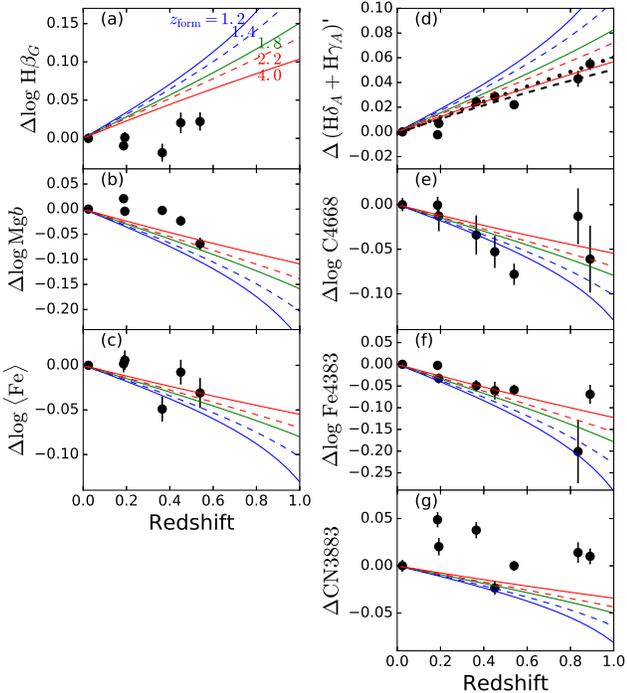}
\caption{
The zero point offsets of the scaling relations for the $z \ge 0.19$ cluster samples relative 
to the local reference sample, shown as a function of redshift. 
Results for the $z=0.5-0.9$ clusters are adopted from J\o rgensen \& Chiboucas (2013).
Predictions from models for passive evolution based on models 
from Thomas et al.\ (2011) 
are overplotted, in panel (a) labeled with the assumed formation redshift $z_{\rm form}$.
Blue solid line -- $z_{\rm form}=1.2$; blue dashed line -- $z_{\rm form}=1.4$;
green line -- $z_{\rm form}=1.8$; red dashed line -- $z_{\rm form}=2.2$; red solid line -- $z_{\rm form}=4.0$.
For CN3883 we adopt the dependence on age established in J\o rgensen \& Chiboucas (2013). 
On panel (d): Black dashed line -- the best fit $z_{\rm form}=6.5$,
black dotted line -- the best fit 1-sigma lower limit $z_{\rm form}=3.1$,
see text for discussion.
\label{fig-zp} }
\end{figure}

\subsection{Zero Points of the Relations \label{SEC-SCALINGZERO} }

On Figure \ref{fig-zp} we show the median zero point differences of the scaling relations for each cluster 
relative to the local reference sample.
The zero point differences shown here for MS0451.6--0305, RXJ0152.7--1357 and RXJ1226.9+3332 are
from J\o rgensen \& Chiboucas (2013).
The zero point differences for the four other clusters (Abell 1689, RXJ0056.2+2622, RXJ0027.6+2616 and RXJ1347.5-1137) 
are based on the new data presented in the current paper. 
The zero points are listed in Table \ref{tab-relations}.
There are no significant zero point differences between the two sub-clusters in RXJ0056.2+2622 or in RXJ0152.7--1357. 
Thus, each of these clusters is treated as one cluster for the purpose of showing the zero point differences.
Predictions of the zero point differences based on passive evolution models with formation redshifts 
between $z=1.2$ and 4 and SSP models from Thomas et al.\ (2011) are overlaid on the figure. 
At fixed metallicity and abundance ratio for a given velocity dispersion, 
these predictions have no significant dependence on the assumed metallicity or abundance ratio,
as shown by the linear parameterization of the models that we derived in J\o rgensen \& Chiboucas (2013).
In the passive evolution models
the Balmer line indices, H$\beta _G$ and  $(\rm{H\delta _A + H\gamma _A})'$, are 
expected to be stronger at higher redshift reflecting the younger stellar populations, while
all the metal indices are expected to be weaker at higher redshift than at present, cf.\ Section \ref{SEC-MODELS}.

The zero point differences for the higher order Balmer lines $(\rm{H\delta _A + H\gamma _A})'$ 
follow the passive evolution model. A $\chi ^2$-fit gives best fit of $z_{\rm form} = 6.5$
with a 1-sigma lower limit of $z_{\rm form} = 3.1$ (Fig.\ \ref{fig-zp}d).  
No upper limit can be established.
However, the zero point differences for the H$\beta _G$-velocity dispersion relation are 
significantly smaller than expected for the passive evolution models (Fig.\ \ref{fig-zp}a).
Equivalently, the H$\beta _G$ indices for the $z=0.19-0.54$ clusters are weaker than expected. 
This may be due to partial emission fill-in, as explored by Concas et al.\ (2017) 
for galaxies in the Sloan Digital Sky Survey (SDSS).
Due to this evidence, we will in the following focus on the higher order Balmer
lines when discussing the age differences.

The scaling relations for the metal line indices have zero point differences
that are less straightforward to interpret.
The indices $\rm \langle Fe \rangle$, C4668 and Fe4383 
are in general weaker at higher redshift than at present, though the zero point differences for the 
scaling relations (Fig.\ \ref{fig-zp}c, e and f) show significantly higher scatter relative to any 
given passive evolution model than is the case for $(\rm{H\delta _A + H\gamma _A})'$.
The data give 1-sigma lower limits
on $z_{\rm form}$ of 2.0 and 4.3 for $\rm \langle Fe \rangle$ and Fe4383, respectively,
consistent with the result based on  $(\rm{H\delta _A + H\gamma _A})'$. 
The best fit result for C4668 is $z _{\rm form} = 1.4_{-0.3}^{+0.8}$, marginally 
inconsistent with the result based on  $(\rm{H\delta _A + H\gamma _A})'$. However,
the result is to a large extent driven by the large zero point difference found for MS0451.6--0305,
for which the scaling relation also has a rather low scatter.
The Mg$b$ and the CN3883 indices are for the majority of the $z\ge 0.19$ clusters significantly stronger
than expected based on the passive evolution models (Fig.\ \ref{fig-zp}b and g).
Thus, these indices cannot be used to constrain $z_{\rm form}$, but instead may be 
an indicator of the limitations of the models.
Part of the differences in behavior of the metal line indices relative to the $(\rm{H\delta _A + H\gamma _A})'$ index 
may also be due to cluster-to-cluster variations in the metallicities and 
abundance ratios. We return to this question in Section \ref{SEC-INDICESENV}.

\begin{figure*}
\begin{center}
\epsfxsize 13.0cm
\epsfbox{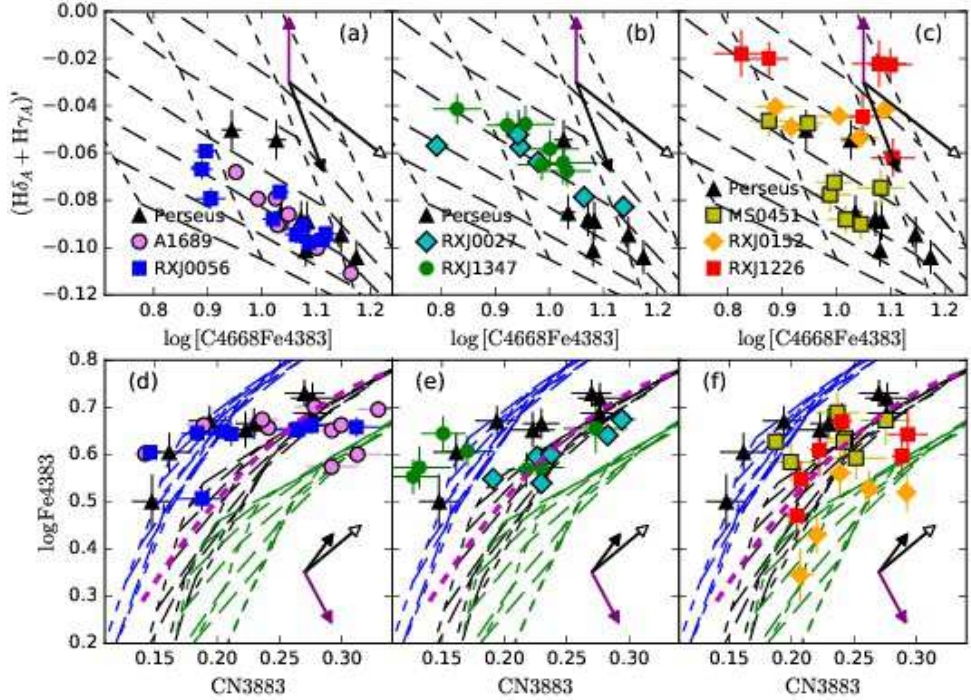}
\end{center}
\caption{ 
Absorption line strengths versus each other for the line indices derived from the composite
spectra, or for the local reference sample the luminosity weighted average indices.
Black triangles -- local reference sample (Perseus), pink circles -- Abell 1689; blue squares --
RXJ0056.2+2622; cyan diamonds -- RXJ0027.6+2616; green circles -- RXJ1347.5--1145; yellow squares --
MS0451.6--0305; orange diamonds -- RXJ0152.7--1357; red squares -- RXJ1226.9+3332.
Overlaid grids show SSP models from Thomas et al.\ (2011).
Short-dashed lines -- constant metallicity [M/H] = --0.33, 0, 0.35, 0.67 shown.
Long-dashed lines -- constant age, ages from 2 to 15 Gyr shown.
Green grid -- $\rm [\alpha / Fe] = 0.0$,
black grid -- $\rm [\alpha / Fe] = 0.3$, and 
blue grid -- $\rm [\alpha / Fe] = 0.5$.
The arrows on each panel show the effect on the line indices from changes 
in age (filled black), metallicity [M/H] (open black) and $\rm [\alpha / Fe]$ (purple). 
In all cases, we show the effect of changes of 0.3 dex. 
Purple lines on panels (d)-(f) show the second order fit to the models 
for $\rm [\alpha / Fe] = 0.3$. These fits are used in the process of determining
$\rm [\alpha / Fe]$, see text.
\label{fig-allline_comp} }
\end{figure*}

\section{Stellar populations: Ages, metallicities and abundance ratios \label{SEC-INDICES} }

In this section we estimate ages, metallicities [M/H], and abundance
ratios $\rm [\alpha / Fe]$ of the galaxies in the clusters. 
We then establish how those parameters depend on velocity dispersions of the galaxies.
We investigate the changes with the redshifts of the clusters and possible
cluster-to-cluster differences.

\subsection{Determination of Ages, Metallicities and Abundance Ratios}
 
In this investigation, we use the line indices derived from the composite spectra for the 
$z=0.19-0.89$ clusters and the luminosity weighted average line indices for the local
reference sample, cf.\ Section \ref{SEC-COMPOSITES}.
This is done primarily to gain S/N as even with our fairly high S/N individual spectra,
ages, [M/H] and  $\rm [\alpha / Fe]$  for the individual galaxies would have uncertainties 
as large as 0.3 dex. When we use the indices from the composite spectra and the luminosity weighted average indices, 
the resulting uncertainties on the ages, [M/H] and  $\rm [\alpha / Fe]$ are typically 0.06 dex.

\begin{figure}
\epsfxsize 7cm
\begin{center}
\epsfbox{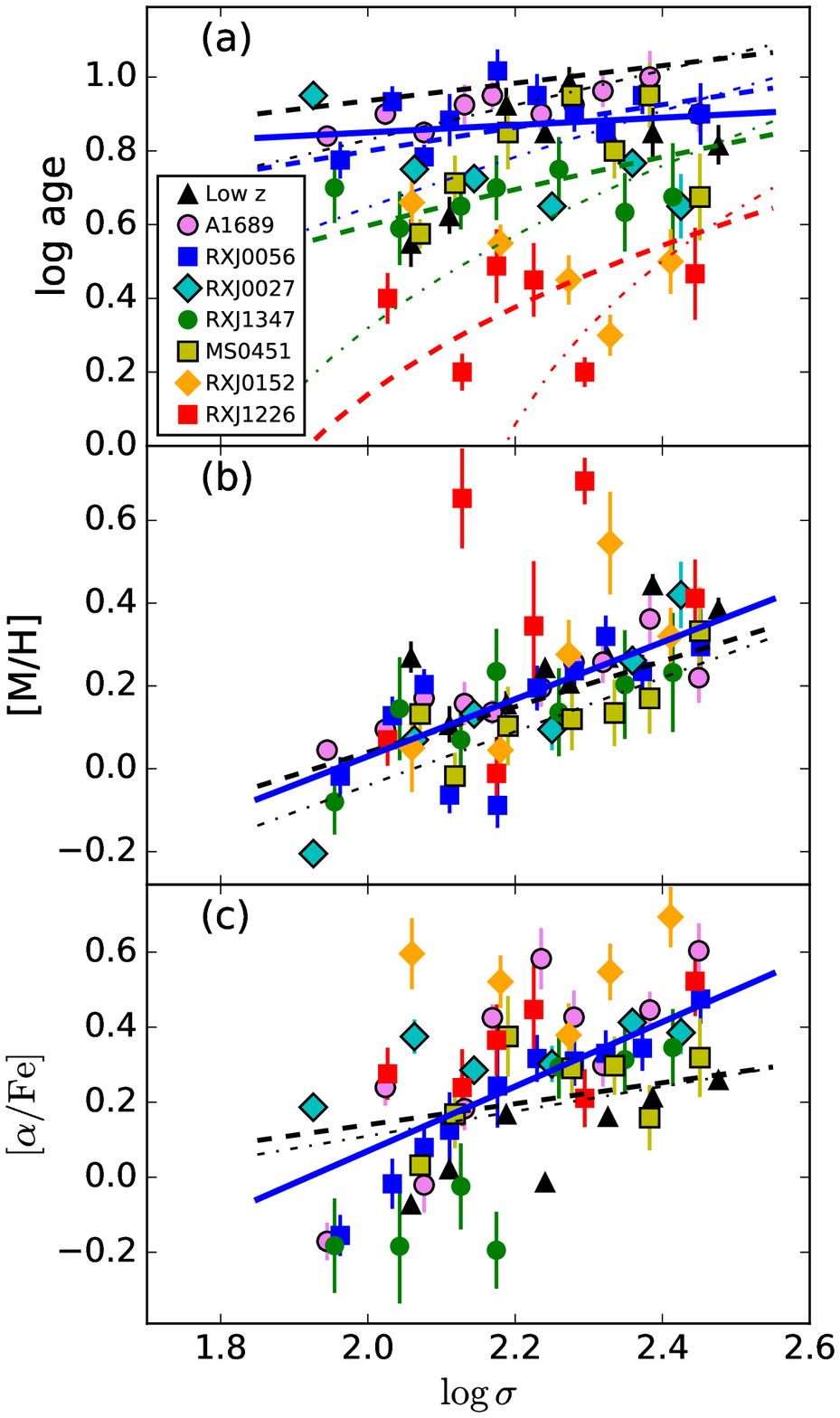}
\end{center}
\caption{ 
Ages, metallicities [M/H], and abundance ratios $\rm [\alpha /Fe]$ derived from the line indices for the composites, 
or for the local reference sample from luminosity weighted average line indices,
shown as a function of the average $\log \sigma$ for the spectra making up the composites or averages.
Symbols as on Fig.\ \ref{fig-allline_comp}.
Blue solid lines -- best least squares fits to the data. In the case of the ages, the zero points were allowed
to vary from cluster to cluster and the relations are shown for the average zero point for the $z \le 0.19$ clusters
(Perseus, Abell 1689, RXJ0056.2+2622).
Black dashed lines -- Thomas et al.\ (2005) relations between age, [M/H] and $\rm [\alpha /Fe]$ and velocity dispersion.
In panel (a), the models are also shown for redshifts $z=0, 0.2, 0.5, 0.9$ as black, blue, green and red, respectively,
under the assumption of passive evolution.
Thin dot-dashed lines -- similar models, but based on relations from Thomas et al.\ (2010).
\label{fig-lsigmaagemetal_allcomp} }
\end{figure}

We limit our analysis to determinations based on indices in the blue wavelength region
($(\rm{H\delta _A + H\gamma _A})'$, [C4668Fe4383], Fe4383, CN3883), as these are available
for all $z=0.19-0.89$ clusters in our sample.
Figure \ref{fig-allline_comp} shows the indices versus each other in the relevant
combinations, and with SSP models from Thomas et al.\ (2011) overlaid.
As done in J\o rgensen \& Chiboucas (2013) we use the higher order Balmer lines versus 
the combination index [C4668Fe4383] to determine ages and [M/H], as the combination index
is constructed to be independent of $\rm [\alpha / Fe]$. 
We use the model grid for $\rm [\alpha / Fe] = 0.3$ for this determination.
To determine $\rm [\alpha / Fe]$ we use CN3883 versus Fe4383. 
We first fit the $\rm [\alpha /Fe] = 0.3$ models with a second order polynomial. 
Then the abundance ratios are determined from the 
distance between a given data point and this polynomial, measured along the 
direction of the $\rm [\alpha / Fe]$ dependency (shown as purple arrows on the figures). 
To estimate the uncertainties we add and subtract the adopted uncertainties
on the line indices and derive age, [M/H] and $\rm [\alpha / Fe]$ at these extreme points
in the parameter spaces. The maximum absolute differences between results from the extreme points
and the result from the measured indices is used as the uncertainty on a given parameter. 
See J\o rgensen \& Chiboucas (2013) for details.

In the following sections we investigate how the derived ages, metallicities and abundance ratios
change with velocity dispersion and with redshift.

\begin{deluxetable}{lrr}
\tablecaption{Relations for Age, Metallicities and Abundance ratios \label{tab-agemetalalpharelations} }
\tablewidth{0pc}
\tabletypesize{\scriptsize}
\tablehead{
\colhead{Relation} & \colhead{$\rm rms_{meas}$ } & \colhead{$\rm rms_{int}$} \\
\colhead{(1)} & \colhead{(2)} & \colhead{(3)} 
}
\startdata
$\log {\rm age} = (0.10 \pm 0.09) \log \sigma +0.65$\tablenotemark{a} &  0.10 & 0.08 \\ 
${\rm [M/H]} = (0.69 \pm 0.09) \log \sigma -1.35$ &  0.13 & 0.12 \\ 
${\rm [\alpha/Fe]} = (0.86 \pm 0.16) \log \sigma -1.65$ &  0.17 & 0.15 \\ 
\enddata
\tablecomments{Column 1: Best fit relation. 
Column 2: Scatter of the relation as rms in the Y-direction. 
Column 3: Estimated intrinsic scatter. 
}
\tablenotetext{a}{Relation fit as parallel lines allowing different the zero points for each cluster. The table
lists the average zero point for $z \le 0.19$ clusters (Perseus, Abell 1689 and RXJ0056.2+2622).}
\end{deluxetable}

\subsection{Correlations with Velocity Dispersions}

In Figure \ref{fig-lsigmaagemetal_allcomp} we show the derived ages, metallicities [M/H] and 
abundance ratios $\rm [\alpha / Fe]$ 
versus the mean velocity dispersion of the spectra making up each composite spectrum or
entering into the luminosity weighted average line indices. 
The best least squares fits to the data are shown as solid blue lines on on the figure.
The relations are summarized in Table \ref{tab-agemetalalpharelations}.
The fit to ages versus velocity dispersion was established as a set of parallel lines, allowing the
zero points for the clusters to be different for each cluster. The fit is shown for the 
average zero point for the $z \le 0.19$ clusters (Perseus, Abell 1689, RXJ0056.2+2622). 
Correlations are present for [M/H] and $\rm [\alpha / Fe]$,
while the ages are consistent with no correlation with the velocity dispersion.

We used the Thomas et al.\ (2005) relations between age and velocity dispersion at $z\approx 0$
to derive predictions for the stellar population ages as a function of velocity dispersion
and redshift, under the assumption of passive evolution.  The predictions are shown on panel
(a), color coded to match the clusters at $z\approx 0.0$, 0.2, 0.5 and 0.9. 
The predictions are in general consistent with the ages, though our results 
are also consistent with flat relations offset from each other as a function of the cluster redshifts.
We have also shown predictions using the steeper age-velocity dispersion relation at 
$z\approx 0$ from Thomas et al.\ (2010), thin dot-dashed lines on the figure.
This relation results in much steeper relations at higher redshift, in disagreement with the data at $z>0.5$. 
In Section \ref{SEC-DISCUSSION}, we address the possible reasons for this disagreement.
The intrinsic scatter in the age-velocity dispersion relations correspond to a scatter in
age of 0.08 dex, cf.\ Table \ref{tab-agemetalalpharelations}.

The relation between the metallicities [M/H] and the velocity dispersions is almost identical
to the one found by Thomas et al.\ (2005, 2010).
The relation for the abundance ratios $\rm [\alpha /Fe]$ is marginally steeper than 
found by Thomas et al. The steeper slope is almost exclusively driven
by a few of the measurements at low velocity dispersions.
The scatter relative to the relations listed in Table \ref{tab-agemetalalpharelations}
implies intrinsic scatter of [M/H] and $\rm [\alpha /Fe]$ at fixed velocity dispersion
of 0.12 dex and 0.15, respectively.  The scatter should be taken as lower limits since
they are based on determinations from composite spectra (or luminosity weighted average line indices), 
the construction of which may have eliminated some of the intrinsic scatter present for the individual galaxies.
Even so, we conclude that the [M/H]--velocity dispersion and $\rm [\alpha /Fe]$--velocity dispersion
relations are steep and tight at all redshifts covered by our sample. 
Neither relation depends on the redshift.

\begin{figure}
\epsfxsize 7cm
\epsfbox{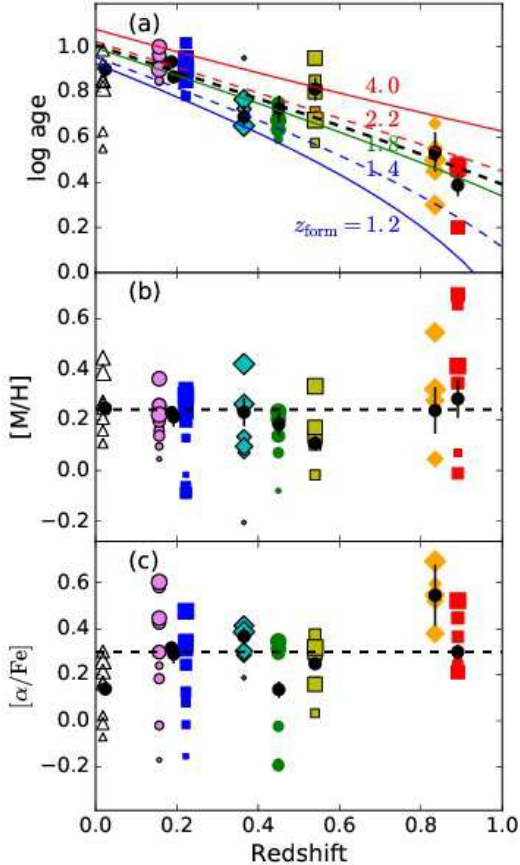}
\caption{
Ages, metallicities and abundance ratios for the clusters, versus redshift. 
Black circles -- median values for each cluster based on the relations between the 
velocity dispersions and the line indices and equivalent to $\log \sigma = 2.24$, see text.
Median values for the $z=0.5-0.9$ clusters are adopted from J\o rgensen \& Chiboucas (2013).
Color coded points are based on the composite spectra ($z=0.19-0.89$ clusters) or the luminosity
weighted average line indices (Perseus). The points are color coded as in Figure \ref{fig-allline_comp}.
The sizes reflect the mean velocity dispersions of the galaxies included in each composite or average.
Predictions for passive evolution are overplotted 
on panel (a), labeled with the assumed formation redshifts $z_{\rm form}$.
The dashed black line on panel (a) shows the best fit passive evolution model, $z_{\rm form} = 1.96$.
The dashed black lines on panels (b), and (c) mark the median [M/H] and $\rm [\alpha /Fe]$ 
for all the clusters based in the individual values equivalent to $\log \sigma = 2.24$ (black circles).
\label{fig-agemetalalpha} }
\end{figure}

\subsection{Variation with Redshift and Cluster-to-Cluster Differences  \label{SEC-INDICESENV} }

Next we investigate the redshift dependency of ages, metallicities [M/H] and abundance ratios
$\rm [\alpha /Fe]$. Figure \ref{fig-agemetalalpha} summarizes the determinations based on the composite 
spectra, as well as determinations based on the index values at $\log \sigma =2.24$ on the scaling
relations for the clusters. Thus, the figure shows both representative median values for each
cluster based on the scaling relations, and the typical scatter in the parameters, 
primarily due to the dependence on the velocity dispersions as reflected in the symbol sizes.
On panel (a) we show the predictions from passive evolution models with formation 
redshifts $z_{\rm form}=1.2-4.0$. 
As expected the dominating trend is the change in age with redshift.
A best fit to the median points for the clusters gives a formation redshift of 
$z_{\rm form}=1.96_{-0.19}^{+0.24}$, shown as the dashed black line on panel (a).
This formation redshift is lower than found from the
zero point differences of the $(\rm{H\delta _A + H\gamma _A})'$--velocity dispersion relation, 
see Figure \ref{fig-zp}d,
for which we find $z_{\rm form} \ge 3.1$ as the 1-sigma lower limit.
However, it is also clear that the local reference sample contains galaxies too young to be the 
descendants of the galaxies in the high redshift clusters, if only passive evolution is at work.
We discuss this in more detail in Section \ref{SEC-DISCREDSHIFT}.

The median value of the metallicities [M/H] (Figure \ref{fig-agemetalalpha}b) is 0.24,
marked with a dashed line.  However, as noted in J\o rgensen \& Chiboucas (2013), 
MS0451.6--0305 at $z=0.54$ has significantly lower metallicity at [M/H]$\approx 0.1$.

The median value of the abundance ratios  $\rm [\alpha /Fe]$ (panel c) is 0.3,
marked with a dashed line.
RXJ0152.7--1357 at $z=0.83$ has significantly higher abundance ratio of $\rm [\alpha /Fe] \approx 0.55$ (see also
J\o rgensen et al.\ 2005). 
RXJ1347.5--1347 at $z=0.45$ and the local reference sample (Perseus and Abell 194)
have  $\rm [\alpha /Fe] = 0.14$.
Additional observations to confirm or refute these results would be very valuable.

\section{Discussion \label{SEC-DISCUSSION} }

Here we discuss our results in relation to the questions of the 
average evolution as a function of redshift, the role of environment and/or galaxy
velocity dispersion in driving the evolution of galaxies, and the dependency
of ages, metallicities  [M/H] and abundance ratios $\rm [\alpha / Fe]$
on velocity dispersions of the galaxies. 

\subsection{The Average Evolution as a Function of Redshift \label{SEC-DISCREDSHIFT} }

The zero point differences of the relations between the galaxy velocity dispersions and the 
line indices support a high formation redshift. The best fit of the zero point differences for
the $(\rm{H\delta _A + H\gamma _A})'$--velocity dispersion relation gives $z_{\rm form}=6.5$ and a 1-sigma
lower limit of $z_{\rm form} \ge 3.1$, cf.\ Figure \ref{fig-zp}. 
However, if we use the ages derived from the composite spectra we find $z_{\rm form} = 1.96_{-0.19}^{+0.24}$,
cf.\ Figure \ref{fig-agemetalalpha}. 
The two results can be reconciled by realizing that the zero point differences
are strongly affected by the fact that the local reference sample of galaxies in the Perseus cluster 
contains galaxies too young to have their progenitors included in the higher redshift samples, i.e.\ by
progenitor bias in the samples. 
See van Dokkum \& Franx (2001) for one of the earliest discussions of progenitor bias.
The difference in look-back time between the two high redshift clusters (RXJ0152.7--1357
and RXJ1226.9+3332 at an average redshift $z=0.86$) and Perseus ($z=0.018$) is 6.9 Gyr for
our adopted cosmology.
To be included in the passive galaxy sample at any redshift a galaxy would most likely 
have to contain stellar populations on average at least 1 Gyr old, 
since at younger ages we expect emission lines strong enough 
that our criteria on EW[\ion{O}{2}] would exclude such galaxies from the sample. 
Thus, as a test we remove from the local reference sample the 34 galaxies for which
$(\rm{H\delta _A + H\gamma _A})'$ versus [C4668Fe4383] indicate ages less than 8 Gyr.
The zero point of the $(\rm{H\delta _A + H\gamma _A})'$-velocity dispersion relation 
for the remaining 31 galaxies is 0.005 lower than found from the full sample.
This in turns affects the zero point differences for the two high redshift clusters.
If we adjust the zero point differences for just those two clusters, realizing that the 
differences for the other clusters also ought to be offset, but with smaller amounts, then
we find a best fit 1-sigma lower limit of $z_{\rm form} \ge 2.0$.
Thus, the difference between the determination of $z_{\rm form}$ from
the scaling relations and the formation redshift derived from the ages, may
be fully explained as an effect of the progenitor bias.
In addition, the median age of the galaxies in the local reference sample older than
8 Gyr is 10 Gyr, placing the median value exactly on the best fit of 
$z_{\rm form}$ from the ages based on the blue indices, Figure \ref{fig-agemetalalpha}.
We also note that the ages of the galaxies in the local reference sample indicate
that about half of such passive cluster galaxies at $z\approx 0$ have become
passive since $z\approx 0.9$. This is similar to the conclusion reached by
S\'{a}nchez-Bl\'{a}zquez et al.\ (2009) in their analysis of line indices of the EDisCS sample of
passive galaxies. 
It is possible that the main stellar mass of such apparently young galaxies was formed 
earlier, and that the measured low ages are caused by later star formation episodes
involving a small fraction of the mass, see discussion by, e.g., Serra \& Trager (2007).

The zero point differences of scaling relations involving the iron indices ($\rm \langle Fe \rangle$, Fe4383)
give formation redshifts consistent with the scaling relation for $(\rm{H\delta _A + H\gamma _A})'$,
while the scaling relation for C4668 gives a marginally lower formation redshift.
We note that some of the largest outliers relative to the expected zero point differences for passive 
evolution are clusters that are also outliers relative to the median
metallicity [M/H] and abundance ratio  $\rm [\alpha / Fe]$ of the clusters.
Specifically, RXJ0152.7--1357 ($z=0.83$), which we found to have very high $\rm [\alpha / Fe]$, is
offset to weaker Fe4383 and stronger C4668 than expected from $z_{\rm form} \approx 2$, 
Figure \ref{fig-zp}e and f.
RXJ0027.6+2616 ($z=0.37$) and MS0451.6--0305 ($z=0.54$), which we found to have lower than median [M/H], is offset
to weaker Mg$b$, $\langle {\rm Fe} \rangle$, and/or C4668 than expected.

In the following, we adopt $z_{\rm form} = 1.96_{-0.19}^{+0.24}$ 
as our best estimate of a common formation redshift for the galaxies, under the assumption of passive evolution.
In J\o rgensen \& Chiboucas (2013) we determined the formation redshift from the Fundamental Plane for
the three highest redshift clusters, and found $z_{\rm form} = 1.95^{+0.3}_{-0.2}$ for the massive galaxies
and a lower $z_{\rm form} = 1.24 \pm 0.05$ for the less massive galaxies.
Our results here based on direct age estimates are in agreement with these results though we cannot put
tight constraints on a possible mass (or velocity dispersion) dependency on $z_{\rm form}$ as done from the FP.
Other studies of the FP for clusters up to $z\approx 1$ give similar results for the massive galaxies, 
see van Dokkum \& van der Marel (2007), and references therein.

\subsection{The Role of Cluster Environment and Cluster-to-Cluster Differences \label{SEC-DISCENV} }

We found that the residuals relative to the Mg$b$-velocity dispersion and the C4668-velocity dispersion relation 
correlate with the cluster center distances and with  
the phase-space parameter $|v_{\rm ||}|/\sigma _{\rm cl} \cdot R_{\rm cl}/R_{500}$.
A weaker correlation may also be present for the residuals of the scaling relation for $(\rm{H\delta _A + H\gamma _A})'$.
However, the slopes of the relations
are very shallow and the correlations can only explain a very small fraction of the 
intrinsic scatter in the relations.
None of the other scaling relations show significant dependences on the cluster environment.
We caution that our coverage in cluster 
center distances is fairly limited reaching only to $R_{500}$ for the majority of the clusters,
and $1.8 R_{500}$ for the two highest redshift clusters in the sample.
In addition, the clusters were on purpose chosen to all
be very massive and thus limit any effect of cluster environment on our results.
Our results are in agreement with the study of four nearby clusters by Smith et al.\ (2006).
These authors also found that the Mg$b$-velocity dispersion and the C4668-velocity dispersion relations 
depend on the cluster center distances, while the scaling relations for the iron lines do not.
They find weaker dependancies on cluster center distances for the Balmer line-velocity dispersion relations,
also in general agreement with our results. Their study reaches cluster center distances of $1.5 R_{500}$.
Harrison et al.\ (2011) in their study of four nearby clusters, comment that the only
difference they found between the outskirts and the cores of these clusters was
that the relations between the velocity dispersions and age, [M/H], and $\rm [\alpha /Fe]$ 
were weaker in the outskirts than the cores. 
This study reaches ten times the virial radius, or $\approx 15\,R_{500}$.
McDermid et al.\ (2015) on the other hand, in
their study of the ATLAS-3D sample found the Virgo cluster galaxies to be older and
have higher $\rm [\alpha / Fe]$ abundance ratios than the field galaxies. 

We do find a few significant cluster-to-cluster differences in the average metallicities [M/H] 
and abundance ratios $\rm [\alpha / Fe]$. 
We cannot tie these differences to specific properties of these clusters.
RXJ0152.7--1357 is a double cluster probably in the process of merging (e.g., Girardi et al.\ 2005) 
and the high $\rm [\alpha / Fe]$ may be related to this event. 
However, RXJ0056.9+2622 is also a double
cluster, presumably in the process of merging (Barrena et al.\ 2007), but its [M/H] and 
$\rm [\alpha / Fe]$ agree with the sample-wide values.
The clusters with lower than average [M/H], MS0451.6--0305, has a
X-ray structure consistent with a relaxed cluster with no sub-structure.
Thus, there are no obvious reasons that the metallicity should be different from the average
and no obvious pathway to increasing the metallicity to be similar to our local reference sample.
We speculate that these cluster-to-cluster differences are stochastic. 
They may have been established in sub-structures that later merged to form the larger clusters.
Moran et al.\ (2007) in their study of the two clusters MS0451.6--0305 and CL0024+17 found 
evidence of cluster-wide differences in the star formation history of the galaxies, 
but also no clear trace of what has caused the differences.
More in-depth studies of massive intermediate redshift clusters are obviously needed to quantify the
frequency of such differences between clusters and maybe understand their origin.

\subsection{The Velocity Dispersion as the Driver of Galaxy Evolution \label{SEC-DISCVELDISP} }

Many authors have found ages, metallicities and abundance ratios of bulge-dominated passive
galaxies to be correlated with the velocity dispersions of the galaxies.
Harrison et al.\ (2011) provide an overview of determinations prior to the publication
of that paper, including the results from Thomas et al.\ (2005, 2010).
Harrison et al.\ also present new results for galaxies
in four nearby clusters. McDermid et al.\ (2015) give results based on the
ATLAS-3D sample of nearby bulge-dominated galaxies.
We have in Section \ref{SEC-INDICES} used the results from Thomas et al.\ (2005, 2010) as 
representative for the many results in the literature (see references in Harrison et al.). 
In general, the studies agree on an $\rm [\alpha / Fe]$ dependency on the velocity dispersion 
with a slope of $\approx 0.3-0.35$ with a median value of 0.33.
The slopes for [M/H] varies in the interval 0.2-0.8, with a median value of $\approx 0.6$.
The slopes for the ages varies between 0.24 (Thomas et al.\ 2005) to very steep slopes of $\approx 1$ 
(Bernardi et al.\ 2006; McDermid et al.\ 2015). The median value is $\approx 0.4$. 
It is beyond the scope of the present paper to investigate all the possible reasons for the disagreement between these studies. 
However, it is clear that authors who find a very steep slope for the age dependency on the velocity dispersion 
include nearby young galaxies (ages less than $\approx 5$ Gyr) whose progenitors would not be in our
$z>0.5$ samples, as they are not expected to be passive galaxies at those redshifts.

Based on ages, [M/H] and $\rm [\alpha / Fe]$ derived from our composite spectra
we find steep and tight relations between [M/H] and velocity dispersions and between
 $\rm [\alpha / Fe]$ and velocity dispersions.
However, the relation between ages and velocity dispersions is very shallow, with the 
slope not being significantly different from zero.
As shallow slopes are found also for the higher redshift clusters in the sample, this result
significantly constrains the slope of the relation at low redshift to be no steeper 
than found by Thomas et al.\ (2005).

From the scatter in the relations between ages, [M/H] and $\rm [\alpha / Fe]$  and velocity dispersions
we find the intrinsic scatter of the three parameters to be 0.08, 0.12 and 0.15 dex, respectively.
The scatter should in all cases be understood as lower limits because of our use of composite spectra, or 
luminosity weighted average indices, in the determination.
Both Harrison et al.\ (2011) and McDermid et al.\ (2015) find the intrinsic scatter in 
the ages to be significantly higher at 0.20 dex.
Presumably this apparent disagreement is due to a combination of our use of composite spectra
and the inclusion of many very young galaxies in the samples of Harrison et al.\ and McDermid et al.
Both studies find the intrinsic scatter in [M/H] and $\rm [\alpha / Fe]$ to be slightly lower than our results, 
0.10 and 0.07, respectively. We here quote the McDermid et al.\ values from 
line indices calibrated to half an effective radius as comparable to our measurements.
The main conclusion from all these results is that the velocity dispersion and the properties 
of the stellar populations are tightly correlated. The star formation history of a bulge-dominated 
galaxy must to a high degree be determined by the velocity dispersion of the galaxy. 
This is to a large extent in agreement with the results presented by Muzzin et al.\ (2012) who, 
based on a study of cluster galaxies at $z=0.8-1.2$, concluded that properties of passive
bulge-dominated galaxies are determined by their mass (rather than their environment).
We caution that mass and velocity dispersion cannot be assumed to be fully equivalent though they are 
tightly correlated.

\section{Summary and Conclusions \label{SEC-CONCLUSION}}

We have presented a joint analysis of stellar populations in passive bulge-dominated
galaxies in seven massive clusters at redshifts of $z=0.19-0.89$. 
The analysis is based on our new deep ground-based optical spectroscopy
for the clusters at $z=0.19-0.45$ combined with our previously published data
for the three clusters at $z=0.54-0.89$ (J\o rgensen et al.\ 2005; J\o rgensen \& Chiboucas 2013).
We have analyzed stellar populations of member galaxies using our measurements of absorption line 
strengths and velocity dispersions.
Our main conclusions from the analysis are as follows:

\begin{enumerate}
\item
The $z=0.19-0.89$ cluster galaxies follow relations between velocity
dispersions and line indices with slopes consistent with the relations 
for our local reference sample. 
The scatter in the relations is in general also consistent
with the local reference sample, except for the double cluster RXJ0152.7--1357.
This cluster has higher scatter in the velocity dispersion-Fe4383 relation,
associated with the unusually weak Fe4383 indices for some of the galaxies
in the cluster.
\item
We determine line indices from composite spectra, stacked according to velocity
dispersion.
We derive ages, metallicities [M/H], and abundance ratios $\rm [\alpha /Fe]$
from the line indices from the composite spectra.
The [M/H]--velocity dispersion and $\rm [\alpha /Fe]$--velocity dispersion relations are steep and tight
at all redshifts and show no significant changes with redshift.
At fixed velocity dispersion, the intrinsic scatter of [M/H] is 0.12 dex, while
the scatter in $\rm [\alpha /Fe]$ is 0.15 dex.
The age dependency on velocity dispersion is very shallow at all redshifts with a slope not statistically
different from zero, while the zero point changes with redshift reflect passive evolution.
The intrinsic age scatter is 0.08 dex.
The low scatter in all three parameters indicate a large degree of synchronization in the 
evolution of the galaxies.
\item 
We have used the zero point differences for the line index--velocity dispersion relations 
as well as the direct age estimates to investigate the mean age 
variation with redshift and derive formation redshifts under the assumption
of passive evolution. 
The zero point differences for the $(\rm{H\delta _A + H\gamma _A})'$-velocity dispersion relation 
give a 1-sigma constraint for the formation redshift of $z _{\rm form} \ge 3.1$, under
the assumption of passive evolution of the galaxies.
Relations for $\langle {\rm Fe} \rangle$, C4668, and Fe4383 are mostly consistent
with this result, with outliers related to possible cluster-to-cluster
differences in metallicities and/or abundance ratios.

The average ages of the stellar populations (derived from 
$(\rm{H\delta _A + H\gamma _A})'$ and [C4668Fe4383])
as a function of redshift give a formation redshift of $z_{\rm form} = 1.96_{-0.19}^{+0.24}$,
under the assumption of passive evolution.
The difference between this result and the result based on the
zero point differences for the $(\rm{H\delta _A + H\gamma _A})'$-velocity dispersion relation 
can be fully explained by progenitor bias and the fact that 
about half of the galaxies in our local reference sample have ages too young
for their progenitors to be part of the passive population at $z\approx 0.8-0.9$.
When correcting for this effect the zero point differences for the 
$(\rm{H\delta _A + H\gamma _A})'$-velocity dispersion relation give a 
1-sigma constraint on the formation redshift of $z_{\rm form} \ge 2.0$.
\item
The median metallicity of the stellar populations is [M/H]=0.24, with no dependence
on redshift. The median abundance ratios is $\rm [\alpha /Fe] = 0.3$, also independent
on redshift.
We do find indications of cluster-to-cluster differences in [M/H] and $\rm [\alpha /Fe]$.
MS0451.6--0305 have significantly lower [M/H] than the median value.
RXJ0152.7--1357 has higher $\rm [\alpha /Fe]$ than the median, while
RXJ1347.5--1145 has lower  $\rm [\alpha /Fe]$.
It is not clear if these are stochastic variations, if the differences are related
to specific cluster properties, or which processes may be able to remove or
establish such differences at times scales equivalent to the look-back times
to these clusters, ie.\ 4-7 Gyr. 
\item 
We find weak and shallow dependencies on the cluster environment of 
the residuals in velocity-dispersion-line index
relations for the Mg$b$, C4668, and $(\rm{H\delta _A + H\gamma _A})'$ indices.
The dependencies account for only 2\% of the intrinsic scatter in the relations.
We caution that our samples cover only cluster center distances out to 
$\approx R_{500}$ in the $z\le 0.54$ clusters, and only out to
$\approx 1.8 R_{500}$ in the two higher redshift clusters.
\end{enumerate}

We conclude that (at least) these four areas of improvements may be needed in order to gain 
a more complete understanding of the evolution
of the stellar populations of passive galaxies over the last half of the age of 
the Universe: (1) Development and use of stellar population models as SEDs,
correctly reproducing the CN, C and Mg features simultaneously (see Conroy et al.\ 2014 for 
an example of how this may be done), (2) 
use of such SEDs to attempt a quantification of the star formation
history through full-spectrum fitting of the available high S/N spectra
of intermediate redshift passive galaxies, 
(3) a more complete investigation into the stochastic variations of stellar populations
between different cluster (and field) galaxy samples and their possible relation
to differences in stellar populations in the original substructures of the clusters, and (4) an in-depth 
comparison of stellar population in passive galaxies in the field and in clusters of 
different richness at intermediate redshifts.

\acknowledgments
Karl Gebhardt is thanked for making his kinematics software available.
We thank Ricardo Demarco and Richard McDermid for comments on an earlier draft of this paper.
The Gemini TACs and the former Director Matt Mountain are thanked for generous time allocations
to carry out these observations.

Based on observations obtained at the Gemini Observatory (processed using the Gemini IRAF package), 
which is operated by the Association of Universities for Research in Astronomy, Inc., 
under a cooperative agreement with the NSF on behalf of the Gemini partnership: the 
National Science Foundation (United States), the National Research Council (Canada), 
CONICYT (Chile), Ministerio de Ciencia, Tecnolog\'{i}a e Innovaci\'{o}n Productiva (Argentina), 
and Minist\'{e}rio da Ci\^{e}ncia, Tecnologia e Inova\c{c}\~{a}o (Brazil). 
The data were obtained while also the Science and Technology Facilities Council (United Kingdom), 
and the Australian Research Council (Australia) contributed to the Gemini Observatory.

The new data presented in this paper originate from the following Gemini programs:
GN-2001B-Q-10, GN-2002B-SV-90, GN-2003B-DD-3, GN-2003B-Q-21, and GS-2005A-Q-27. 

This research has made use of the NASA/IPAC Extragalactic Database (NED) which is 
operated by the Jet Propulsion Laboratory, California Institute of Technology, 
under contract with the National Aeronautics and Space Administration. 
Photometry from SDSS is used for comparison purposes. Funding for the SDSS and SDSS-II 
has been provided by the Alfred P. Sloan Foundation, the Participating Institutions, 
the National Science Foundation, the U.S. Department of Energy, the National 
Aeronautics and Space Administration, the Japanese Monbukagakusho, the Max 
Planck Society, and the Higher Education Funding Council for England. 
The SDSS Web Site is http://www.sdss.org/.
This research has made use of the VizieR catalog access tool, CDS, Strasbourg, France.

Observations have been used that were obtained with {\it XMM-Newton}, an ESA science mission 
funded by ESA member states and NASA, and with the {\it Chandra X-ray Observatory}, and 
obtained from these missions data archives.
In part, based on observations made with the NASA/ESA {\it Hubble Space Telescope}, 
obtained from the data archive at the Space Telescope Science Institute (STScI). 
STScI is operated by the Association of Universities for Research in Astronomy, 
Inc. under NASA contract NAS 5-26555.

\appendix

\section{Photometric Data \label{SEC-PHOTOMETRY}}

\subsection{Photometric Parameters, Calibration and External Comparison}

\begin{deluxetable*}{llllrccc}
\tablecaption{GMOS-N and GMOS-S Imaging Data \label{tab-imdata} }
\tablewidth{0pc}
\tabletypesize{\scriptsize}
\tablehead{
\colhead{Cluster} & \colhead{Program ID\tablenotemark{a}} & \colhead{Dates} & \colhead{Filter} & \colhead{Exposure time} & \colhead{FWHM\tablenotemark{b}} & \colhead{Sky brightness} & \colhead{$A_n$\tablenotemark{c}} \\
\colhead{} & \colhead{} & \colhead{(UT)} & \colhead{} & \colhead{} & \colhead{(arcsec)} & \colhead{(mag arcsec$^{-2}$)} & \colhead{(mag)} }
\startdata
Abell 1689 F1\tablenotemark{d}  & GN-2003B-DD-3 & 2003 Dec 24 & $g'$ & 4 $\times$ 180sec   & 0.54 & 21.55 & 0.089 \\  
               & GN-2001B-Q-10 & 2001 Dec 24 & $r'$     & 5 $\times$ 300sec   & 1.05 & 20.32 & 0.062 \\  
               & GN-2001B-Q-10 & 2001 Dec 25 & $i'$     & 3 $\times$ 300sec   & 1.04 & 19.49 & 0.046 \\  
Abell 1689 F2\tablenotemark{d}  & GN-2003B-DD-3 & 2003 Dec 24 & $g'$     & 4 $\times$ 180sec   & 0.57 & 21.54 & 0.089 \\  
               & GN-2001B-Q-10 & 2001 Dec 24 & $r'$     & 6 $\times$ 300sec   & 1.05 & 19.80 & 0.062 \\  
               & GN-2001B-Q-10 & 2001 Dec 25 & $i'$     & 3 $\times$ 300sec   & 0.81 & 19.51 & 0.046 \\  
RXJ0056.2+2622 F1\tablenotemark{e} & GN-2003B-Q-21 & 2003 Jul 2 & $g'$  & 4 $\times$ 120sec   & 0.79 & 21.94 & 0.192 \\ 
               & GN-2003B-Q-21 & 2003 Jul 2 & $r'$     & 4 $\times$ 120sec   & 0.70 & 21.24 & 0.133 \\
               & GN-2003B-Q-21 & 2003 Jul 2 & $i'$     & 4 $\times$ 120sec   & 0.62 & 20.34 & 0.099 \\  
RXJ0056.2+2622 F2\tablenotemark{e} & GN-2003B-Q-21 & 2003 Jul 30 & $g'$  & 4 $\times$ 120sec   & 1.04 & 22.06 & 0.192 \\ 
               & GN-2003B-Q-21 & 2003 Jul 30 & $r'$     & 4 $\times$ 120sec   & 0.85 & 21.20 & 0.133 \\
               & GN-2003B-Q-21 & 2003 Jul 30 & $i'$     & 5 $\times$ 120sec   & 1.05 & 19.98 & 0.099 \\   
RXJ0027.6+2616 & GN-2003B-Q-21 & 2003 Jul 1 to 2003 Jul 2 & $g'$     & 4 $\times$ 360sec   & 0.60 & 21.97 & 0.134 \\  
               & GN-2003B-Q-21 & 2003 Jul 1 to 2003 Jul 2 & $r'$     & 4 $\times$ 300sec   & 0.47 & 21.22 & 0.099 \\
               & GN-2002B-SV-90 &2002 Sep 30 &\\
               & GN-2003B-Q-21 & 2003 Jul 1 & $i'$  & 32$\times$ 120sec   & 0.65 & 20.02 & 0.069 \\
RXJ1347.5--1145& GS-2005A-Q-27 & 2005 Apr 13 & $g'$     & 4 $\times$ 450sec   & 1.16 & 22.16 & 0.204 \\  
               & GS-2005A-Q-27\tablenotemark{f} & 2005 Jan 11 & $g'$  & 1 $\times$ 450sec & 0.99 & 19.58 & 0.204 \\ 
               & GS-2005A-Q-27 & 2005 Jan 11 & $r'$     & 4 $\times$ 300sec   & 0.72 & 20.94 & 0.141 \\
               & GS-2005A-Q-27 & 2005 Jan 11 & $i'$     & 4 $\times$ 300sec   & 0.73 & 20.39 & 0.105 \\
\enddata
\tablenotetext{a}{Observations with program IDs starting with GN and GS were executed with GMOS-N and GMOS-S, respectively.}
\tablenotetext{b}{Image quality measured as the average FWHM of 7-10 stars in the field from the final stacked images.}
\tablenotetext{c}{Galactic extinction at cluster center, Schlafly \& Finkbeiner (2011) as provided through the NASA/IPAC Extragalactic Database}
\tablenotetext{d}{F1 pointing Eastern field (RA,DEC)$_{\rm J2000}$ = (13\,11\,37.0, --1\,20\,29),
F2 pointing Western field  (RA,DEC)$_{\rm J2000}$ = (13\,11\,23.5, --1\,20\,29) }
\tablenotetext{e}{F1 pointing Southern field (RA,DEC)$_{\rm J2000}$ = (0\,55\,59.0, 26\,20\,30), 
F2 pointing Northern field  (RA,DEC)$_{\rm J2000}$ = (0\,56\,00.5, 26\,25\,10) }
\tablenotetext{f}{Observation obtained in twilight, used for photometric calibration, only.}
\end{deluxetable*}

Table \ref{tab-imdata} summarizes the GMOS-N and GMOS-S imaging of the $z=0.19-0.45$ clusters.
The derived photometric parameters (total magnitudes and colors) are 
listed in Table \ref{tab-phot}.
Colors were derived as total colors based on the total magnitudes and as aperture colors.
We used both the image quality of the data and the expected sizes of the galaxies 
to optimize the choice of aperture size. Specifically, the aperture diameter in arcsec was derived as
\begin{equation}
D_{\rm app} = 2 \cdot 2.355 \left ( ({\rm FWMM}/2.355)^2 + r_{\rm galaxy}^2 \right )^{0.5}
\end{equation}
where $r_{\rm galaxy}$ is the angular size of a member galaxy with a size (effective radius) of 2.5 kpc.
The resulting apertures sizes for Abell 1689, RXJ0056.2+2622, and RXJ0027.6+2616 are similar in the three filters.
Thus, we used the same aperture size for all three filters;
$D_{\rm app} = 4.35$, 4.18, and 2.66 arcsec for the three clusters.
For RXJ1347.5--1145 we used $D_{\rm app} = 3.09$ arcsec for the $g'$-band, and 2.51 arcsec for the $r'$- and $i'$-band.

\begin{figure*}
\plotone{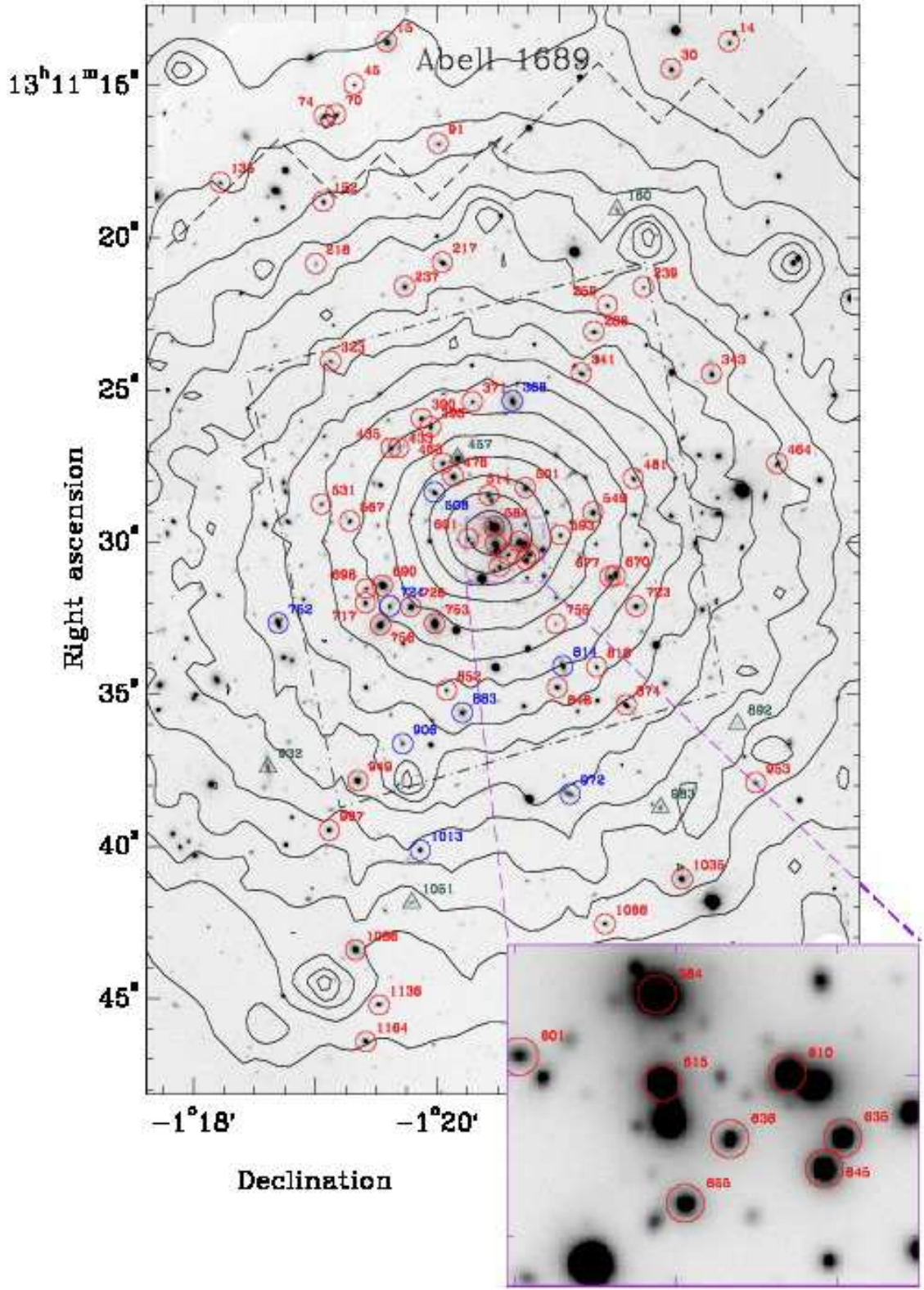}
\caption{
GMOS-N $r'$-band images of Abell 1689 with the spectroscopic sample marked.
Contours of the {\it XMM-Newton} X-ray data are overlaid.
Red circles -- confirmed bulge-dominated members with EW[\ion{O}{2}]$\le 5${\AA}. 
Blue circles -- confirmed members with EW[\ion{O}{2}] $> 5${\AA} and/or disk-dominated.
Dark green triangles -- confirmed non-members. 
Dot-dashed line shows the coverage by {\it HST}/ACS imaging, while {\it HST}/WCPC2 imaging
covers the field east of the dashed line.
The X-ray image is the sum of the images from the two {\it XMM-Newton} EPIC-MOS cameras. 
The X-ray image was smoothed; any structure seen is significant at 
the 3$\sigma$ level or higher. The spacing between the contours is logarithmic with a factor of 
1.5 between each contour.
\label{fig-A1689grey} }
\end{figure*}

\begin{figure*}
\plotone{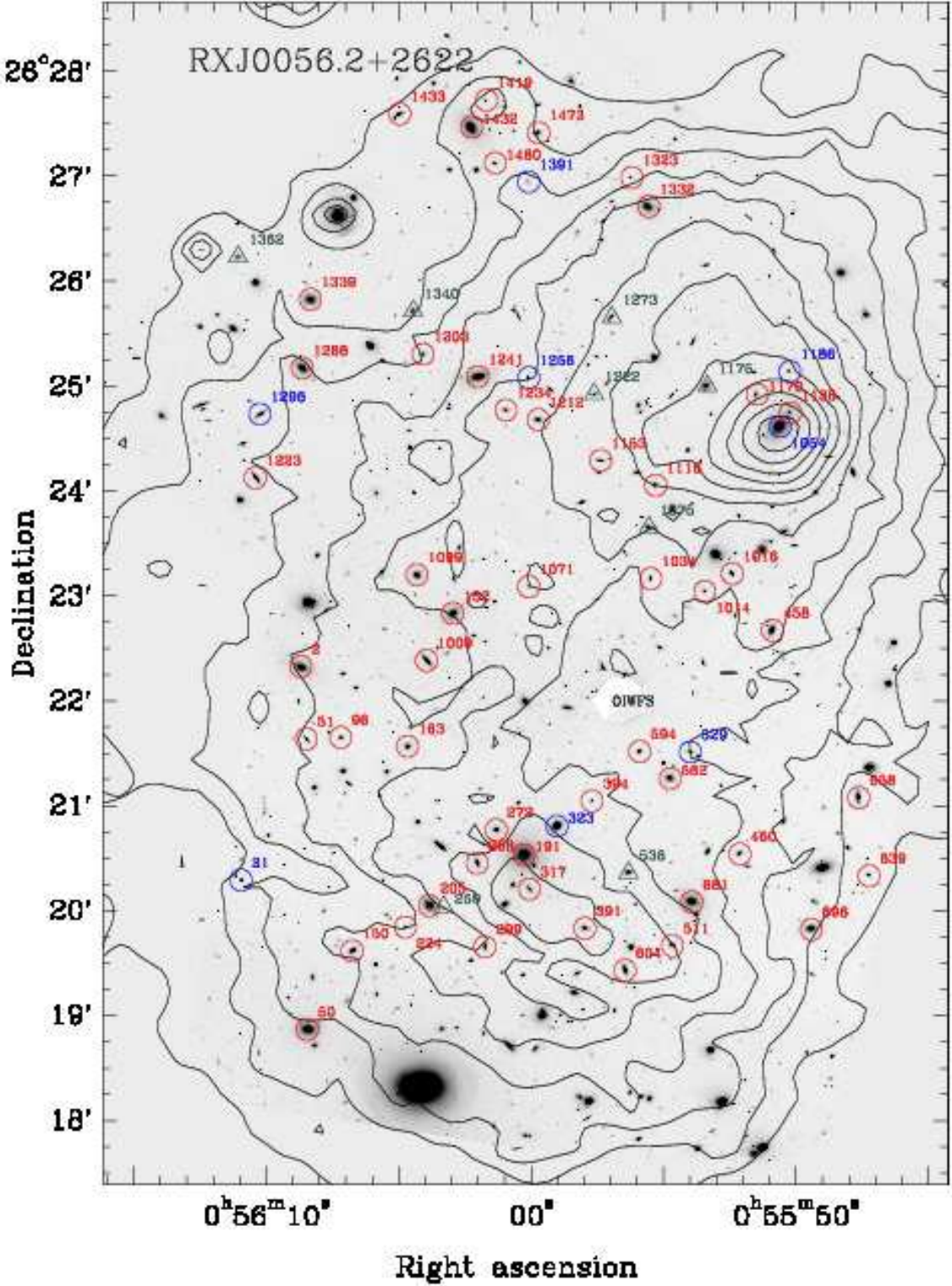}
\caption{
GMOS-N $r'$-band images of RXJ0056.2+2622 with the spectroscopic sample marked.
Contours of the {\it XMM-Newton} X-ray data are overlaid.
Red circles -- confirmed bulge-dominated members with EW[\ion{O}{2}]$\le 5${\AA}. 
Blue circles -- confirmed members with EW[\ion{O}{2}] $> 5${\AA} and/or disk-dominated. 
including ID 1296, which has strong [\ion{O}{3}] but for which the wavelength coverage does 
not include the [\ion{O}{2}] line.
Dark green triangles -- confirmed non-members. 
The location of the vignetting from the OIWFS is marked.
The X-ray image is the sum of the images from the two {\it XMM-Newton} EPIC-MOS cameras. 
The X-ray image was smoothed; any structure seen is significant at 
the 3$\sigma$ level or higher. The spacing between the contours is logarithmic with a factor of 
1.5 between each contour.
\label{fig-RXJ0056grey} }
\end{figure*}

\begin{figure*}
\plotone{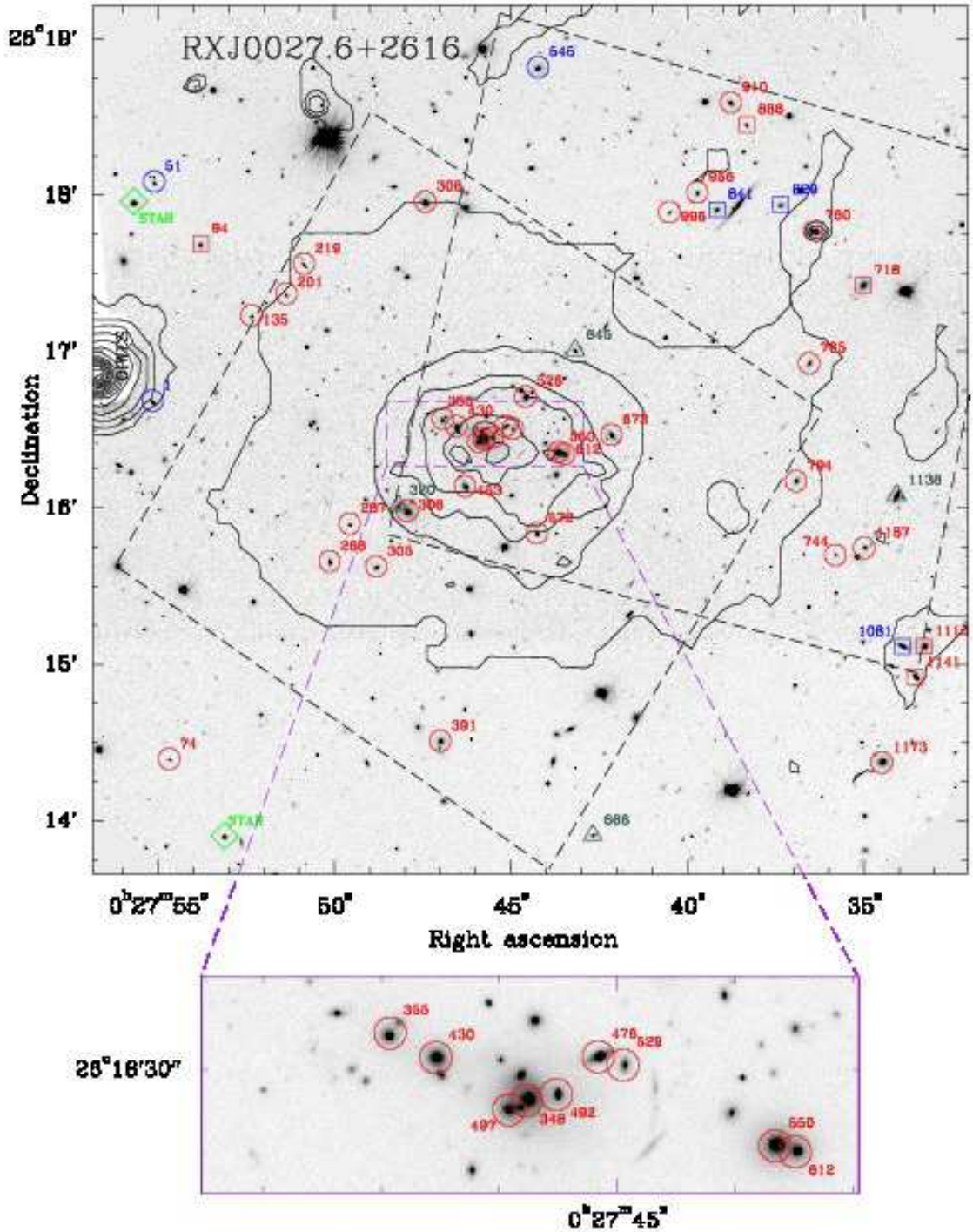}
\caption{
GMOS-N $i'$-band image of RXJ0027.6+2616 with the spectroscopic sample marked.
Contours of the {\it Chandra} X-ray data are overlaid.
Red circles -- confirmed bulge-dominated members with EW[\ion{O}{2}]$\le 5${\AA}. 
Blue circles -- confirmed members with EW[\ion{O}{2}] $> 5${\AA} and/or disk-dominated. 
Red boxes -- confirmed bulge-dominated members of the foreground group, with EW[\ion{O}{2}]$\le 5${\AA}.
Blue boxes -- confirmed members of the foreground group, with EW[\ion{O}{2}] $> 5${\AA} and/or disk dominated.
Dark green triangles -- confirmed non-members. 
Green diamonds -- blue stars included in the mask to facilitate correction for telluric absorption lines. 
The approximate location of the two {\it HST}/ACS fields are marked with dashed black lines.
The location of the vignetting from the OIWFS is marked.
The X-ray image is from the {\it Chandra} ACIS camera (\dataset [ADS/Sa.CXO\#obs/14012] {Chandra ObsId 14012}).
The X-ray image was smoothed; any structure seen is significant at 
the 3$\sigma$ level or higher. The spacing between the contours is logarithmic with a factor of 
1.5 between each contour.
\label{fig-RXJ0027grey} }
\end{figure*}

\begin{figure*}
\plotone{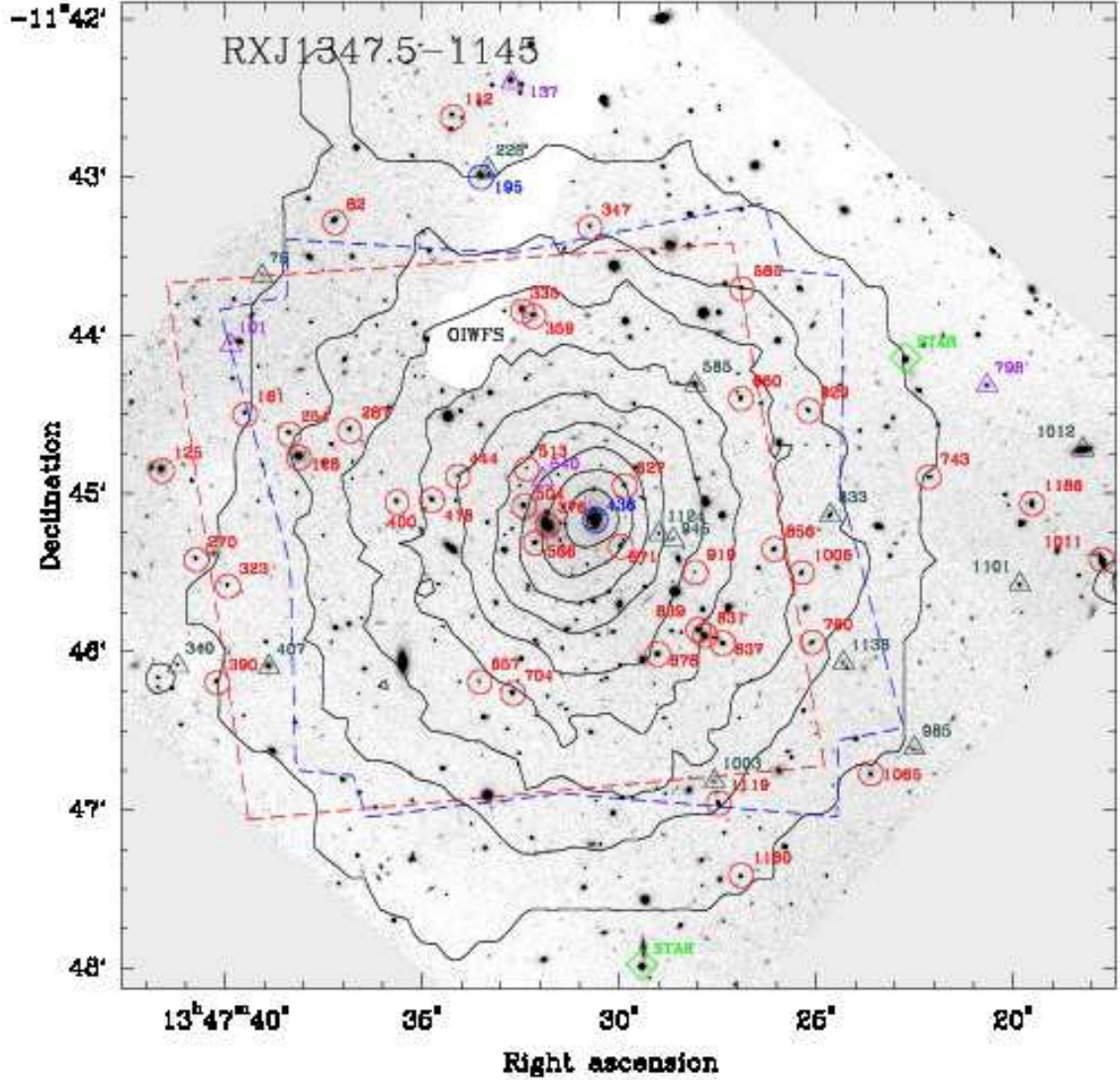}
\caption{
GMOS-S $i'$-band image of RXJ1347.5--1145 with the spectroscopic sample marked.
Contours of the {\it XMM-Newton} X-ray data are overlaid.
Red circles -- confirmed bulge-dominated members with EW[\ion{O}{2}]$\le 5${\AA}. 
Blue circles -- confirmed members with EW[\ion{O}{2}] $> 5${\AA} and/or disk-dominated. 
Dark green triangles -- confirmed non-members. 
Purple triangles -- targets for which the spectra do not allow redshift determination. 
Green diamonds -- blue stars included in the mask to facilitate correction for telluric absorption lines. 
The approximate location of the {\it HST}/ACS fields are marked with dashed lines; 
red line -- field observed in F814W and F850LP; blue line -- field observed in F606W, F625W, and F775W.
The X-ray image is the sum of the images from the two {\it XMM-Newton} EPIC-MOS cameras. 
The X-ray image was smoothed; any structure seen is significant at 
the 3$\sigma$ level or higher. The spacing between the contours is logarithmic with a factor of 
1.5 between each contour.
\label{fig-RXJ1347grey} }
\end{figure*}

Table \ref{tab-phot} includes only the spectroscopic samples. 
Photometry used for the sample selection described in Section \ref{SEC-SAMPLES} covers to $g' \approx 25$ mag in Abell 1689 and
RXJ0056.6+2622 and $r' \approx 25.5$ mag in RXJ0027.6+2616 and RXJ1347.5--1145.
Figures \ref{fig-A1689grey}-\ref{fig-RXJ1347grey} show the spectroscopic samples overlaid on greyscale images of
the clusters and with X-ray contours overlaid.

The photometry from GMOS-N was calibrated using the magnitude zero points and color terms from J\o rgensen (2009),
while the calibration of the GMOS-S photometry is based on observations of a standard field the night of the 
science observations combined with the color terms from the Gemini web page.
To achieve internal consistency between the two fields observed in RXJ0056.2+2622,
an offset of 0.095 mag were added to the $i'$-band magnitudes for galaxies in field 1 
(F1 in Table \ref{tab-imdata}). 
In addition, we used SDSS DR12 for external comparison and
in some cases calibration of our magnitude zero points. This was done consistently
for the full GCP sample of 14 clusters (and one field that is not a cluster).
Based on this comparison, the Abell 1689 $r'$ and $i'$-band magnitudes were offset with $-0.147$ and $-0.117$, respectively.
We will describe the calibration of full GCP sample in detail in 
J\o rgensen et al.\ (in prep.).

The GMOS-N data of Abell 1689 were also processed by Houghton et al.\ (2012).
We compared our photometry in $g'$ and $r'$ for Abell 1689 with that from 
Houghton et al. These authors used slightly different zero points than 
those used in this paper and may not have used any color terms in their calibration.
They did not publish photometry in the $i'$-band. 
While Houghton et al.\ use the same spectroscopic
ID numbering as we do (the numbering originates from our original mask design), 
there is some confusion in both the published tables and the VizieR catalog version of 
Houghton et al.'s data. It also appears that for the slitlet containing both ID 584 and 
ID 615, Houghton et al.\ extracted only ID 615, but by mistake assigned the spectrum
to ID 584. In Table \ref{tab-A1689houghtonids} we list our unique IDs 
together with Houghton et al.\ photometry IDs and the coordinates for the galaxies in question.
We take the correct numbering into account in our comparisons. 

\begin{figure}
\begin{center}
\epsfxsize 8.5cm
\epsfbox{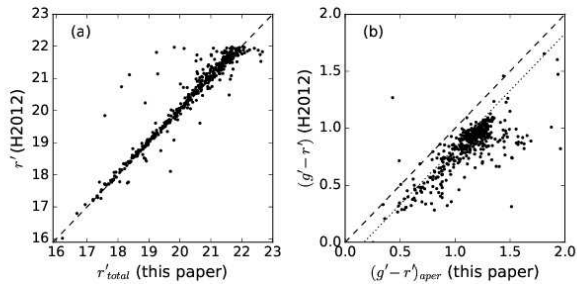}
\end{center}
\caption{Comparison of our photometry to that of Houghton et al.\ (2012).
The median offset in the $r'$-band total magnitudes is $-0.05$ with a scatter of 0.41,
with our magnitudes being brighter. 
The median offset in the the $(g'-r')$ color is 0.24 with a scatter of 0.21. 
Dashed lines - one-to-one relations.
Dotted line on panel (b) -- the expected offset of 0.17 in the color due to differences in adopted zero points
and the inclusion of a color term in our calibration.
\label{fig-a1689photcomp} }
\end{figure}

Figure \ref{fig-a1689photcomp}
and Table \ref{tab-a1689comp} summarize the comparisons of the $r'$-band total magnitudes 
and the aperture colors $(g'-r')$ with the data from Houghton et al. 
The majority of the galaxies with absolute
differences larger than 0.5 mag are located in the center of the cluster. 
Omitting these galaxies from the comparisons reduces the scatter but does not
significantly affect the median differences, see Table \ref{tab-a1689comp}.
Only two of those galaxies are in the spectroscopic sample, ID 636 and ID 655.
An offset in the total magnitudes of $-0.07$ is expected due to the difference in 
adopted zero points for the standard calibration.
An offset in $(g'-r')$ of 0.17 is expected due to the difference in adopted zero points
and the use of a color term in our calibration.
The remainder of the offset is may be due to the difference in the aperture
sizes. We use aperture size of 4.35 arcsec while Houghton et al.\ use 2.9 arcsec.
However, we do not convolve the $g'$-band to the $r'$-band resolution, since our
main color determination comes from the total magnitudes. 

\startlongtable
\begin{deluxetable*}{rrrrrrrrrrrr}
\tablecaption{GMOS Photometric Data for the Spectroscopic Samples \label{tab-phot} }
\tablewidth{0pc}
\tabletypesize{\scriptsize}
\tablehead{
\colhead{Cluster} & \colhead{ID} & \colhead{RA (J2000)} & \colhead{DEC (J2000)} &
\colhead{$g'_{\rm total}$} & \colhead{$r'_{\rm total}$} & \colhead{$i'_{\rm total}$} & 
\colhead{$(g'-r')_{\rm total}$} &
\colhead{$(r'-i')_{\rm total}$} &
\colhead{$(g'-r')_{\rm aper}$} &
\colhead{$(r'-i')_{\rm aper}$} &
\colhead{$n_{\rm ser}$} \\
(1) & (2) & (3) & (4) & (5) & (6) & (7) & (8) & (9) & (10) & (11) & (12)
}
\startdata
A1689 &    14&  13 11 13.54&   -1 22 23.9&  20.52&  19.15&  18.76&  1.372&  0.387&  1.314&  0.410&   \nodata\\
A1689 &    15&  13 11 13.54&   -1 19 35.1&  18.77&  17.72&  17.17&  1.057&  0.541&  1.178&  0.493&   \nodata\\
A1689 &    30&  13 11 14.44&   -1 21 55.4&  19.72&  18.47&  18.07&  1.251&  0.394&  1.219&  0.432&   \nodata\\
A1689 &    45&  13 11 14.95&   -1 19 19.0&  21.09&  20.00&  19.57&  1.097&  0.423&  1.098&  0.422&   \nodata\\
A1689 &    70&  13 11 15.94&   -1 19 10.1&  20.11&  18.91&  18.56&  1.207&  0.352&  1.191&  0.406&   \nodata\\
A1689 &    74&  13 11 15.97&   -1 19 04.4&  19.83&  18.56&  18.17&  1.273&  0.384&  1.259&  0.451&   \nodata\\
A1689 &    91&  13 11 16.88&   -1 20 00.5&  20.46&  19.39&  18.84&  1.067&  0.551&  1.157&  0.540&   \nodata\\
A1689 &   135&  13 11 18.17&   -1 18 13.0&  20.74&  19.61&  19.24&  1.128&  0.371&  1.158&  0.418&   \nodata\\
A1689 &   152&  13 11 18.80&   -1 19 03.9&  20.50&  19.21&  18.76&  1.296&  0.448&  1.309&  0.440&   \nodata\\
A1689 &   160&  13 11 19.06&   -1 21 28.8&  20.05&  19.53&  19.27&  0.516&  0.264&  0.477&  0.257&   \nodata\\
\enddata
\tablecomments{Column 1: Galaxy cluster. Column 2: Object ID. Columns 3--4; Right ascension in  hours, minutes, and seconds,
declination in degrees, arcminutes, and arcseconds. 
Positions are consistent with USNO (Monet et al.\ 1998), with an rms scatter of $\approx 0.5$ arcsec. 
Columns 4--7: Total magnitudes in $g'$, $r'$ and $i'$. 
Columns 8--9: Colors derived from the total magnitudes.
Columns 10--11: Colors derived from aperture magnitudes.
Column 12: Adopted  S\'{e}rsic indices from Houghton et al.\ (2012) used for sample selection, see text.
Table 8 is published in its entirety in the machine-readable format.
      A portion is shown here for guidance regarding its form and content.
}
\end{deluxetable*}

\begin{deluxetable*}{rrrl}
\tablecaption{Abell 1689: Cross-references with Houghton et al.\ \label{tab-A1689houghtonids} }
\tablewidth{0pt}
\tabletypesize{\scriptsize}
\tablehead{
\colhead{ID} & \colhead{H2012 phot-ID} & \colhead{Redshift} & \colhead{Comments} \\
\colhead{(1)} & \colhead{(2)} &\colhead{(3)} & \colhead{(4)}  
}
\startdata
584 & 493 & 0.1831 & Object shares slit with ID 615. H2012 extracted 615 but mislabeled it 584.\\
615 & 620 & 0.1771 & Object shares slit with ID 584. H2012 extracted 615 but mislabeled it 584.\\
972 & 546 & 0.1956 & Object listed in H2012 Table 6, cross referenced incorrectly in their Table 3.\\
983 & 315 & 0.2149 & Object listed in H2012 Table 3, but not in their Table 6. \\
\enddata
\tablecomments{Column 1: Galaxy ID from this paper. 
Column 2: Photometric ID from Houghton et al.\ (2012, H2012).
Column 3: Redshift from our data. Column 4: Comments.
 } 
\end{deluxetable*}

\begin{deluxetable}{rrrr}
\tablecaption{Abell 1689: Comparison with Houghton et al.\label{tab-a1689comp} }
\tablewidth{0pt}
\tabletypesize{\scriptsize}
\tablehead{
\colhead{Parameter} & \colhead{$\Delta$} & \colhead{rms} &  \colhead{N} }
\startdata
$r'$ & -0.05 & 0.41 & 471 \\
$r'$\tablenotemark{a} & -0.05 & 0.14 & 434 \\
$(g'-r')$ & 0.24 & 0.21 & 471 \\
$(g'-r')$\tablenotemark{a} & 0.25 & 0.16 & 434 \\
Redshift & -0.0001 & 0.0001 & 71 \\
$\log \sigma$ & -0.047 & 0.160 & 71 \\
$\log \sigma$\tablenotemark{b} & -0.051 & 0.061 & 64 \\
\enddata
\tablenotetext{a}{Excluding galaxies for which the $r'$-band magnitudes from 
Houghton et al.\ deviate with more than 0.5 mag from our determination. }
\tablenotetext{b}{Excluding galaxies for which Houghton et al.\ find velocity dispersions less
than $75\,{\rm km\,s^{-1}}$.}
\tablecomments{Median differences $\Delta$ = ``our data'' -- ``Houghton et al.''. } 
\end{deluxetable}

\begin{deluxetable*}{lcl}
\tablecaption{Selection criteria for spectroscopic samples in the $z=0.18-0.45$ clusters \label{tab-spsel} }
\tablewidth{0pc}
\tabletypesize{\scriptsize}
\tablehead{
\colhead{Cluster} & \colhead{Obj.Class} & \colhead{Selection criteria} }
\startdata
Abell 1689 & 1 & Confirmed member based on redshift $\wedge ~ 17 \le g' \le 21.2 ~ \wedge ~ 0.9 \le (g'-r') \le 1.5 ~ \wedge ~ 0.25 \le (r'-i') \le 0.6 $ \\
& 2 & No redshift $\wedge ~ 17 \le g' \le 21.2 ~ \wedge ~ 0.9 \le (g'-r') \le 1.5 ~ \wedge ~ 0.25 \le (r'-i') \le 0.6 $ \\
& 3 & $21.1 < g' \le 22.2 ~ \wedge ~ 0.9 \le (g'-r') \le 1.5 ~ \wedge ~ 0.25 \le (r'-i') \le 0.6 $ \\
& 4 & $18 \le g' \le 22.2 ~ \wedge ~ 0.0 \le (g'-r') < 0.9 $ \\
RXJ0056.2+2622 & 1 & Confirmed member based on redshift $ \wedge ~ 17 \le g' \le 21.2 ~ \wedge ~ 0.8 \le (g'-r') \le 1.4 ~ \wedge ~ 0.2 \le (r'-i') \le 0.55 $  \\
               & 2 & No redshift $ \wedge ~ 17 \le g' \le 21.2 ~ \wedge ~ 0.8 \le (g'-r') \le 1.4 ~ \wedge ~ 0.2 \le (r'-i') \le 0.55 $ \\
               & 3 & $21.1 < g' \le 22.2 ~ \wedge ~ 0.8 \le (g'-r') \le 1.4 ~ \wedge ~ 0.2 \le (r'-i') \le 0.55 $ \\
               & 4 & $20 \le g' \le 22.2 ~ \wedge ~ 0.0 \le (g'-r') < 0.8 $ \\
RXJ0027.6+2616  & 1 & $18 \le r' \le 20.1 ~ \wedge ~ 0.4 \le (r'-i')\le 0.8 ~ \wedge ~ 1.5 \le (g'-r')\le 1.9$ \\
 & 2 &  $20.1 < r' \le 20.7 ~ \wedge ~ 0.4 \le (r'-i')\le 0.8 ~ \wedge ~ 1.5 \le (g'-r')\le 1.9$ \\
 & 3 &  $20.7 < r' \le 21.5 ~ \wedge ~ 0.4 \le (r'-i')\le 0.8 ~ \wedge ~ 1.5 \le (g'-r')\le 1.9$ \\
 & 4 &  $18 < r' \le 21.5 ~ \wedge ~ 0.1 \le (r'-i') < 0.4 $ \\
RXJ1347.5--1145\tablenotemark{a}  & 1 & Confirmed member based on redshift $ ~ \wedge 18 \le r' \le 21 ~ \wedge ~ 0.6 \le (r'-i')\le 0.9$  \\
 & 2 & Confirmed member based on redshift $ ~ \wedge 21 < r' \le 22.4 ~ \wedge ~ 0.6 \le (r'-i')\le 0.9$  \\
 & 3 & No redshift $ ~ \wedge 18 \le r' \le 22.4 \wedge ~ 0.6 \le (r'-i')\le 0.9$ \\
 & 4 & $19 \le i' < 22.4 ~ \wedge ~ 0.2 \le (r'-i') < 0.6 $ \\
\enddata
\tablenotetext{a}{At the time of sample selection only imaging in $r'$ and $i'$ was available.}
\end{deluxetable*}

\section{Spectroscopic Data \label{SEC-SPECTROSCOPY}}

\subsection{Observations and Reductions}

The selection criteria for the spectroscopic samples are summarized in Table \ref{tab-spsel}.
The spectroscopic data were processed using the same techniques as described
for the MS0451.6--0306 data in J\o rgensen \& Chiboucas (2013). 
The processing produces 1-dimensional spectra fully calibrated and on a relative flux-scale.

The masks for Abell 1689 contained slits titled to be aligned with the major axis of the galaxies.
As we want to subtract off the sky signal before we interpolate the spectra, the tilted slits
in the Abell 1689 data were first semi-rectified by oversampling the spectra by a factor five
and then shifting the spectra by integer rows in the oversampled spectra. 
The sky signal was then subtracted and the rows shifted back to their original 
position to before processing the data through the remainder of the steps.
See Barr et al.\ (2005) for details on this process.

The observations of Abell 1689 were obtained with the detectors unbinned, while all other observations
used binning in the spectral direction. These unbinned observations unnecessarily oversample the spectra.
Thus, after basic processing and extraction the Abell 1689 observations
were rebinned to match the binning of the other observations. 

For RXJ0027.6+2616 and RXJ1347.5--1145 two blue stars were included in the mask in order 
to obtain a good correction for the telluric absorption lines. 
The masks for Abell 1689 and RXJ0056.2+2622 did not include blue stars. Instead the telluric
correction was established from a stack of all the spectra in each mask. This is the same
technique used by Barr et al.\ (2005) for the GCP cluster RXJ0142.0+2131.

The observations of  RXJ1347.5--1145 were obtained with GMOS-S with the E2V CCDs.
These detectors have fairly strong fringing in the red. Therefore, the observations were
obtained in pairs of exposures with dithers along the slit. We used these to determine
the fringe correction in a similar way as done for RXJ0152.7--1157, see J\o rgensen et al.\ (2005).
Two frames were excluded from further processing at this point as the fringes were too 
strong to allow a useful correction and the stack of the better reduced frames gave
sufficient S/N for our purpose.

All flux calibrated 1-dimensional spectra
were median filtered with a 5-pixel filter in the spectral direction to limit the effect of residuals
from the sky subtraction. The filtering was taken into account in the determination of the 
instrumental resolution.

\subsection{Spectroscopic Parameters}

The calibrated spectra were fit with stellar templates as described in J\o rgensen \& Chiboucas (2013).
This results in determination of the redshifts and the velocity dispersions.
As in J\o rgensen \& Chiboucas, we use three template stars with spectral types K0III, G1V and B8V.
The fits were performed with the kinematics fitting software made available by Karl Gebhardt, see
Gebhardt et al.\ (2000, 2003) for a description of the software.
The software performs the fitting in pixel space. Thus, it is straight forward to mask wavelength
intervals affected by either emission lines or strong residuals from the sky subtraction. In particular,
the residuals from the strong skyline at 5577 {\AA} were masked in all cases.
For all four clusters the fits were done for the wavelength region 3750--5500{\AA} in the 
rest frame of the clusters.
Aperture correction of the velocity dispersions were performed using the technique from 
J\o rgensen et al.\ (1995b).

The strengths of the absorption lines were measured using the definitions of the Lick/IDS indices
(Worthey et al.\ 1994) as well as the higher order Balmer line indices H$\delta _A$ and H$\gamma _A$ (Worthey \& Ottaviani 1997).
We measure the H$\beta _G$ index as described in J\o rgensen (1999). 
The original passband definition for this index is from Gonz\'{a}lez (1993).
The indices CN3883 and CaHK were measured using the passband definitions from Davidge \& Clark (1994).
The definition of the D4000 index is from Bruzual (1983) and Gorgas et al.\ (1999).
We have adopted the bandpass definition for the higher order Balmer line H$\zeta_{\rm A}$ from Nantais et al.\ (2013).

The line indices were aperture corrected and corrected for the velocity dispersions of the galaxies using the 
techniques and corrections detailed in J\o rgensen et al.\ (2005), see also J\o rgensen \& Chiboucas (2013).
As in J\o rgensen et al.\ (2014) we assume that H$\zeta _A$ has no aperture correction.

The residuals from the strong skyline at 5577 {\AA} in some cases affect the determination
of the line indices. Where such effects are significant, the line index determinations have been
removed from the tables of our measurements. The details are as follows.
For members of Abell 1689 and RXJ0056.2+2622 the 5577 {\AA} skyline falls within the passbands of C4668.
The passbands of this index are fairly wide. We have therefore evaluated whether the 
C4668 measurements are affected by the 5577 {\AA} residuals
by interpolating across the wavelength region that may be affected (5565--5589 {\AA})
and repeating the line measurement on the interpolated spectra. For those galaxies where 
the difference in C4668 derived from the spectra before interpolation and after is larger than
0.065 in log space (approximately twice the typical measurement uncertainty)
we have deemed the index unreliable and omitted the measurements from the tables and analysis.

For members of RXJ0027.6+2616 the skyline falls within the passbands
of the H$\delta _A$ index. As the line is fairly weak and the passbands narrow, 
we have chosen not to attempt to measure this line.
Since we want to use the combination index 
$({\rm H\delta _A + H\gamma _A})' \equiv -2.5~\log \left ( 1.-({\rm H\delta _A + H\gamma _A})/(43.75+38.75) \right )$ \\
(Kuntschner 2000),
in the analysis, we solve this problem by deriving an empirical relation between fully corrected
measurements of H$\delta _A$ and H$\gamma _A$ based on the data for the 94 members of the clusters 
MS0451.6--0305, RXJ0152.7--1357, and RXJ1226.9+2226.
We find ${\rm H}\delta _{\rm A,cor} = (0.689 \pm 0.045) {\rm H}\gamma _{\rm A,cor} + 2.707 $
with an rms scatter of 1 {\AA}.
We then use that relation to derive H$\delta _A$ for the RXJ0027.6+2616 galaxies, and subsequently
use those values together with the H$\gamma _A$ measurements to derive $(\rm{H\delta _A + H\gamma _A})'$.
This in effect means that $(\rm{H\delta _A + H\gamma _A})'$ for this cluster is based on H$\gamma _A$, only.

For members of RXJ1347.5+1145 the indices, H$\zeta _A$, CN3883 and D4000 may be affected by the 
residuals from the 5577 {\AA} skyline. As H$\zeta _A$ is a fairly weak index and the passbands are narrow, 
we have chosen not to attempt to measure this line if for the individual galaxy redshift the 
5577 {\AA} line is within 3 {\AA} of any of the passbands.
For CN3883 and D4000 we take the same approach as used for C4668 for the Abell 1689 and RXJ0056.2+2622 members.
We use limits of 0.025 and 0.035 for changes in CN3883 and D4000, respectively.
Thus, these indices are only listed for galaxies for which the measurements are not 
significantly affected by the sky residuals.

The formal uncertainties on the indices were determined based on the S/N following Cardiel et al.\ (1998).
However, as done in J\o rgensen \& Chiboucas (2013) for the $z=0.54-0.89$ clusters, we also 
performed sub-stacking of the frames to evaluate the uncertainties, see Section \ref{SEC-INTCOMP}
and Table \ref{tab-intcomp}.

For galaxies with detectable emission from  [\ion{O}{2}] we determined the equivalent width 
of the [\ion{O}{2}]$\lambda\lambda$3726,3729 doublet.
With an instrumental resolution of $\sigma = 1.5-2.5$\,{\AA} (FWHM $= 3.5-5.8$\,{\AA}) and 
galaxy velocity dispersions typically $\ge 100\,{\rm km\,s^{-1}}$, 
the doublet is not resolved in our spectra and we refer to it simply as the ``[\ion{O}{2}] line''. 

Table \ref{tab-kin} lists the results from the template fitting 
(redshifts and velocity dispersions),
while Table \ref{tab-line} gives  the derived absorption line strengths.
The emission line equivalent widths are listed in Table \ref{tab-EWOII}.

\begin{figure*}
\epsfxsize 17.5cm
\epsfbox{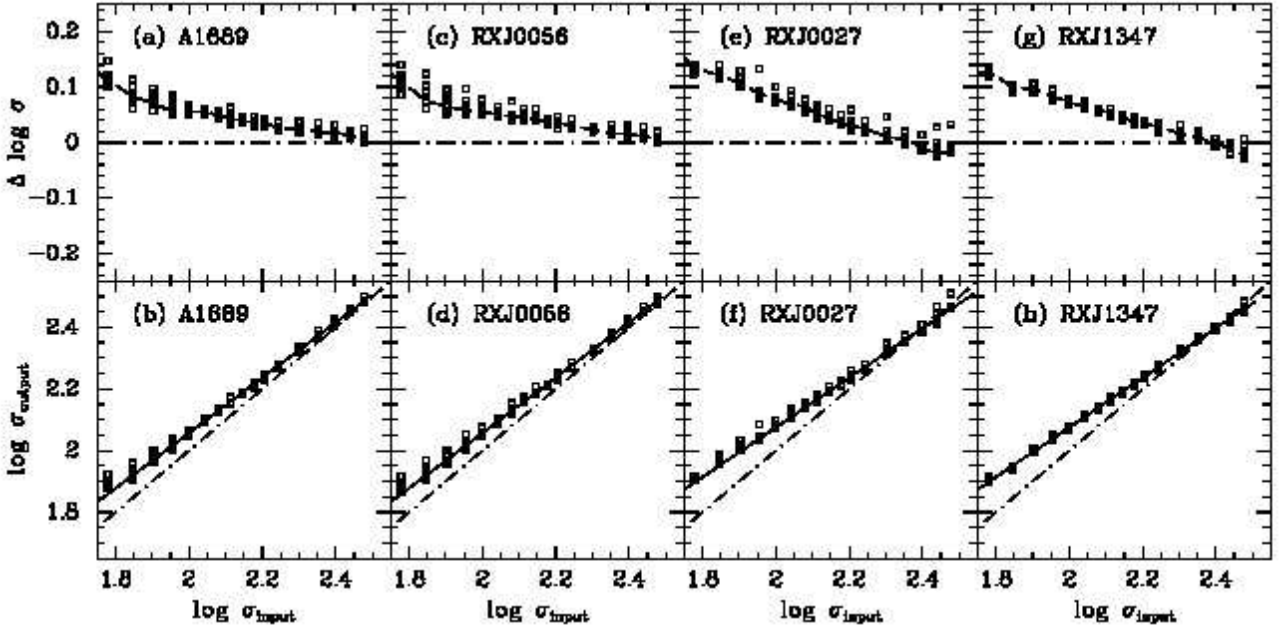}
\caption{Results from  of the velocity dispersions. The top row of panels show the 
systematic error on $\log \sigma$, 
$\Delta \log \sigma = \log \sigma _{\rm output} - \log \sigma _{\rm input}$ as a function of the 
input values $\log \sigma _{\rm input}$. The bottom row of panels show input versus
output values. Simulations apply to the four clusters as follows: (a) and (b) Abell 1689,
(c) and (d) RXJ0056.2+2622, (e) and (f) RXJ0027.6+2616, (g) and (h) RXJ1347.5--1145.
Dashed lines on panels (a), (c), (e), and (g) -- the median values of $\Delta \log \sigma$. 
Dot-dashed lines on  panels (b), (d), (f), and (h) -- one-to-one relations.
Solid lines on panels (b), (d), (f), and (h) -- the adopted correction 
for the systematic errors, see Eq.\ \ref{eq-sigmasys1} and \ref{eq-sigmasys2}.
\label{fig-velsim} }
\end{figure*}

\subsection{Systematic Effects in Derived Velocity Dispersions \label{SEC-SIM}}

As in J\o rgensen \& Chiboucas (2013), we performed simulations to evaluate 
any systematic effects on the derived velocity dispersions. Model spectra were 
constructed using the average of the K0III and G1V template stars. 
Random noise was then added using information from the real noise spectra from the spectra. We made
100 realizations of each of the combinations covering velocity dispersions
from 50 to 300 $\rm km\,s^{-1}$  and S/N per {\AA} in the rest frame from 10 to 50 . 
The S/N of our data are in general higher than 50, but as it turns out only for the 
very low S/N is the S/N a factor in recovering the velocity dispersion. 
We then fit the model spectra and compared the recovered velocity dispersions with
the input values. The simulations are summarized in Figure \ref{fig-velsim}.
Simulations were run for all four clusters, as the location of the higher noise 
wavelength intervals due to sky lines depend on the redshift of the cluster.
As can be seen from the figure, in all cases there is a small systematic offset
to higher velocity dispersion. We correct for this by fitting a first order polynomial
to the difference as a function of output velocity dispersion, constraining the 
fits to $\log \sigma > 1.9$. The simulations for Abell 1689 and RXJ0056.2+2622 are nearly
identical. Thus, we fit those simulations together and derive
\begin{equation}
\log \sigma _{\rm corrected} = \log \sigma _{\rm out} - 0.0519 + 0.1142 ( \log \sigma _{\rm out} -2.1)
\label{eq-sigmasys1}
\end{equation}
The simulations for RXJ0027.6+2616 and RXJ1347.5--1145 are different with at most 0.003 for $\log \sigma \ge 2.0$.
Thus, we also fit these simulations together and derive
\begin{equation}
\log \sigma _{\rm corrected} = \log \sigma _{\rm out} - 0.0690 + 0.2407 ( \log \sigma _{\rm out} -2.1)
\label{eq-sigmasys2}
\end{equation}
The fits were derived for input velocity dispersions of $\log \sigma > 1.9$. Deviations 
from the fits for velocity dispersions smaller than this limit is of the order 
0.01-0.05 on $\log \sigma$.

As the size of the systematic effects in the determination of the velocity 
dispersions is of the same size or larger than the systematic errors expected
between different datasets, we choose to correct the derived velocity dispersions
using the formulae in Eq. \ref{eq-sigmasys1} and \ref{eq-sigmasys2}.
The velocity dispersions listed in Table \ref{tab-kin} as $\log \sigma _{\rm cor}$ have
been corrected for this effect, as well as aperture corrected. 
The raw measurements are listed in the tables as $\log \sigma$.

\subsection{Internal Comparisons \label{SEC-INTCOMP} }

\begin{figure*}
\epsfxsize 17.5cm
\epsfbox{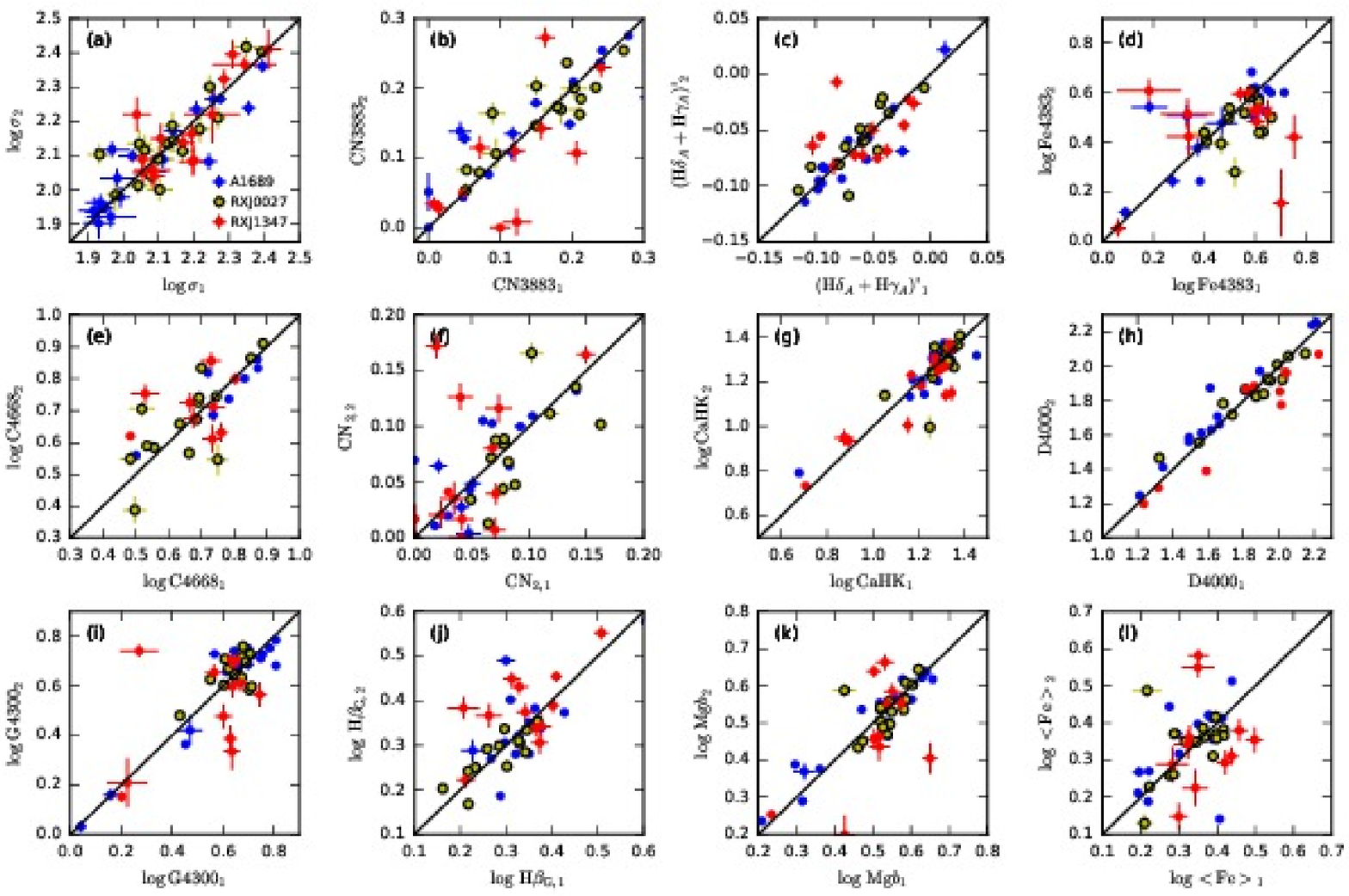}
\caption{Internal comparison of velocity dispersions and line indices derived from 
sub-stacks of the available data.
Blue -- Abell 1689; yellow -- RXJ0027.6+2616; red -- RXJ1347.5--1145.
Solid lines show the one-to-one relations. See Table \ref{tab-intcomp} for scatter
and the adopted uncertainties on the line indices.
\label{fig-intcomp} }
\end{figure*}

The mask designs for Abell 1689, RXJ0027.6+2616, and RXJ1347.5--1145 are such that 
for each cluster 15--19 of the galaxies were observed in two masks, each with sufficient S/N to derive 
velocity dispersions and line indices. Thus, we effectively have repeat observations
of these galaxies and use these for internal comparisons of the parameters and to
assess the uncertainties.
Figure \ref{fig-intcomp} and Table \ref{tab-intcomp} summarize the comparisons.
In general, the uncertainties on the velocity dispersions from the kinematics fitting
are in good agreement with the internal comparisons. For the line indices the uncertainties
derived from the S/N of the spectra are underestimated most likely due to systematics
from the sky subtraction, as we also noted in J\o rgensen \& Chiboucas (2013). 
The ratios between the scatter in the comparisons and the expected scatter based on
the uncertainties range between 2 and 10. We therefore adopt global uncertainties on the line indices.
The adopted uncertainties are listed in Table \ref{tab-intcomp}, and shown
on the figures of these parameters as the typical error bars.
The uncertainties for RXJ0056.2+2622 measurements are assumed to be the same as for Abell 1689 measurements,
as the two clusters have similar redshifts and were observed with identical instrument configurations 
and to similar S/N ratios.

\begin{deluxetable*}{l rrrrr rrrrr rrrrr}
\tablecaption{Internal comparison \label{tab-intcomp} }
\tablewidth{0pc}
\tabletypesize{\scriptsize}
\tablehead{
\colhead{Parameter} & \multicolumn{5}{c}{ A1689 } 
& \multicolumn{5}{c}{ RXJ0027.6+2616 } 
& \multicolumn{5}{c}{ RXJ1347.5--1145 } \\
& \colhead{N} & \colhead{rms} & \colhead{Ratio} & \colhead{$\sigma_{\rm median}$} & \colhead{$\sigma _{\rm adopt}$}
& \colhead{N} & \colhead{rms} & \colhead{Ratio} & \colhead{$\sigma_{\rm median}$} & \colhead{$\sigma _{\rm adopt}$}
& \colhead{N} & \colhead{rms} & \colhead{Ratio} & \colhead{$\sigma_{\rm median}$} & \colhead{$\sigma _{\rm adopt}$}  \\
\colhead{(1)} & \colhead{(2)} & \colhead{(3)} & \colhead{(4)} & \colhead{(5)} & \colhead{(6)} &
\colhead{(7)} & \colhead{(8)} & \colhead{(9)} & \colhead{(10)} & \colhead{(11)} & \colhead{(12)} &
\colhead{(13)} & \colhead{(14)} & \colhead{(15)} & \colhead{(16)} 
}
\startdata
$\log \sigma$  &  19 & 0.064 &  1.5 & 0.014 & 0.020 & 17 & 0.064 &  1.4 & 0.024 & 0.035 & 15 & 0.072 &  1.2 & 0.030 & 0.036 \\
CN3883 &  15 & 0.048 &  4.4 & 0.006 & 0.028 & 15 & 0.033 &  1.9 & 0.008 & 0.016 & 11 & 0.065 &  3.5 & 0.010 & 0.036 \\
$(\rm{H\delta _A + H\gamma _A})'$ &  19 & 0.013 &  4.0 & 0.001 & 0.006 & 16 & 0.018 &  4.0 & 0.001 & 0.006 & 15 & 0.031 &  3.7 & 0.004 & 0.016 \\
$\log\, {\rm Fe4383}$ &  17 & 0.117 &  3.7 & 0.010 & 0.036 & 15 & 0.089 &  2.7 & 0.014 & 0.038 & 12 & 0.233 &  2.7 & 0.033 & 0.090 \\
$\log\, {\rm C4668}$ &  8 & 0.188 &  6.4 & 0.006 & 0.036 & 15 & 0.092 &  3.5 & 0.008 & 0.030 & 11 & 0.275 &  4.6 & 0.023 & 0.105 \\
CN$_2$ &  19 & 0.025 &  5.9 & 0.003 & 0.016 & 16 & 0.029 &  4.9 & 0.003 & 0.017 & 15 & 0.051 &  4.1 & 0.007 & 0.029 \\
$\log\, {\rm CaHK}$ &  17 & 0.060 &  4.6 & 0.004 & 0.020 & 17 & 0.074 &  4.2 & 0.006 & 0.027 & 15 & 0.083 &  2.7 & 0.013 & 0.034 \\
D4000 &  14 & 0.064 &  9.7 & 0.004 & 0.037 & 14 & 0.065 &  6.7 & 0.005 & 0.032 & 10 & 0.091 &  7.9 & 0.007 & 0.054 \\
$\log\, {\rm G4300}$ &  18 & 0.063 &  3.4 & 0.006 & 0.020 & 15 & 0.063 &  3.3 & 0.008 & 0.028 & 13 & 0.184 &  3.3 & 0.019 & 0.061 \\
log H$_{\beta}$ &  17 & 0.066 &  7.1 & 0.004 & 0.030 & 14 & 0.098 &  7.9 & 0.005 & 0.042 & 12 & 0.071 &  2.5 & 0.014 & 0.035 \\
log Mg{\it b}  &  18 & 0.040 &  4.4 & 0.004 & 0.016 & 16 & 0.051 &  4.1 & 0.004 & 0.018 & 13 & 0.108 &  2.7 & 0.016 & 0.043 \\
log $\rm \langle Fe \rangle$ &  17 & 0.122 &  10.5 & 0.004 & 0.040 & 16 & 0.080 &  4.9 & 0.006 & 0.031 & 12 & 0.125 &  2.9 & 0.014 & 0.041 \\
\enddata
\tablecomments{Column 1: Parameter. Columns 2--6: for Abell 1689, number of galaxies in comparison, rms scatter of comparison,
ratio between the rms scatter and the expected scatter based on nominal individual uncertainties, median of nominal individual uncertainties
for all cluster members included in the analysis, and adopted uncertainty on the parameter (except for the velocity dispersion we use the
individual uncertainties from the kinematics fitting. 
Columns 7--11: Same information for RXJ0027.6+2616. 
Columns 12--16: Same information for RXJ1347.5--1145.
}
\end{deluxetable*}

\subsection{External Comparisons}

\begin{figure}
\begin{center}
\epsfxsize 8.5cm
\epsfbox{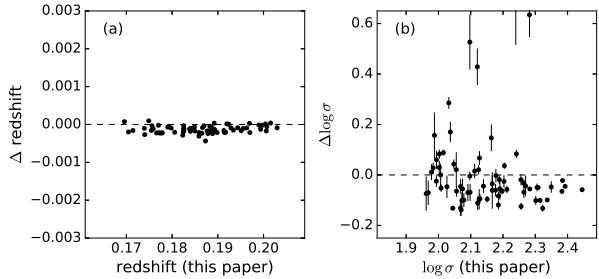}
\end{center}
\caption{Comparison of our redshifts and velocity dispersions to the 
values from Houghton et al.\ (2012).
Differences are $\Delta$ = ``our data'' -- ``Houghton et al.''. 
The median offsets and scatter are summarized in Table \ref{tab-a1689comp},
see text for discussion.
\label{fig-a1689speccomp} }
\end{figure}

In this section we compare our redshifts and velocity dispersions to those published by Houghton et al.\ (2012).
We converted the relative radial velocities published by Houghton et al.\ to redshift, using the cluster redshift of 0.183 assumed
by Houghton et al. Figure \ref{fig-a1689speccomp} and Table \ref{tab-a1689comp} summarize the comparisons.
The very small difference in redshifts may originate from the accuracy of the cluster redshift stated by
Houghton et al., and is in any case of no importance for our results.
The offset between the velocity dispersions is $-0.05$ in $\log \sigma$ with our measurements being smaller.
Houghton et al.\ used 1.4 arcsec extraction aperture while we use 2.0 arcsec. However, adopting the aperture correction
from J\o rgensen et al.\ (1995b) that would only explain an offset of 0.003 in $\log \sigma$.
Houghton et al.\ use the arc spectra for determining the 
instrumental resolution and state this as $\approx 75 {\rm km\,s^{-1}}$ at 5000 {\AA} (equivalent
to $\approx 70 {\rm km\,s^{-1}}$ at 4500 {\AA} in the rest frame of the cluster).
We measure the instrumental resolution from sky spectra stacked the same way
as the galaxy spectra and find $\approx 85 {\rm km\,s^{-1}}$ at 4500 {\AA} in the rest frame of the cluster. 
Thus, a systematic difference is expected of $-0.03$ to $-0.01$ on $\log \sigma$ for
galaxies with $\sigma=100 {\rm km\,s^{-1}}$ and $200 {\rm km\,s^{-1}}$, respectively.
The remainder of the difference between the two sets of measurements is most likely
due to differences in choice of template spectra.
The scatter in the comparison confirms our estimates of the random measurement
uncertainties on $\log \sigma$ of 0.02 for the highest velocity dispersion galaxies,
rising to 0.04 for the lower velocity dispersion galaxies.

\section{Average spectral parameters and spectral parameters from composite spectra}

Table \ref{tab-composites} gives luminosity weighted average spectra parameters
for the local reference sample and spectral parameters from the composite spectra of the $z=0.19-0.89$ sub-samples.
The table also lists the average values of the velocity dispersions for the individual galaxies 
included in each average or composite.

\section{Presentation of the spectra \label{SEC-SPECTRA}}

\begin{figure*}
\figurenum{22}
\plotone{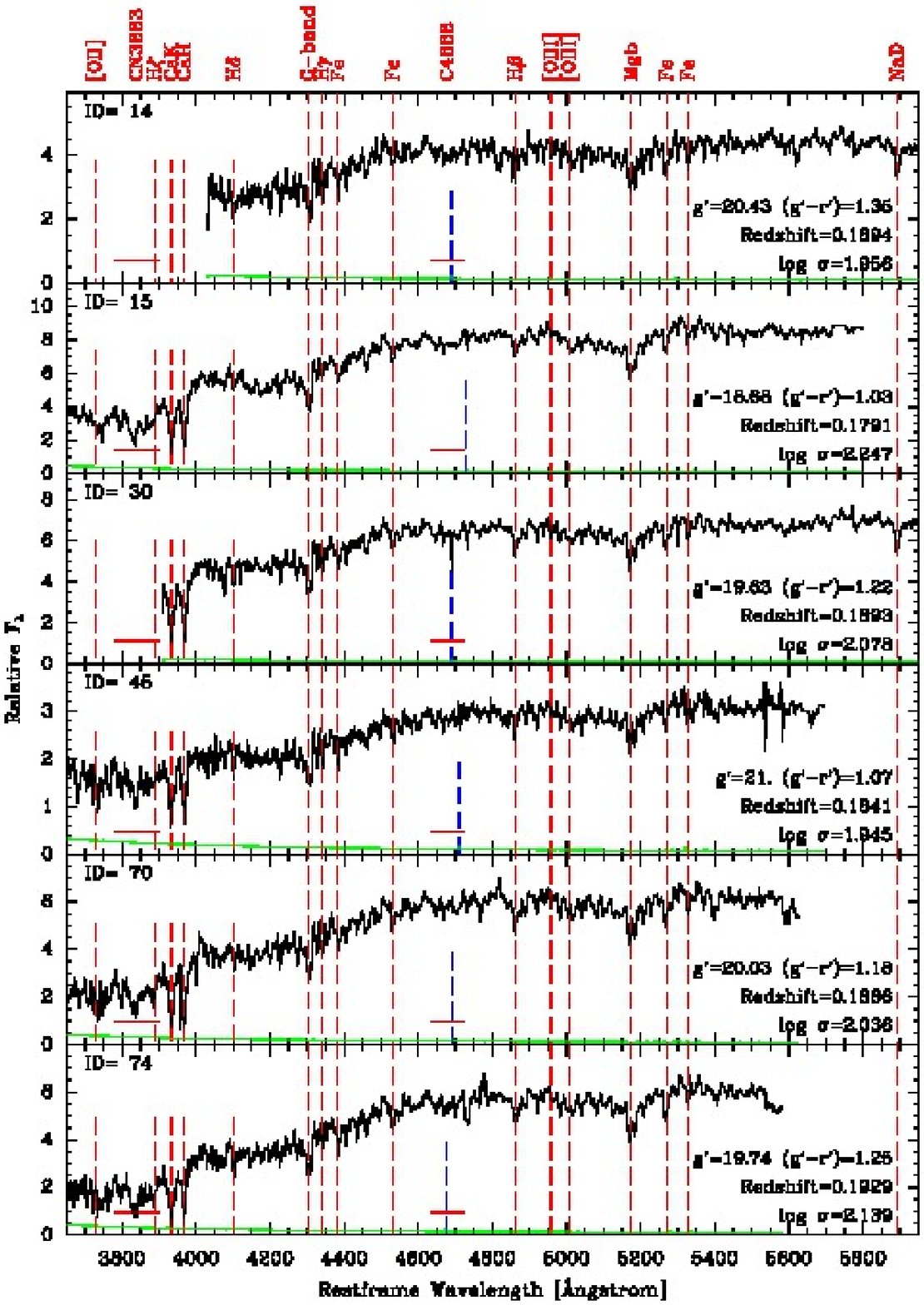}
\caption{Spectra of the galaxies that are considered members of Abell 1689.
Black lines -- the galaxy spectra, green lines -- the random noise multiplied by four.
At the strong sky lines, the random noise underestimates the real noise due to systematic
errors in the sky subtraction.
Some of the absorption lines are marked (Table \ref{tab-lines}). The locations of the emission lines from [\ion{O}{2}]
and [\ion{O}{3}] are also marked, though emission is only present in some of the galaxies.
The location of the possible residuals from the strong skyline at 5577{\AA} is marked with
blue dashed lines.
The spectra shown in this figure have been processed as described in the text.
Only the first page of this figure appears in main journal. All component figures are available in the Figure set. 
\label{fig-specA1689}
}
\end{figure*}

\begin{figure*}
\figurenum{24}
\plotone{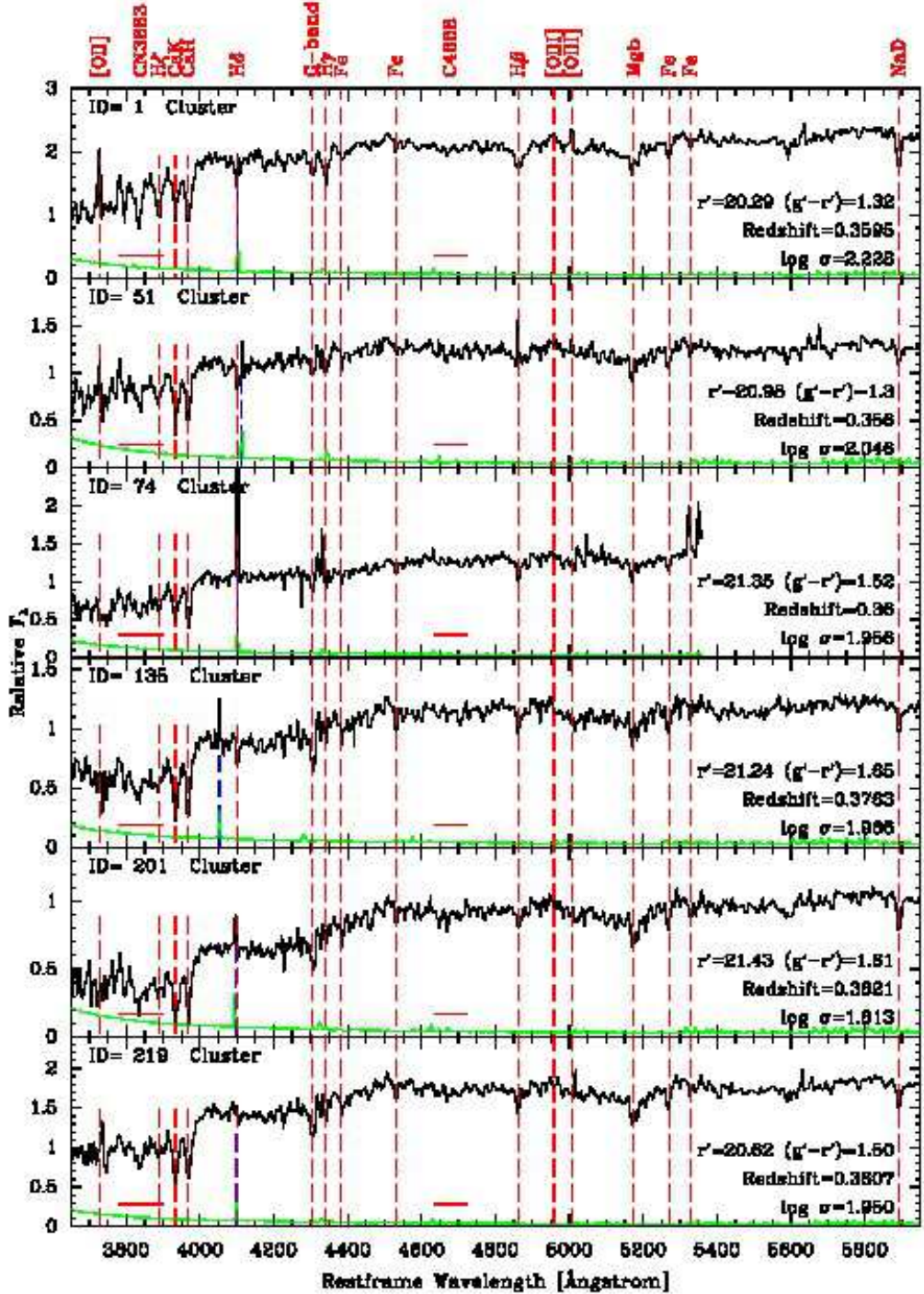}
\caption{Spectra of the galaxies that are considered members of RXJ0027.6+2616 or of the foreground group.
The latter is labelled ``group'' and displayed last in the panels.
Black lines -- the galaxy spectra, green lines -- the random noise multiplied by four.
At the strong sky lines, the random noise underestimates the real noise due to systematic
errors in the sky subtraction.
Some of the absorption lines are marked (Table \ref{tab-lines}). The locations of the emission lines from [\ion{O}{2}]
and [\ion{O}{3}] are also marked, though emission is only present in some of the galaxies.
The location of the possible residuals from the strong skyline at 5577{\AA} is marked with
blue dashed lines.
The spectra shown in this figure have been processed as described in the text.
Only the first page of this figure appears in main journal. All component figures are available in the Figure set. 
\label{fig-specRXJ0027}
}
\end{figure*}

\begin{figure*}
\figurenum{25}
\plotone{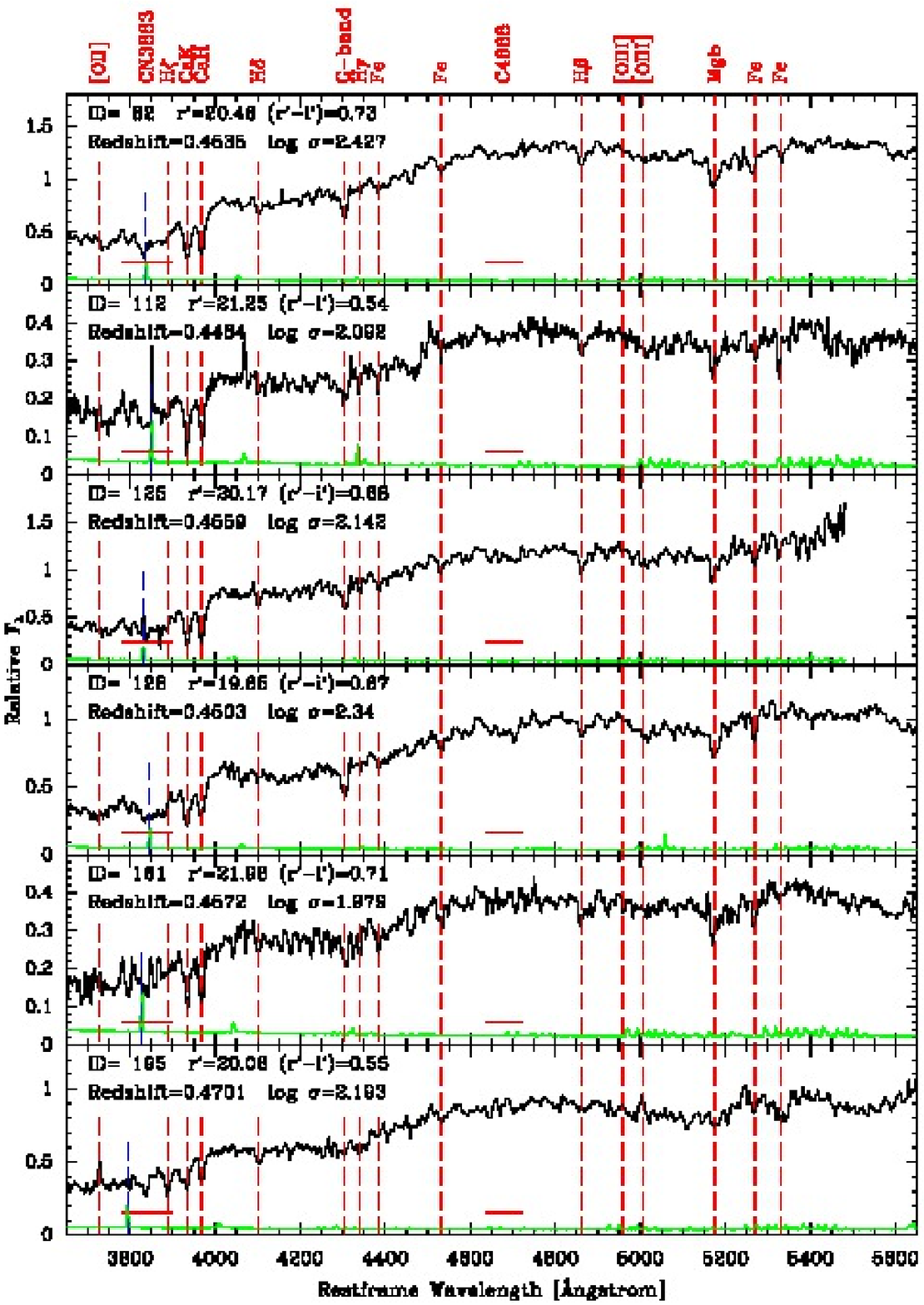}
\caption{Spectra of the galaxies that are considered members of RXJ1347.5--1137.
Black lines -- the galaxy spectra, green lines -- the random noise multiplied by four.
At the strong sky lines, the random noise underestimates the real noise due to systematic
errors in the sky subtraction.
Some of the absorption lines are marked (Table \ref{tab-lines}). The locations of the emission lines from [\ion{O}{2}]
and [\ion{O}{3}] are also marked, though emission is only present in some of the galaxies.
The location of the possible residuals from the strong skyline at 5577{\AA} is marked with
blue dashed lines.
The spectra shown in this figure have been processed as described in the text.
Only the first page of this figure appears in main journal. All component figures are available in the Figure set. 
\label{fig-specRXJ1347}
}
\end{figure*}

The spectra of the galaxies are shown in Figures \ref{fig-specA1689}--\ref{fig-specRXJ1347}.
Major spectra features labeled on the figures are listed in Table \ref{tab-lines}.
Full figures are available in the online journal.

\clearpage
\startlongtable
\begin{deluxetable*}{rrrrrrrrrrrr}
\tablecaption{Results from Template Fitting \label{tab-kin} }
\tabletypesize{\scriptsize}
\tablehead{
\colhead{Cluster} & \colhead{ID} & \colhead{Redshift} & \colhead{Member\tablenotemark{a}} &  
\colhead{$\log \sigma$} & \colhead{$\log \sigma _{\rm cor}$\tablenotemark{b}} & 
\colhead{$\sigma _{\log \sigma}$} & \multicolumn{3}{c}{Template fractions} & \colhead{$\chi ^2$} & \colhead{S/N\tablenotemark{c}} \\
\colhead{} & \colhead{}&\colhead{} &\colhead{} &\colhead{} &\colhead{} &\colhead{} & \colhead{B8V} & \colhead{G1V} & \colhead{K0III} & }
\startdata
A1689 &    14& 0.1894&  1&  2.003&  1.956&  0.001&   0.00&   0.57&   0.43&    3.5&    142\\
A1689 &    15& 0.1791&  1&  2.264&  2.247&  0.014&   0.00&   0.41&   0.59&   18.8&    247\\
A1689 &    30& 0.1893&  1&  2.112&  2.078&  0.028&   0.00&   0.76&   0.24&   10.7&    264\\
A1689 &    45& 0.1841&  1&  1.994&  1.945&  0.003&   0.00&   0.68&   0.32&    7.1&    130\\
A1689 &    70& 0.1886&  1&  2.074&  2.036&  0.040&   0.00&   0.48&   0.52&    5.9&    197\\
A1689 &    74& 0.1929&  1&  2.167&  2.139&  0.029&   0.00&   0.62&   0.38&   13.7&    180\\
A1689 &    91& 0.1943&  1&  2.127&  2.095&  0.011&   0.00&   0.76&   0.24&    6.1&     82\\
A1689 &   135& 0.1867&  1&  1.994&  1.946&  0.029&   0.00&   0.58&   0.42&    3.7&    126\\
A1689 &   152& 0.1879&  1&  2.068&  2.029&  0.030&   0.00&   0.44&   0.56&   11.3&    215\\
A1689 &   160& 0.0817&  0&  \nodata&  \nodata&  \nodata&  \nodata&  \nodata&  \nodata&  \nodata&     56\\
\enddata
\tablecomments{Table 13 is published in its entirety in the machine-readable format.
      A portion is shown here for guidance regarding its form and content. }
\tablenotetext{a}{Adopted membership: 1 -- galaxy is a member of the cluster; 
0 -- galaxy is not a member of the cluster.}
\tablenotetext{b}{Velocity dispersions corrected to a standard size aperture equivalent to a
circular aperture with diameter of 3.4 arcsec at the distance of the Coma cluster,
and corrected for systematic effects using Equation \ref{eq-sigmasys1}.}
\tablenotetext{c}{S/N per {\AA}ngstrom in the rest frame of the galaxy. The wavelength
interval 4100-5500 {\AA} was used for all galaxies. }
\end{deluxetable*}

\startlongtable
\begin{deluxetable*}{r rrrr rrr  rrr rrr  rrr}
\tablecaption{Line Indices for Cluster Members\label{tab-line} }
\tabletypesize{\scriptsize}
\tablewidth{0pc}
\tablehead{
\colhead{Cluster/ID} & \colhead{H$\zeta _{\rm A}$} & \colhead{CN3883}& \colhead{CaHK}& \colhead{D4000}
& \colhead{H$\delta _{\rm A}$}
& \colhead{CN$_1$} & \colhead{CN$_2$} & \colhead{G4300} & \colhead{H$\gamma _{\rm A}$}
& \colhead{Fe4383} & \colhead{C4668} & \colhead{H$\beta$} & \colhead{H$\beta _{\rm G}$} & \colhead{Mg$b$} 
& \colhead{Fe5270} & \colhead{Fe5335} 
} 
\startdata
{\bf A1689:} \\
 14& \nodata& \nodata& \nodata& \nodata& 3.17& \nodata& \nodata& 4.95& -4.14& 4.48& 7.04& \nodata & \nodata& 3.43& 3.54& 2.65\\
 14& \nodata& \nodata& \nodata& \nodata& 0.13& \nodata& \nodata& 0.10& 0.11& 0.13& 0.12& \nodata & \nodata& 0.05& 0.05& 0.05\\
 15& 0.72& 0.269& 23.42& 2.084& -2.49& 0.126& 0.149& 5.70& -6.24& 4.39& 6.02& 1.75 & 1.92& 4.59& 2.55& 3.25\\
 15& 0.13& 0.005& 0.20& 0.003& 0.08& 0.002& 0.002& 0.06& 0.07& 0.08& 0.07& 0.03 & 0.02& 0.03& 0.03& 0.03\\
 30& \nodata& \nodata& \nodata& \nodata& -0.12& \nodata& 0.030& 4.98& -4.37& 4.39& \nodata& 2.06 & 2.22& 3.62& 3.00& 2.28\\
 30& \nodata& \nodata& \nodata& \nodata& 0.07& \nodata& 0.002& 0.05& 0.06& 0.07& \nodata& 0.03 & 0.02& 0.03& 0.03& 0.03\\
 45& 0.72& 0.151& 16.27& 1.630& -0.83& 0.046& 0.069& 4.82& -4.73& 2.74& 4.90& 0.77 & 0.83& 3.33& 2.10& 2.90\\
 45& 0.19& 0.007& 0.39& 0.004& 0.15& 0.004& 0.005& 0.12& 0.13& 0.16& 0.14& 0.05 & 0.04& 0.05& 0.05& 0.06\\
 70& 3.00& 0.238& 26.56& 2.124& -1.48& 0.083& 0.107& 5.20& -4.49& 4.90& 6.02& 2.21 & 2.66& 3.91& 3.36& 2.78\\
 70& 0.15& 0.007& 0.26& 0.004& 0.11& 0.003& 0.003& 0.08& 0.08& 0.09& 0.09& 0.03 & 0.02& 0.03& 0.04& 0.04\\
 74& 2.57& 0.238& 23.37& 2.316& -0.83& 0.050& 0.098& 5.60& -5.90& 5.12& 6.00& 2.18 & 2.29& 4.80& 3.06& 2.67\\
 74& 0.19& 0.008& 0.33& 0.006& 0.13& 0.003& 0.004& 0.08& 0.09& 0.11& 0.10& 0.04 & 0.02& 0.04& 0.04& 0.04\\
 91& 3.22& 0.109& 22.30& 2.244& -0.62& 0.023& 0.060& 5.72& -6.25& 5.28& 6.06& 1.74 & 1.78& 3.92& 3.51& 2.35\\
 91& 0.33& 0.016& 0.63& 0.011& 0.24& 0.006& 0.007& 0.19& 0.21& 0.24& 0.22& 0.08 & 0.05& 0.08& 0.08& 0.09\\
 135& 1.27& 0.120& 21.00& 1.878& -1.04& 0.021& 0.043& 5.51& -4.75& 4.35& 5.52& 1.97 & 2.34& 3.96& 2.34& 3.32\\
 135& 0.22& 0.009& 0.42& 0.005& 0.16& 0.004& 0.004& 0.11& 0.12& 0.14& 0.13& 0.05 & 0.03& 0.05& 0.05& 0.06\\
 152& 1.28& 0.219& 21.18& 1.757& -1.99& 0.078& 0.105& 6.31& -5.33& 4.84& \nodata& 1.76 & 2.05& 4.30& 3.25& 2.67\\
 152& 0.14& 0.005& 0.26& 0.003& 0.11& 0.003& 0.003& 0.07& 0.08& 0.09& \nodata& 0.03 & 0.02& 0.03& 0.03& 0.04\\
 217& 1.15& 0.217& 22.64& 2.207& -1.65& 0.039& 0.070& 5.89& -6.21& 3.95& 6.41& 1.79 & 2.01& 4.51& 2.54& 2.47\\
 217& 0.13& 0.005& 0.20& 0.003& 0.08& 0.002& 0.002& 0.06& 0.07& 0.08& 0.07& 0.03 & 0.02& 0.03& 0.03& 0.03\\
\enddata
\tablecomments{The indices have been corrected for galaxy velocity dispersion and aperture corrected. 
Table 14 is published in its entirety in the machine-readable format.
In the portion shown here for guidance regarding its form and content, the lines are wrapped such
that the second line for each galaxy lists the uncertainties.
}
\end{deluxetable*}

\begin{deluxetable}{l rrr}
\tablecaption{Equivalent Widths of [\ion{O}{2}] for Cluster Members \label{tab-EWOII} }
\tabletypesize{\scriptsize}
\tablewidth{0pc}
\tablehead{
\colhead{Cluster} & \colhead{ID} & \colhead{EW [\ion{O}{2}]} &  \colhead{$\sigma _{\rm EW[OII]}$}
} 
\startdata
 A1689& 368& 6.6& 1.1\\
 A1689& 508& 10.6& 2.3\\
 A1689& 724& 24.1& 2.2\\
 A1689& 752& 6.3& 1.2\\
 A1689& 814& 5.8& 0.4\\
 A1689& 906& 7.6& 1.7\\
 A1689& 972& 6.2& 0.3\\
 A1689& 1013& 36.1& 0.7\\
 RXJ0056.2+2622& 323& 8.7& 1.2\\
 RXJ0056.2+2622& 1054& 30.9& 1.5\\
 RXJ0056.2+2622& 1256& 29.6& 4.8\\
 RXJ0056.2+2622& 1391& 91.5& 12.3\\
 RXJ0027.6+2616& 1& 8.2& 0.4\\
 RXJ0027.6+2616& 760& 3.9& 0.4\\
 RXJ0027.6+2616& 841& 30.6& 3.7\\
 RXJ0027.6+2616& 1081& 5.4& 0.8\\
 RXJ1347.5--1145& 195& 4.7& 0.4\\
 RXJ1347.5--1145& 436& 103.4& 3.4\\
\enddata
\end{deluxetable}


\startlongtable
\begin{deluxetable*}{r rr rrrr rrr rr}
\tablecaption{Average and Composite Spectral Parameters \label{tab-composites} }
\tabletypesize{\scriptsize}
\tablewidth{0pc}
\tablehead{
\colhead{$\log \sigma$} 
& \colhead{$\langle \log \sigma \rangle$} & \colhead{rms} & \colhead{$N_{\rm gal}$}
& \multicolumn{2}{c}{CN3883} & \multicolumn{2}{c}{ $({\rm H\delta _A + H\gamma _A})'$} & \multicolumn{2}{c}{log Fe4383} & \multicolumn{2}{c}{log C4668} \\
\colhead{(1)} & \colhead{(2)} & \colhead{(3)} & \colhead{(4)} & \colhead{(5)} & \colhead{(6)} &
\colhead{(7)} & \colhead{(8)} & \colhead{(9)} & \colhead{(10)}  & \colhead{(11)}  & \colhead{(12)}
} 
\startdata
\multicolumn{2}{l}{\bf Perseus:} & & & & & & & & \\
$2.05-2.10$ & 2.058 & 0.056 & 3 & 0.162 & 0.013 & -0.054 & 0.006 & 0.605 & 0.029 & 0.825 & 0.015 \\
$2.10-2.15$ & 2.111 & 0.016 & 3 & 0.148 & 0.009 & -0.050 & 0.005 & 0.500 & 0.033 & 0.777 & 0.015 \\
$2.15-2.20$ & 2.188 & 0.017 & 3 & 0.229 & 0.008 & -0.085 & 0.004 & 0.665 & 0.019 & 0.813 & 0.012 \\
$2.20-2.25$ & 2.241 & 0.014 & 8 & 0.194 & 0.007 & -0.088 & 0.003 & 0.671 & 0.013 & 0.848 & 0.007 \\
$2.25-2.30$ & 2.273 & 0.016 & 12 & 0.276 & 0.005 & -0.101 & 0.002 & 0.687 & 0.010 & 0.852 & 0.006 \\
$2.30-2.35$ & 2.326 & 0.011 & 14 & 0.223 & 0.005 & -0.089 & 0.002 & 0.652 & 0.010 & 0.866 & 0.005 \\
$2.35-2.40$ & 2.387 & 0.012 & 8 & 0.270 & 0.007 & -0.104 & 0.003 & 0.730 & 0.012 & 0.931 & 0.006 \\
$\ge2.40$ & 2.476 & 0.049 & 10 & 0.276 & 0.006 & -0.094 & 0.003 & 0.719 & 0.011 & 0.906 & 0.006 \\
\multicolumn{2}{l}{\bf Abell 1689:} \\
$<2.00$ & 1.945 & 0.020 & 12 & 0.142 & 0.004 & -0.068 & 0.001 & 0.601 & 0.009 & 0.752 & 0.006 \\
$2.00-2.05$ & 2.024 & 0.014 & 8 & 0.241 & 0.004 & -0.079 & 0.001 & 0.656 & 0.007 & 0.774 & 0.005 \\
$2.05-2.10$ & 2.076 & 0.017 & 7 & 0.189 & 0.005 & -0.079 & 0.001 & 0.662 & 0.009 & 0.805 & 0.006 \\
$2.10-2.15$ & 2.131 & 0.016 & 5 & 0.236 & 0.007 & -0.090 & 0.002 & 0.673 & 0.010 & 0.810 & 0.007 \\
$2.15-2.20$ & 2.169 & 0.012 & 9 & 0.292 & 0.005 & -0.091 & 0.001 & 0.651 & 0.007 & 0.812 & 0.005 \\
$2.20-2.25$ & 2.235 & 0.011 & 4 & 0.292 & 0.007 & -0.086 & 0.002 & 0.574 & 0.014 & 0.858 & 0.007 \\
$2.25-2.30$ & 2.280 & 0.019 & 5 & 0.299 & 0.005 & -0.098 & 0.001 & 0.662 & 0.008 & 0.876 & 0.004 \\
$2.30-2.35$ & 2.319 & 0.021 & 3 & 0.278 & 0.006 & -0.100 & 0.002 & 0.700 & 0.009 & 0.868 & 0.006 \\
$2.35-2.40$ & 2.383 & 0.006 & 3 & 0.329 & 0.008 & -0.111 & 0.002 & 0.695 & 0.009 & 0.932 & 0.005 \\
$\ge2.40$ & 2.449 & 0.000 & 2 & 0.312 & 0.009 & -0.093 & 0.002 & 0.600 & 0.016 & 0.871 & 0.008 \\
\multicolumn{2}{l}{\bf RXJ0056.2+2622:} \\
$<2.00$ & 1.962 & 0.023 & 10 & 0.146 & 0.003 & -0.059 & 0.001 & 0.604 & 0.006 & 0.695 & 0.006 \\
$2.00-2.05$ & 2.076 & 0.019 & 8 & 0.204 & 0.002 & -0.076 & 0.001 & 0.648 & 0.005 & 0.817 & 0.004 \\
$2.05-2.10$ & 2.033 & 0.011 & 5 & 0.184 & 0.004 & -0.088 & 0.001 & 0.644 & 0.007 & 0.805 & 0.005 \\
$2.10-2.15$ & 2.111 & 0.011 & 2 & 0.211 & 0.005 & -0.067 & 0.001 & 0.643 & 0.010 & 0.674 & 0.010 \\
$2.15-2.20$ & 2.176 & 0.007 & 2 & 0.188 & 0.005 & -0.079 & 0.001 & 0.506 & 0.013 & 0.737 & 0.009 \\
$2.20-2.25$ & 2.230 & 0.019 & 3 & 0.265 & 0.004 & -0.095 & 0.001 & 0.653 & 0.007 & 0.846 & 0.005 \\
$2.25-2.30$ & 2.280 & 0.015 & 6 & 0.263 & 0.003 & -0.091 & 0.001 & 0.652 & 0.005 & 0.859 & 0.003 \\
$2.30-2.35$ & 2.323 & 0.012 & 4 & 0.273 & 0.003 & -0.094 & 0.001 & 0.662 & 0.006 & 0.897 & 0.004 \\
$2.35-2.40$ & 2.373 & 0.017 & 5 & 0.276 & 0.002 & -0.098 & 0.001 & 0.661 & 0.005 & 0.866 & 0.003 \\
$\ge2.40$ & 2.452 & 0.071 & 3 & 0.311 & 0.003 & -0.097 & 0.001 & 0.658 & 0.007 & 0.893 & 0.004 \\
\multicolumn{2}{l}{\bf RXJ0027.6+2616:} \\
$<2.00$ & 1.926 & 0.051 & 10 & 0.191 & 0.004 & -0.057 & 0.001 & 0.548 & 0.008 & 0.610 & 0.007 \\
$2.00-2.10$ & 2.063 & 0.021 & 3 & 0.230 & 0.007 & -0.057 & 0.001 & 0.539 & 0.013 & 0.767 & 0.008 \\
$2.10-2.20$ & 2.144 & 0.021 & 9 & 0.226 & 0.004 & -0.064 & 0.001 & 0.595 & 0.006 & 0.781 & 0.004 \\
$2.20-2.30$ & 2.250 & 0.022 & 2 & 0.237 & 0.008 & -0.052 & 0.002 & 0.600 & 0.015 & 0.742 & 0.011 \\
$2.30-2.40$ & 2.359 & 0.023 & 5 & 0.283 & 0.005 & -0.079 & 0.001 & 0.640 & 0.009 & 0.851 & 0.005 \\
$\ge2.40$ & 2.425 & 0.023 & 2 & 0.294 & 0.006 & -0.083 & 0.001 & 0.673 & 0.010 & 0.913 & 0.006 \\
\multicolumn{2}{l}{\bf RXJ1347.5--1145:} \\
$<2.00$ & 1.955 & 0.051 & 6 & 0.127 & 0.003 & -0.041 & 0.001 & 0.553 & 0.013 & 0.646 & 0.010 \\
$2.00-2.10$ & 2.043 & 0.036 & 4 & 0.132 & 0.003 & -0.048 & 0.001 & 0.572 & 0.010 & 0.765 & 0.006 \\
$2.10-2.15$ & 2.126 & 0.018 & 8 & 0.170 & 0.003 & -0.048 & 0.001 & 0.607 & 0.008 & 0.721 & 0.006 \\
$2.15-2.20$ & 2.174 & 0.014 & 6 & 0.151 & 0.003 & -0.068 & 0.001 & 0.644 & 0.008 & 0.818 & 0.006 \\
$2.20-2.30$ & 2.259 & 0.043 & 6 & 0.218 & 0.004 & -0.065 & 0.001 & 0.567 & 0.011 & 0.797 & 0.006 \\
$2.30-2.40$ & 2.349 & 0.006 & 4 & 0.230 & 0.004 & -0.058 & 0.001 & 0.572 & 0.011 & 0.810 & 0.006 \\
$\ge2.40$ & 2.414 & 0.012 & 3 & 0.273 & 0.004 & -0.064 & 0.001 & 0.654 & 0.011 & 0.807 & 0.007 \\
\multicolumn{2}{l}{\bf MS0451.3--0306:} \\
$2.00-2.10$ & 2.071 & 0.019 & 8 & 0.187 & 0.003 & -0.048 & 0.001 & 0.627 & 0.007 & 0.738 & 0.007 \\
$2.10-2.20$ & 2.118 & 0.009 & 5 & 0.199 & 0.004 & -0.046 & 0.001 & 0.583 & 0.010 & 0.681 & 0.009 \\
$2.20-2.25$ & 2.190 & 0.021 & 4 & 0.251 & 0.004 & -0.078 & 0.001 & 0.593 & 0.009 & 0.791 & 0.006 \\
$2.25-2.30$ & 2.277 & 0.013 & 5 & 0.243 & 0.004 & -0.088 & 0.001 & 0.635 & 0.007 & 0.806 & 0.005 \\
$2.30-2.35$ & 2.335 & 0.012 & 5 & 0.242 & 0.003 & -0.072 & 0.001 & 0.627 & 0.006 & 0.787 & 0.005 \\
$2.35-2.40$ & 2.383 & 0.020 & 4 & 0.237 & 0.004 & -0.090 & 0.001 & 0.688 & 0.006 & 0.816 & 0.005 \\
$\ge2.40$ & 2.451 & 0.032 & 3 & 0.276 & 0.004 & -0.075 & 0.001 & 0.672 & 0.007 & 0.858 & 0.005 \\
\multicolumn{2}{l}{\bf RXJ0152.7--1357:} \\
$2.10-2.20$ & 2.060 & 0.034 & 2 & 0.207 & 0.007 & -0.049 & 0.004 & 0.344 & 0.056 & 0.802 & 0.027 \\
$2.20-2.25$ & 2.180 & 0.018 & 4 & 0.220 & 0.005 & -0.040 & 0.002 & 0.429 & 0.031 & 0.745 & 0.022 \\
$2.25-2.30$ & 2.273 & 0.012 & 3 & 0.239 & 0.005 & -0.044 & 0.002 & 0.560 & 0.023 & 0.818 & 0.020 \\
$2.30-2.35$ & 2.329 & 0.018 & 5 & 0.262 & 0.004 & -0.042 & 0.002 & 0.527 & 0.021 & 0.913 & 0.012 \\
$\ge2.40$ & 2.411 & 0.052 & 4 & 0.292 & 0.004 & -0.054 & 0.002 & 0.519 & 0.025 & 0.870 & 0.015 \\
\multicolumn{2}{l}{\bf RXJ1226.9+3332:} \\
$2.05-2.10$ & 2.027 & 0.042 & 9 & 0.207 & 0.002 & -0.020 & 0.001 & 0.547 & 0.014 & 0.693 & 0.018 \\
$2.10-2.15$ & 2.128 & 0.012 & 7 & 0.222 & 0.004 & -0.022 & 0.002 & 0.609 & 0.024 & 0.876 & 0.021 \\
$2.15-2.20$ & 2.175 & 0.011 & 4 & 0.204 & 0.004 & -0.018 & 0.002 & 0.470 & 0.031 & 0.668 & 0.032 \\
$2.20-2.30$ & 2.225 & 0.018 & 3 & 0.293 & 0.004 & -0.045 & 0.002 & 0.642 & 0.018 & 0.835 & 0.018 \\
$2.30-2.40$ & 2.295 & 0.038 & 4 & 0.240 & 0.004 & -0.023 & 0.002 & 0.670 & 0.020 & 0.877 & 0.019 \\
$\ge2.40$ & 2.444 & 0.078 & 5 & 0.288 & 0.003 & -0.062 & 0.001 & 0.597 & 0.020 & 0.906 & 0.013 \\
\enddata
\tablecomments{Column 1: Selection criteria for $\log \sigma$. 
Column 2: Average $\log \sigma$ for selected galaxies. 
Column 3: Rms scatter in $\langle \log \sigma \rangle$.
Column 4: Number of selected galaxies. 
Columns 5--12: Line indices and uncertainties, for Perseus the luminosity weighted average of individual measurements, for $z=0.19-0.89$ 
clusters line indices measured from the composite spectra. }
\end{deluxetable*}

\begin{deluxetable*}{lrl}
\tablecaption{Spectral Lines Marked on Figures \label{tab-lines} }
\tablewidth{0pc}
\tablehead{
\colhead{Line} & \colhead{Wavelength(s) [{\AA}]} & \colhead{Notes} 
} 
\startdata
 H$\beta$  & 4861.7 & Balmer line \\
 H$\delta$ & 4340.8 & Balmer line \\
 H$\gamma$ & 4102.1 & Balmer line \\
 H$\zeta$  & 3889.4 & Balmer line \\
 $[$\ion{O}{2}$]$  & 3726.0, 3728.8 & Oxygen emission line doublet marked at 3727{\AA} \\
 $[$\ion{O}{3}$]$  & 4958.9, 5006.8 & Oxygen emission lines \\
 CN3883 & 3780.0--3900.0 & Broad absorption band, horizontal line marks the index passband \\
 CaH, CaK & 3933.7, 3968.5 & Main calcium (CaII) lines \\  
 G-band & 4282.6--4317.6 & Broad absorption band marked at center of Lick/IDS index passband \\
 Fe  & 4383.3, 4531.1, 5270.4, 5328.5 & Main iron lines \\
 C4668  & 4635.2--4721.5 & Broad absorption band, horizontal line marks the Lick/IDS index passband \\
 Mg  & 5167.3, 5172.7, 5183.6 & Magnesium (MgI) triplet marked at 5175{\AA} \\
 NaD & 5889.9, 5895.9   & Sodium doublet marked at 5893{\AA} \\
\enddata
\end{deluxetable*}


\clearpage

\begin{figure*}
\figurenum{23}
\plotone{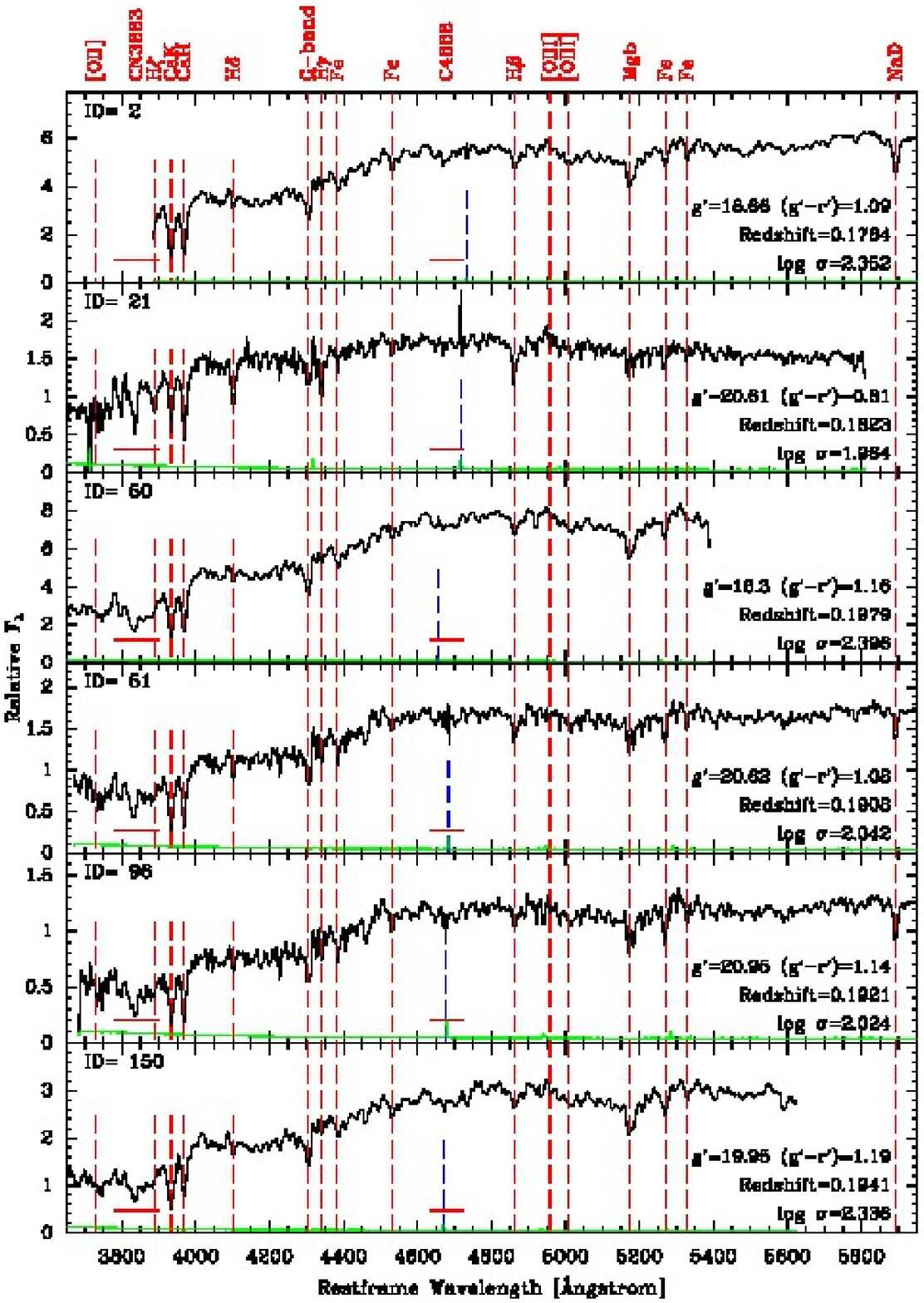}
\caption{Spectra of the galaxies that are considered members of RXJ0056.2+2622.
Black lines -- the galaxy spectra, green lines -- the random noise multiplied by four.
At the strong sky lines, the random noise underestimates the real noise due to systematic
errors in the sky subtraction.
Some of the absorption lines are marked (Table \ref{tab-lines}). The locations of the emission lines from [\ion{O}{2}]
and [\ion{O}{3}] are also marked, though emission is only present in some of the galaxies.
The location of the possible residuals from the strong skyline at 5577{\AA} is marked with
blue dashed lines.
The spectra shown in this figure have been processed as described in the text.
Only the first page of this figure appears in main journal. All component figures are available in the Figure set. 
}
\end{figure*}

\clearpage

\begin{thebibliography}{}


\bibitem[Abell et al.(1989)]{abell:1989}
Abell, G.\ O., Corwin Jr., H.\ G., Olowin, R.\ P. 1989, ApJS, 70, 1


\bibitem[Andersson \& Madejski(2004)]{andersson:2004}
Andersson, K.\ E., \& Madejski, G.\ M. 2004, ApJ, 607, 190


\bibitem[Barr et al.(2005)]{barr:2005}
Barr, J., Davies, R., J\o rgensen, I., Bergmann, M., \& Crampton, D. 2005, AJ, 130, 445


\bibitem[Barrena et al.(2007)]{barrena:2007}
Barrena, R., Boschin, W., Garirdi, M., \& Spolaor, M. 2007, A\&A, 469, 861

\bibitem[Barro et al.(2014)]{barro:2014}
Barro, G., Faber, S.\ M., P\'{e}rez-Gonz\'{a}lez, P.\ G., et al. 2014, ApJ, 791, 52

\bibitem[Beers et al.(1990)]{beers:1990}
Beers, T.\ C., Flynn, K., \& Gebhardt, K. 1990, AJ, 100, 32



\bibitem[Bernardi et al.(2006)]{bernardi:2006}
Bernardi, M., Nichol, R.\ C., Sheth, R.\ K., Miller, C.\ J., \& Brinkmann, J. 2006, AJ, 131, 1288
 
\bibitem[Bertin \& Arnouts(1996)]{bertin:1996}
Bertin, E., \& Arnouts, S. 1996, A\&AS, 117, 393

\bibitem[Biviano \& Poggianti(2009)]{biviano:2009}
Biviano, A., \& Poggianti, B.\ M. 2009, A\&, 501, 419


\bibitem[B\"{o}hringer et al.(2000)]{bohringer:2000}
B\"{o}hringer, H., Voges, W., Huchra, J.\ P., et al. 2000, ApJS, 129, 435





\bibitem[Brooks \& Christensen(2016)]{brooks:2016}
Brooks, A., \& Christensen, C. 2016, in Galactic Bulges, Astrophysics and Space Science Library, Vol.\ 418, 371, Springer 


\bibitem[Bruzual(1983)]{bruzual:1983}
Bruzual A., G. 1983, ApJ, 273, 105


 



\bibitem[Cardiel et al.(1998)]{cardiel:1998}
Cardiel, N., Gorgas, J., Cenarro, J.\ \& Gonz\'{a}lez, J.\ J. 1998, A\&AS, 127, 597


\bibitem[Czoske(2004)]{czoske:2004}
Czoske, O. 2004, in IAU Colloq. 195, Outskirts of Galaxy Clusters: Intense Life
in the Suburbs, ed.\ A.\ Diaferio (Cambridge: Cambridge Univ. Press), 183

\bibitem[Chiboucas et al.(2009)]{Chiboucas:2009}
Chiboucas, K., Barr, J., Flint, K., et al. 2009, ApJS, 184, 271

\bibitem[Choi et al.(2014)]{choi:2014}
Choi, J., Conroy, C., Moustakas, J., et al. 2014, ApJ, 792, 95

\bibitem[Cohen \& Kneib(2002)]{cohen:2002}
Cohen, J.\ G., \& Kneib, J.-P. 2002, ApJ, 573, 524

\bibitem[Concas et al.(2017)]{concas:2017}
Concas, A., Pozzetti, L., Moresco, M., \& Cimatti, A. 2017, MNRAS, 468, 1747

\bibitem[Conroy et al.(2014)]{conroy:2014}
Conroy, C., Graves, G.\ J., \& van Dokkum, P.\ G. 2014, ApJ, 780, 33

\bibitem[Correa et al.(2015)]{correa:2015}
Correa, C.\ A., Wyithe, J.\ S.\ B., Schaye, J., \& Duffy, A.\ R. 2015, MNRAS, 452, 1271





\bibitem[Davidge \& Clark(1994)]{davidge:1994}
Davidge, T.\ J., \& Clark, C.\ C. 1994, AJ, 107, 946




\bibitem[Dressler(1980)]{dressler:1980}
Dressler, A. 1980, ApJ, 236, 351


\bibitem[Dressler et al.(1997)]{dressler:1997}
Dressler, A., Oemler Jr., A., Couch, W.\ J., et al. 1997, ApJ, 490, 577

\bibitem[Dressler et al.(1999)]{dressler:1999}
Dressler, A., Smail, I., Poggianti, B.\ M., et al. 1999, ApJS, 122, 51



 

\bibitem[Evans et al.(2010)]{evans:2010}
Evans, I.\ N., Primini, F.\ A., Glotfelty, K.\ J., et al. 2010, ApJS, 189, 37

\bibitem[Ettori et al.(2004)]{ettori:2004}
Ettori, S., Tozzi, P., Borgani, S., \& Rostati, P. 2004, A\&A, 417, 13

\bibitem[Ettori et al.(2009)]{ettori:2009}
Ettori, S., Morandi, A., Tozzi, P., et al. 2009, A\&A, 501, 61


\bibitem[Faber et al.(2007)]{faber:2007}
Faber, S.\ M., Willmer, C.\ N.\ A., Wolf, C. et al.\ 2007, ApJ, 665, 265

\bibitem[Fakhouri et al.(2010)]{fakhouri:2010}
Fakhouri, O., Ma, C.-P., Boylan-Kolchin, M. 2010, MNRAS, 406, 2267



\bibitem[Gebhardt et al.(2000)]{gebhardt:2000}
Gebhardt, K., Richstone, D., Kormendy, J., et al. 2000, AJ, 119, 1157

\bibitem[Gebhardt et.al.(2003)]{gebhardt:2003}
Gebhardt, K., Faber, S.\ M.; Koo, D.\ C., et al. 2003, ApJ, 597, 239


\bibitem[Girardi et al.(2005)]{girardi:2005}
Girardi, M., Demarco, R., Rosati, P., \& Borgani, S. 2005, A\&A, 442, 29




\bibitem[Gonz\'{a}lez(1993)]{gonzalez:1993}
Gonz\'{a}lez, J.\ J. 1993, PhD thesis, Univ.\ California, Santa Cruz

\bibitem[Gorgas et al.(1999)]{gorgas:1999}
Gorgas J., Cardiel N., Pedraz S., \& Gonz\'{a}lez J.\ J. 1999, A\&AS, 139, 29


\bibitem[Haines et al.(2013)]{haines:2012}
Haines, C.\ P., Pereira, M.\ J., Sanderson, J.\ R., et al. 2012, ApJ, 754, 97

\bibitem[Haines et al.(2015)]{haines:2015}
Haines, C.\ P., Pereira, M.\ J., Smith, G.\ P., et al. 2015, ApJ, 806, 101

\bibitem[Harrison et al.(2011)]{harrison:2011}
Harrison, C., D., Colless, M., Kuntschner, H., et al. 2011, MNRAS, 413, 1036



\bibitem[Hook et al.(2004)]{hook:2004}
Hook, I.\ M., J{\o}rgensen, I. Allington-Smith, J.\ R., et al. 2004, PASP, 116, 425

\bibitem[Houghton et al.(2012)]{houghton:2012}
Houghton, R.\ C.\ W., Davies, R.\ L., Dalla Bont\`{a}, E., \& Masters, R. 2012, MNRAS, 423, 256





\bibitem[J{\o}rgensen(1997)]{IJ:1997}
J{\o}rgensen, I. 1997, MNRAS, 288, 161

\bibitem[J{\o}rgensen(1999)]{IJ:1999}
J{\o}rgensen, I. 1999, MNRAS, 306, 607

\bibitem[J{\o}rgensen(2009)]{IJ:2009}
J{\o}rgensen, I. 2009, PASA, 26, 17

\bibitem[J{\o}rgensen, Chiboucas (2013)]{IJ:2013}
J{\o}rgensen, I., \& Chiboucas, K. 2013, AJ, 145, 77

\bibitem[J{\o}rgensen et al.(1995a)]{IJ:1995a} 
J{\o}rgensen, I., Franx, M., \& Kj{\ae}rgaard, P. 1995a, MNRAS, 273, 1097

\bibitem[J{\o}rgensen et al.(1995b)]{IJ:1995b} 
J{\o}rgensen, I., Franx, M., \& Kj{\ae}rgaard, P. 1995b, MNRAS, 276, 1341


\bibitem[J{\o}rgensen et al.(2005)]{IJetal:2005}
J{\o}rgensen, I., Bergmann, M., Davies, R., Barr, J., Takamiya, M., \& Crampton, D. 2005, AJ, 129, 1249



\bibitem[J{\o}rgensen et al.(2014a)]{IJetal:2014a}
J{\o}rgensen, I., Chiboucas, K., Toft, S., Bergmann, M., Zirm, A., Schiavon, R., Gr\"{u}tzbauch, R. 2014, AJ, 148, 117


\bibitem[Kelson et al.(2006)]{kelson:2006}
Kelson, D.\ D., Illingworth, G.\ D., Franx, M., \& van Dokkum, P.\ G. 2006, ApJ, 653, 159




\bibitem[Kreisch et al.(2016)]{kreisch:2016}
Kreisch, C.\ D., Machacek, M.\ E., Jones, C., \& Randall, S.\ W. 2016, ApJ, 830, 39

\bibitem[Kriek et al.(2016)]{kriek:2016}
Kriek, M., Conroy, C., van Dokkum, P.\ G., et al. 2016, Nature, 540, 248



\bibitem[Kuntschner(2000)]{kuntschner:2000}
Kuntschner, H. 2000, MNRAS, 315, 184





\bibitem[Lemze et al.(2009)]{lemze:2009}
Lemze, D., Broadhurst, T., Rephaeli, Y, Barkana, R., \& Umetsu, K. 2009, AJ, 701, 1336

\bibitem[Lu et al.(2010)]{lu:2010}
Lu, T., Gilbank, D.\ G., Balogh, M.\ L., et al. 2010, MNRAS, 403, 1787

\bibitem[Mahdavi \& Geller(2001)]{mahdavi:2001}
Mahdavi, A., \& Geller, M.\ J. 2001, ApJL, 554, L129

\bibitem[Mahdavi et al.(2013)]{mahdavi:2013}
Mahdavi, A., Hoekstra, H., Babul, A., et al. 2013, ApJ, 767, 116

\bibitem[Mahdavi et al.(2014)]{mahdavi:2014}
Mahdavi, A., Hoekstra, H., Babul, A., et al. 2014, ApJ, 794, 175



\bibitem[Mateus et al.(2007)]{mateus:2007}
Mateus, A., Sodr\'{e} Jr., L., Fernandes, R.\ C., \& Stasi\'{n}ska, G. 2007, MNRAS, 374, 1457

\bibitem[Maughan et al.(2003)]{maughan:2003}
Maughan, B.\ J., Jones, L.\ R., Ebeling, H., et al. 2003, ApJ, 587, 589


\bibitem[McDermid et al.(2015)]{mcdermid:2015}
McDermid, R.\ M., Alatalo, K., Blitz, L., et al. 2015, MNRAS, 448, 3484


\bibitem[Monet et al.(1998)]{monet:1998}
Monet, D., et al. 1998, VizieR Online Data Catalog, 1252, \\
http://vizier.cfa.harvard.edu/viz-bin/VizieR?-source=I/252



\bibitem[Moran et al.(2007b)]{moran:2007}
Moran, S., Ellis, R.\ S., Treu, T., et al. 2007b, ApJ, 671, 1503

\bibitem[Morandi et al.(2011)]{morandi:2011}
Morandi, A., Pedersen, K., \& Limousin, M. 2011, ApJ, 729, 37

\bibitem[Muzzin et al.(2012)]{muzzin:2012}
Muzzin, A., Wilson, G., Yee, H.\ K.\ C., et al. 2012, ApJ, 746, 188

\bibitem[Nantais et al.(2013)]{nantais:2013}
Nantais, J.\ B., Rettura, A., Lidman, C., et al. 2013, A\&A, 556, 112






\bibitem[Pascut \& Ponman(2015)]{pascut:2015}
Pascut, A., \& Ponman, T.\ J. 2015, MNRAS, 447, 3723


\bibitem[Piffaretti et al.(2011)]{piffaretti:2011}
Piffaretti, R., Arnaud, M., Pratt, G.\ W., Pointecouteau, \& Melin, J.-B. 2011, A\&A, 534, A109



\bibitem[Ravindranath \& Ho(2002)]{ravindranath:2002}
Ravindranath, S., \& Ho, L.\ C. 2002, ApJ, 577, 133

\bibitem[Renzini(2006)]{renzini:2006}
Renzini, A. 2006, ARA\&A, 44, 141

\bibitem[Roediger et al.(2011)]{roediger:2011}
Roediger, J.\ C., Courteau, S., MacArthur, L.\ A., \& McDonald, M. 2011, MNRAS, 416, 1996


\bibitem[Saglia et al.(2010)]{saglia:2010}
Saglia, R.\ P., S\'{a}nchez-Bl\'{a}zquez, P., Bender, R., et al. 2010, A\&A, 524, A6

\bibitem[Salpeter(1955)]{salpeter:1955}
Salpeter, E.\ E. 1955, ApJ, 121, 161


\bibitem[S\'{a}nchez-Bl\'{a}zquez et al.(2009)]{sanchez:2009}
S\'{a}nchez-Bl\'{a}zquez, P., Jablonka, P., Noll, S., et al. 2009, A\&A, 499, 47


\bibitem[Sato \& Martin(2006)]{sato:2006}
Sato, T., \& Martin, C.\ L. 2006, ApJ, 647, 946

\bibitem[Schiavon(2007)]{schiavon:2007}
Schiavon, R.\ P. 2007, ApJS, 171, 146

\bibitem[Schindler et al.(1997)]{schindler:1997}
Schindler, S., Hattori, M, Neumann, D.\ M., \& Boehringer, H. 1997, A\&A, 317, 646

\bibitem[Schlafly \& Finkbeiner(2011)]{schlafly:2011}
Schlafly, E.\ J., \& Finkbeiner, D.\ P. 2011, ApJ, 737, 103


\bibitem[Serra \& Trager(2007)]{serra:2007}
Serra, P, \& Trager, S.\ C. 2007, MNRAS, 374, 769

\bibitem[Sersic(1968)]{sersic:1968}
S\'{e}rsic, J.\ L. 1968, Atlas de Galaxias Australes (Cordoba: Observatorio Astronomico)




\bibitem[Smith et al.(2005)]{smith:2005}
Smith, G.\ P., Treu, T., Ellis, R.\ S., Moran, S.\ M., \& Dressler, A. 2005, ApJ, 620, 78


\bibitem[Smith et al.(2006)]{smith:2006}
Smith, R.\ J., Hudson, M.\ J., Lucey, J.\ R., et al. 2006, MNRAS, 369, 1419

\bibitem[Springel et al.(2005)]{springel:2005}
Springel, V., White, S.\ D.\ M., Jenkins, A., et al. 2005, Nature, 435, 629






\bibitem[Thomas et al.(2005)]{thomas:2005}
Thomas, D., Maraston, C., Bender, R., \& de Oliviera, C.\ M. 2005, ApJ, 621, 673

\bibitem[Thomas et al.(2011)]{thomas:2011}
Thomas, D., Maraston, C., \& Johansson, J. 2011, MNRAS, 412, 2183
(http://www.icg.port.ac.uk/\~{ }thomasd)

\bibitem[Thomas et al.(2010)]{thomas:2010}
Thomas, D., Maraston, C., Schawinski, K., Sarzi, M., \& Silk, J. 2010, MNRAS, 404, 1775



\bibitem[Trager et al.(2000)]{trager:2000}
Trager, S.\ C., Faber, S.\ M., Worthey, G., Gonz\'{a}lez, J.\ J. 2000, AJ, 120, 165

\bibitem[Treu et al.(2005)]{treu:2005}
Treu, T., Ellis, R.\ S., Liao, T.\ X., van Dokkum, P.\ G., 2005, ApJ, 622, L5

\bibitem[Tripicco \& Bell(1995)]{tripicco:1995}
Tripicco, M.\ J., \& Bell, R.\ A. 1995, AJ, 110, 3035


\bibitem[Umetsu et al.(2015)]{umetsu:2015}
Umetsu, K., Sereno, M., Medezinski, E., et al. 2015, ApJ, 806, 207

\bibitem[van den Bosch(2002)]{vandenbosch:2002}
van den Bosch, F.\ C.\ 2002, MNRAS, 331, 98





\bibitem[van Dokkum \& Franx(2001)]{vandokkum:2001}
van Dokkum, P.\ G., \& Franx, M. 2001, ApJ, 553, 90

\bibitem[van Dokkum \& van der Marel(2007)]{vandokkum:2007}
van Dokkum, P.\ G., \& van der Marel, R.\ P. 2007, ApJ, 655, 30



\bibitem[Vazdekis et al.(2015)]{vazdekis:2015}
Vazdekis, A., Coelho, P., Cassisi, S., et al. 2015, MNRAS, 449, 1177 
(http://http://www.iac.es/proyecto/miles/)




\bibitem[Worthey(1994)]{worthey:1994}
Worthey, G. 1994, ApJS, 95, 107

\bibitem[Worthey et al.(1994)]{wortheyetal:1994}
Worthey, G., Faber, S.\ M., Gonz\'{a}lez, J.\ J., \& Burstein, D. 1994, ApJS, 94, 687

\bibitem[Worthey \& Ottaviani(1997)]{worthey:1997}
Worthey, G., \& Ottaviani, D.\ L. 1997, ApJS, 111, 377



\bibitem[Zabludoff et al.(1990)]{zabludoff:1990}
Zabludoff, A., Huchra, J.\ P., \& Geller, M.\ J. 1990, ApJS, 74, 1


 

\end{thebibliography}
\end{document}